\input{epsf}

\documentclass[journal]{IEEEtran}

\makeatletter
\def\ps@headings{%
\def\@oddhead{\mbox{}\scriptsize\rightmark \hfil \thepage}%
\def\@evenhead{\scriptsize\thepage \hfil \leftmark\mbox{}}%
\def\@oddfoot{}%
\def\@evenfoot{}}
\makeatother
\usepackage{epsf}
\usepackage{graphicx}
\usepackage[cmex10]{amsmath}
\usepackage{amsthm}
\usepackage{amssymb}
\usepackage{epsfig,latexsym,amsmath,epsf,amssymb,amsfonts}
\usepackage{algorithm, algorithmic}
\usepackage{placeins}
\usepackage{cite}
\usepackage{comment}
\usepackage{subfigure}
\usepackage{lipsum}
\usepackage{color}
\usepackage{csquotes}
\usepackage{tabularx}
\usepackage{multirow}
\usepackage{bbm}
\usepackage[american]{babel}

\usepackage{mathrsfs}
\usepackage{array}
\usepackage{setspace}
\usepackage{makecell}
\usepackage{url}

\usepackage{tikz}
\usetikzlibrary{trees,shapes,arrows, arrows.meta}

\renewcommand{\theequation}{\arabic{equation}}
\newtheorem{Definition}{\hskip 0pt Definition}

\begin{document}

\title{A Survey on Consensus Mechanisms and Mining Strategy Management in Blockchain Networks}

\author{
  \IEEEauthorblockN{Wenbo Wang,~\IEEEmembership{Member,~IEEE,}
  Dinh Thai Hoang,~\IEEEmembership{Member,~IEEE,}
  Peizhao Hu,~\IEEEmembership{Member,~IEEE,}
  Zehui Xiong,~\IEEEmembership{Student Member,~IEEE,}
  Dusit Niyato,~\IEEEmembership{Fellow,~IEEE,}
  Ping Wang,~\IEEEmembership{Senior Member,~IEEE}
  Yonggang Wen,~\IEEEmembership{Senior Member,~IEEE}
  and
  Dong In Kim,~\IEEEmembership{Fellow,~IEEE}
  \vspace*{-4mm}}
\vspace*{-4mm}

\thanks{Wenbo Wang, Zehui Xiong, Dusit Niyato and Yonggang Wen are with the School of Computer Science and Engineering, Nanyang Technological University, Singapore 639798
(email: wbwang@ntu.edu.sg, zxiong002@e.ntu.edu.sg, dniyato@ntu.edu.sg, ygwen@ntu.edu.sg).}
\thanks{Dinh Thai Hoang is with the Faculty of Engineering and Information Technology, University of Technology Sydney, NSW 2007, Australia (e-mail: hoang.dinh@uts.edu.au).}
\thanks{Peizhao Hu is with the Department of Computer Science, Rochester Institute of Technology, Rochester, NY, USA 14623 (email: ph@cs.rit.edu).}
\thanks{Ping Wang is with the Department of Electrical Engineering \& Computer Science, Lassonde School of Engineering, York University, 4700 Keele St., LAS 2016
Toronto, ON M3J 1P3, Canada (email: pingw@yorku.ca).}
\thanks{Dong In King is wit the School of Information and Communication Engineering, Sungkyunkwan University, Suwon 16419, Korea (e-mail: dikim@skku.ac.kr).}
}

\maketitle
\begin{abstract}
The past decade has witnessed the rapid evolution in blockchain technologies, which has attracted tremendous interests from both the research communities and industries. The blockchain network was originated from the Internet financial sector as a decentralized, immutable ledger system for transactional data ordering. Nowadays, it is envisioned as a powerful backbone/framework for decentralized data processing and data-driven self-organization in flat, open-access networks. In particular, the plausible characteristics of decentralization, immutability and self-organization are primarily owing to the unique decentralized consensus mechanisms introduced by blockchain networks. This survey is motivated by the lack of a comprehensive literature review on the development of decentralized consensus mechanisms in blockchain networks. In this survey, we provide a systematic vision of the organization of blockchain networks. By emphasizing the unique characteristics of incentivized consensus in blockchain networks, our in-depth review of the state-of-the-art consensus protocols is focused on both the perspective of distributed consensus system design and the perspective of incentive mechanism design. From a game-theoretic point of view, we also provide a thorough review on the strategy adoption for self-organization by the individual nodes in the blockchain backbone networks. Consequently, we provide a comprehensive survey on the emerging applications of the blockchain networks in a wide range of areas. We highlight our special interest in how the consensus mechanisms impact these applications. Finally, we discuss several open issues in the protocol design for blockchain consensus and the related potential research directions.
\end{abstract}


\section{Introduction}
\label{sec_introduction}
In the past decade, blockchain networks have gained tremendous popularity for their capabilities of distributively providing immutable ledgers as well as platforms for data-driven autonomous organization. {Proposed by the famous grassroot cryptocurrency project ``Bitcoin''~\cite{nakamoto2008bitcoin}, the blockchain network was originally adopted as the backbone of a public, distributed ledger system to process asset transactions in the form of digital tokens between Peer-to-Peer (P2P) users. Blockchain networks, especially those adopting open-access policies, are distinguished by their inherent characteristics of disintermediation, public accessibility of network functionalities (e.g., data transparency) and tamper-resilience~\cite{dinh2017untangling}.} Therefore, they have been hailed as the foundation of various spotlight FinTech applications that impose critical requirement on data security and integrity (e.g., cryptocurrencies~\cite{7423672,7163021}). Furthermore, with the distributed consensus provided by blockchain networks, blockchains are fundamental to orchestrating the global state machine\footnote{Distributed consensus orchestrates the states of replicated program execution on decentralized notes. It provides the runtime environment for distributively verifying the output of the same program. Therefore, the blockchain network is also known as a distributed Virtual Machine (VM) in the literature~\cite{7467408}.} for general-purpose bytecode execution. Therefore, blockchains are also envisaged as the backbone of the emerging open-access, trusted virtual computers~\cite{7546538} for decentralized, transaction-driven resource management in communication networks and distributed autonomous systems~\cite{8168250, 7467408}. For these reasons, blockchain technologies have been heralded by both the industry and academia as the fundamental ``game changer''~\cite{glaser2017pervasive} in decentralization of digital infrastructures ranging from the financial industry~\cite{7163021} to a broad domain including Internet of Things (IoTs)~\cite{kshetri2017can} and self-organized network orchestration~\cite{8242003}.

Generally, the term ``blockchain networks'' can be interpreted from two levels, namely, the ``blockchains'' which refer to a framework of immutable data organization, and the ``blockchain networks'' on top of which the approaches of data deployment and maintenance are defined. The two aspects are also considered as the major innovation of blockchain technologies. {For data organization, blockchain technologies employ a number of off-the-shelf cryptographic techniques~\cite{Merkle1988, mohr2007survey, goldreich2002zero} and cryptographically associate the users' on-chain identities with the transactions of their tokenized assets.} Thus, blockchains are able to provide the proofs of authentication for asset (i.e., token) transfer and then the proofs of asset ownerships. Furthermore, a blockchain maintains an arbitrary order of the transactional records by cryptographically chaining the record subsets in the form of data ``blocks''  to their chronic predecessors. With the help of cryptographic references, any attempt of data tampering can be immediately detected.
From the perspective of network organization, the problem of replicated agreement~\cite{raynal2010communication,Schneider:1990:IFS:98163.98167} on a single/canonical transaction history among trustless nodes is creatively tackled by the blockchain consensus protocols in an open-access, weakly synchronized network.
Blockchain consensus protocols are able to offer the agreement on the global blockchain-data state among a large number of trustless nodes with no identity authentication and low messaging overhead~\cite{bano2017consensus}. To achieve this, a number of blockchain networks, e.g., Bitcoin, choose to incorporate an incentive-based block creation process known as ``block mining'' in their protocols. With distributed consensus, the blockchain can be viewed as a universal memory of the blockchain network. Meanwhile, the blockchain network can be viewed as a virtual computer (i.e., distributed VM) comprised by every node therein.

With the rapid evolution in blockchain technologies, the demand for the higher-level quality of services by blockchain-based applications presents more critical challenges in designing blockchain protocols. Particularly, the performance of blockchain networks significantly relies on the performance of the adopted consensus mechanisms, e.g., in terms of data consistency, speed of consensus finality, robustness to arbitrarily behaving nodes (i.e., Byzantine nodes~\cite{Schneider:1990:IFS:98163.98167}) and network scalability. Compared with the classical Byzantine consensus protocols allowing very limited network scalability in distributed systems~\cite{Schneider:1990:IFS:98163.98167,Castro:2002:PBF:571637.571640}, most of the existing consensus protocols in open-access blockchain networks (e.g., Bitcoin) guarantee the better network scalability at the cost of limited processing throughput. Also, to achieve decentralized consensus among poorly synchronized, trustless nodes, a number of these protocols incur huge consumption of physical resources (e.g., computing power)~\cite{7423672}. Moreover, to ensure a high probability of consensus finality, the protocols may also impose high latency for transaction confirmation. Out of such concerns, a large volume of research has been conducted with the aim of improving the performance of the open-access blockchain consensus protocols in specific aspects. However, in spite of a few short surveys~\cite{bano2017consensus,Vukolic2016}, a comprehensive study on the development of these consensus protocols and the related problems is still missing. Especially, there is a lack of a concise overview on how such a development can be interpreted under a uniform framework and how it impacts the potential applications of blockchain networks.

During the past decade, the scope of blockchain networks has been expanded way further from tamper-evident distributed ledgers. However, due to the recent market frenzy about cryptocurrencies, most of the existing general reviews and surveys on blockchains emphasize narrowly the scenarios of using blockchain networks as the backbone technologies for cryptocurrencies, especially the market-dominant ones such as Bitcoin and Ethereum~\cite{dinh2017untangling, 7163021, 7423672, zheng2016blockchain, Vukolic2016, conti2017survey, Atzei2017, 7467408}. For example, the issues regarding the client (user)-side application (i.e., wallet), P2P network protocols, consensus mechanisms and user privacy in the scope of Bitcoin are discussed in~\cite{7163021, 7423672}. In~\cite{zheng2016blockchain}, a brief summary of the emerging blockchain-based applications ranging from finance to IoTs is provided. A systematic survey is conducted in~\cite{conti2017survey} with respect to the security in the Bitcoin network including the identified attacks on the consensus mechanisms and the privacy/anonymity issues of the Bitcoin clients. In~\cite{7467408, Atzei2017}, the special issues regarding the design, application and security of the smart contracts\footnote{A smart contract is a deterministic program stored as executable bytecode on the blockchain~\cite{7467408, Atzei2017}. Its replicas are independently executed in the local VMs/containers on some or all nodes in the network, where the same triggering transactions produce the same output on all the honest nodes.} are reviewed in the context of the Ethereum network. In~\cite{bano2017consensus, 8168250}, two brief surveys on consensus protocols in blockchain networks are provided.

The existing surveys on the fast-developing studies of blockchain technologies rarely provide a global view on the issues related to consensus protocols. Our work aims to fill this gap by providing a comprehensive survey on this specific topic. To distinguish our study from the existing works, we present our survey on blockchain networks from the perspective of consensus formation, especially in open-access\footnote{We consider the property of opens access to all network functionalities instead of only open-access blockchain data. Throughout the survey, we use the terms  ``opens-access'' and ``permissionless'' interchangeably.} P2P networks. In analogy to the distributed database, blockchain consensus is perceived as a process of collaborative state transitions among distributed nodes in the framework of blockchain-specified data organization. We emphasize that such a viewpoint brings the taxonomy of blockchain networks into a paradigm that is comparable to the classical problems of global state maintenance in distributed systems\cite{ghosh2014distributed}. Therefore, we are able to cast our analysis of blockchain networks into the context of classical fault-tolerant studies by focusing on the standard consensus properties in distributed systems (i.e., the Agreement-Validity-Termination properties~\cite[Chapter 13.1]{ghosh2014distributed}). We provide a uniform view of blockchain networks by presenting a number of implementation stacks and revealing the interconnection between different protocol components therein. We align our survey on blockchain consensus protocols with a uniform framework based on Zero-Knowledge (ZK) prover-verifier systems~\cite{mohr2007survey, goldreich2002zero} in Section~\ref{sec_consensus}. By focusing on the blockchain protocols for data organization, network organization, and consensus maintenance, our survey contributes in the following aspects:
\begin{itemize}
  \item [(1)] providing a brief overview on the data organization and network protocols of blockchain networks,
  \item [(2)] providing a generic paradigm for the consensus mechanisms using cryptographic techniques in open-access blockchain networks,
  \item [(3)] reviewing the studies on the behaviors of the rational (profit-driven) nodes in the consensus processes of blockchain networks,
  \item [(4)] providing an in-depth review on the research effort toward addressing the concerns (e.g., performance vs. scalability) for blockchain networks with different roadmaps of consensus protocol design, and
  \item [(5)] providing an outlook of the research in the emerging decentralized applications built on top of the consensus layer, which may not be limited to the framework of the prevalent blockchain technologies (cf. our discussion in Sections~\ref{sec_consensus}-\ref{sec_consensus_III}).
\end{itemize}

\begin{figure*}[t]
\centering     
\tikzstyle{atomic_block_basic}=[draw, fill=gray!30, text width=3.5cm, text centered, minimum height=1.8em]
\tikzstyle{atomic_block_type2}=[draw, fill=gray!30, text width=1.4cm, text centered, minimum height=1.8em]
\tikzstyle{atomic_block_type3}=[draw, fill=gray!30, text width=2.5cm, text centered, minimum height=1.8em]
\tikzstyle{atomic_block_type4}=[draw, fill=gray!30, text width=6cm, text centered, minimum height=1.8em]

\tikzstyle{atomic_block_active_phy_layer}=[draw, fill=gray!30, text width=3.5cm, text centered, minimum height=2.0em]

\begin{tikzpicture}[font=\scriptsize, scale=0.8, every node/.style={transform shape}]
  \pgfdeclarelayer{background}
  \pgfdeclarelayer{foreground}
  \pgfsetlayers{background,main,foreground}
  \definecolor{myblue}{RGB}{30,144,255} 
  \definecolor{mygreen}{RGB}{147,193,26} 
  \definecolor{myred}{RGB}{204,35,20} 
  \definecolor{mygray}{RGB}{128,128,128} 

    \node [scale=1] (transaction) at (-2, -6.3) [atomic_block_basic] {Atomic Data Record (i.e., Transactions)};
    \node [scale=1] (block) at (-2,-5.42) [atomic_block_basic] {Data Aggregation: Block (e.g., Blockchain~\cite{Garay2015}) vs. Transaction (e.g., IOTA Tangle~\cite{popov2016tangle})};
    \node [scale=1] (chain) at (-2, -4.3) [atomic_block_active_phy_layer] {Data Ordering: Linear (i.e., Linear List~\cite{Garay2015}) vs. Nonlinear (i.e., Tree~\cite{Sompolinsky2015} and Directed Acyclic Graph~\cite{sompolinsky2016spectre})};
    \node [scale=1] (storage) at (-2,-3.0) [atomic_block_basic] {Storage of Ledger Replica (Local Database)};

    \node [scale=1] (Data) at (0.5, -2.4) [text width=3.0cm] {Data Organization Protocols};

    \node [scale=1] (encryption) at (1.5,-4.0) [atomic_block_type2] {Asymmetric Encryption};
    \node [scale=1] (hash) at (1.5,-5.0) [atomic_block_type2] {Hash Function};
    \node [scale=1] (merkle) at (3.3,-5.0) [atomic_block_type2] {Merkle Tree~\cite{Merkle1988}};
    \node [scale=1] (homo) at (3.3,-4.0) [atomic_block_type2] {Homomorphic Encryption};
    \node [scale=1] (ZKP) at (2.05,-6) [draw, fill=gray!30, text width=3.0cm, text width=2.5cm, text centered, minimum height=1.8em] {Zero-Knowledge Proof~\cite{mohr2007survey, goldreich2002zero}};

    \node (Crypto) [text width=3.0cm] at (2.3,-3.3) {{Cryptographic Functionality Components:}};

    \node [scale=1] (OSI) at (6.5, -6.3) [atomic_block_type3] {Lower OSI Protocol Layers};
    \node [scale=1] (routing) at (6.5,-5.42) [atomic_block_type3] {Peer Discovery and Routing Protocols (e.g., Kademlia~\cite{Maymounkov2002})};
    \node [scale=1] (chain) at (6.5, -4.4) [atomic_block_type3] {Cryptographic Transport Protocols (e.g. Ethereum Wire Protocol~\cite{ehereumwire})};
    \node [scale=1] (p2p) at (6.5,-3.3) [atomic_block_type3] {Overlay P2P Protocols (e.g., Whisper~\cite{etheremwhisper}, Telehash~\cite{telehashprotocol}, JSON-RPC~\cite{json2012json}, etc)};

    \node [scale=1] (Data) at (7.15, -2.4) [text width=3.0cm] {Network Protocols};

    \draw [blue,thick,dashed] (-4.1,-1.8) -- (9.1, -1.8);
    \node [scale=1] (layer0) at (9.2, -2.4) [text width=2.0cm] {Data and Network Organization Layer (see Section~\ref{sec_preliminary})};
    \node [scale=1] (PBFT) at (-1.75, -1) [atomic_block_basic, fill=gray!30, text width=4.0cm] {Byzantine Fault-tolerant Replication Protocols (e.g., Practical BFT~\cite{Castro:2002:PBF:571637.571640} and Ripple~\cite{schwartz2014ripple}) and Hybrid Protocols};
    \node [scale=1] (nakamoto) at (2.4, -1) [atomic_block_basic, fill=gray!30] {Consensus Protocols with Proof of Concept (e.g., Proof of Work~\cite{Garay2015} and Proof of Stake~\cite{Bentov2016})};

    \node [scale=1] (incentive) at (6.5, -1) [atomic_block_type3, fill=gray!30] {Incentive Mechanisms (e.g., Rewards for Block Mining and Uncle Block Reference~\cite{7930224})};

    \node [scale=1] (consensus) at (0.9, -0.25) [text width=3.0cm] {Consensus Protocols};

    \draw [blue,thick,dashed] (-4.1, 0.3) -- (9.1, 0.3);
    \node [scale=1] (layer1) at (9.2, -0.3) [text width=2.0cm] {Consensus Layer (Core Layer, see Sections~\ref{sec_consensus}-\ref{sec_consensus_III})};
    \node [scale=1] (smart_contract) at (-1.2, 1) [atomic_block_type4, fill=gray!30, text width=5.0cm] {Smart Contract Executed in Distributed VMs (e.g., Ethereum VM~\cite{buterin2014ethereum})};
    \node [scale=1] (dstorage) at (4.9, 1) [atomic_block_type4, fill=gray!30, text width=5.0cm] {Service Provision by Distributed Consensus Nodes (e.g., Distributed Data Storage~\cite{Filecoin})};

    \node [scale=1] (DVM) at (2.4, 1.6) [text width=3.4cm] {Distributed Virtual Computers};

    \draw [blue,thick,dashed] (-4.1, 2.1) -- (9.1, 2.1);
    \node [scale=1] (layer2) at (9.2, 1.4) [text width=2.0cm] {Global State Machine Layer (Inter-Op APIs, See Section~\ref{sub_sec_outlook})};

    \node [scale=1] (cryptocurrency) at (-2.8, 2.8) [atomic_block_type2, text width=1.6cm, minimum height=2.2em] {Cryptocurrency};
    \node [scale=1] (dapp) at (-0.1, 2.8) [atomic_block_type2, minimum height=2.2em,] {DApp};
    \node [scale=1] (Webapp) at (3.0, 2.8) [atomic_block_type2, text width=2.5cm, minimum height=2.2em] {Distributed Intermediary for Service Provision};
    \node [scale=1] (market) at (6.5, 2.8) [atomic_block_type2, text width=2.0cm, minimum height=2.2em,] {Distributed Access Control (e.g.,~\cite{8315203})};

    \node [scale=1] (app) at (3.5, 3.4) [text width=4.0cm] {Applications};

    \node [scale=1] (layer3) at (9.2, 3.1) [text width=2.0cm] {Application Layer (See Section~\ref{sub_sec_outlook})};

    \begin{pgfonlayer}{background}
        \path (transaction.west |- storage.north)+(-0.2,0.4) node (a) {};
        \path (transaction.south -| homo.east)+(+0.4,-0.1) node (b) {};
        \path[fill=white!20,rounded corners, draw=black!50, dashed] (a) rectangle (b);

        \path (Crypto.north west |- Crypto.north)+(-0.2,0.2) node (a) {};
        \path (ZKP.south -| homo.east)+(+0.2,-0.15) node (b) {};
        \path[fill=white!10,rounded corners, draw=black!50]
            (a) rectangle (b);

        \path (p2p.north west |- p2p.north)+(-0.2,0.42) node (a) {};
        \path (OSI.south -| OSI.east)+(+0.2,-0.1) node (b) {};
        \path[fill=white!10,rounded corners, draw=black!50, dashed]
            (a) rectangle (b);

        \path (PBFT.west |- PBFT.north)+(-0.2,0.4) node (a) {};
        \path (PBFT.south -| nakamoto.east)+(+0.25,-0.1) node (b) {};
        \path[fill=white!20,rounded corners, draw=black!50, dashed] (a) rectangle (b);

        \path (smart_contract.west |- smart_contract.north)+(-0.25,0.4) node (a) {};
        \path (dstorage.south -| dstorage.east)+(+0.53,-0.1) node (b) {};
        \path[fill=white!20,rounded corners, draw=black!50, dashed] (a) rectangle (b);

        \path (cryptocurrency.west |- cryptocurrency.north)+(-0.35,0.4) node (a) {};
        \path (dapp.south -| market.east)+(+0.45,-0.1) node (b) {};
        \path[fill=white!20,rounded corners, draw=black!50, dashed] (a) rectangle (b);

        \draw [-latex, line width=.5mm, mygray,] (0.45,-3.1) -- (-0.1, -3.1);
        \draw [-latex, line width=.5mm, mygray,] (0.45,-5.9) -- (-0.1, -5.9);
        \draw [-latex, line width=.4mm, mygray,] (-2,-4.2) -- (-2, -3.3);

        \draw [-latex, line width=.3mm, mygray,] (4.8, -1) -- (4.5, -1);
        \draw [-latex, line width=.3mm, mygray,] (4.8, -1) -- (5.15, -1);


    \end{pgfonlayer}
\end{tikzpicture}

\caption{An overview of the blockchain network implementation stacks. The arrow direction indicates the influence on protocol component selection.}
\label{fig_blockchain_protocol_layer}
\end{figure*}
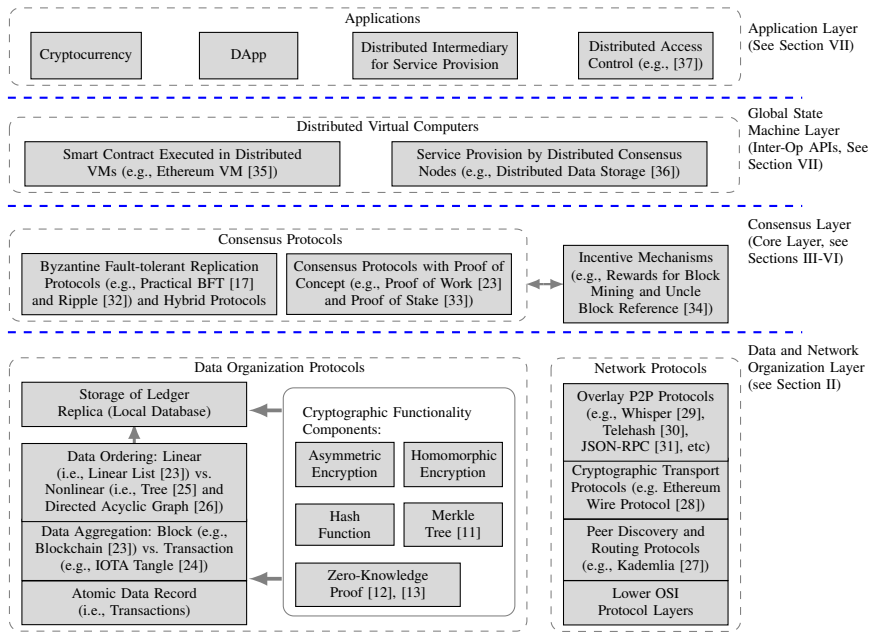

The rest of this survey is organized as follows. Section~\ref{sec_preliminary} provides an introductory overview on the protocol organization of blockchain networks. Section~\ref{sec_consensus} provides an in-depth survey on the popular approaches of consensus protocol design for open-access networks using linear blockchains. Consequently, Section~\ref{sec:AMS} provides a survey on the studies of the rational nodes' strategies in these consensus processes and their impact on the performance of blockchain networks. Section~\ref{sec_consensus_II} extends our survey on blockchain consensus protocols to the emerging fields including virtual block-mining (i.e., blockchain-extension) mechanism and hybrid consensus. Section~\ref{sec_consensus_III} briefly reviews the emerging cross-layer design regarding the data organization and consensus protocols, namely, the ``next-generation blockchains'' which may have different roadmaps for scalability and performance other than the prevalent blockchain paradigm. Section~\ref{sub_sec_outlook} provides a short review of the emerging applications of blockchains as well as an outlook of the potential research directions in the context of telecommunication networks. Section~\ref{sec_conclusion} concludes this survey by summarizing the contributions.

\section{Protocol Overview and Preliminaries}
\label{sec_preliminary}

\subsection{Overview of Blockchain Network Protocols}
The core task of a blockchain network is to ensure that the trustless nodes in the network reach the agreement upon a single tamper-proof record of transactions. The network is expected to tolerate a portion of the nodes deviating from this canonical record with their local views of data (i.e., replica).
From the perspective of system design, a blockchain network can be abstracted into four implementation levels. These levels are the protocols of data and network organization, the protocols of distributed consensus, the framework of autonomous organization relying on smart contracts~\cite{7467408} executed in distributed VMs and the implementation of human-machine interfaces (i.e., application). Following the approach of protocol layer definition in the Open Systems Interconnection (OSI) model, we provide in Figure~\ref{fig_blockchain_protocol_layer} an overview of these layers in blockchain networks and the related ingredient technologies.

The data organization protocols provide a number of ingredient cryptographic functionalities~\cite{Merkle1988, mohr2007survey, goldreich2002zero} to establish unique and secured node identities in a blockchain network. The protocols also define the approaches to form the cryptographic dependence among all the records, e.g., transaction records and account balances, in a local blockchain replica for ordering and tamper proof. From the perspective of data representation, the term ``blockchain'' is named as such partly for historical reason. In early networks such as Bitcoin~\cite{nakamoto2008bitcoin}, the digitally signed transactional records are arbitrarily ``packed up'' into a cryptographically tamper-evident data structure known as the ``block''. The blocks are then organized in a chronological order as a ``chain of blocks'', or more precisely, a linear list of blocks linked by tamper-evident hash pointers. Nevertheless, to improve the  processing efficiency, network scalability and security, the linear data organization framework has been expanded into the nonlinear forms such as trees and graphs of blocks~\cite{kiayias2016trees,sompolinsky2016spectre}. As in linear blockchains, the partial orders are also determined by the chaining direction between blocks. Furthermore, block-less, nonlinear data structures are also adopted in recent protocol design~\cite{popov2016tangle}. Despite the different forms of block organization, cryptographic data representation provides the fundamental protection of privacy and data integrity for blockchain networks. When compared with conventional database, it also provides more efficient on-chain storage without harming the data integrity.

On the other hand, the network protocols provide the means of P2P network organization, namely, peer/route discovery and maintenance as well as encrypted data transmission/synchronization over P2P links. Given reliable data synchronization over P2P connections, the consensus layer provides the core functionality to maintain the originality, consistency and order of the blockchain data across the network. From the perspective of distributed system design, the consensus protocols provide Byzantine agreement~\cite{Schneider:1990:IFS:98163.98167} in blockchain networks. More specifically, the nodes in the network expect to agree on a common update, i.e., consensus, of the blockchain state that they copy as the local replicas even in the presence of possible conflicting inputs and arbitrary faulty (Byzantine) behaviors of some nodes. When choosing the permissoned access-control schemes of network functionalities, blockchain networks usually adopt the well-studied Byzantine Faulty-Tolerant (BFT) consensus protocols such as Practical BFT (PBFT)~\cite{Castro:2002:PBF:571637.571640} for reaching the consensus among a small group of authenticated nodes (e.g., HyperLedger Fabric v0.5~\cite{cachin2016architecture}). On the contrary, in open-access/permissionless blockchain networks, probabilistic Byzantine agreement is achieved by combining a series of cryptographic techniques, e.g., cryptographic puzzle systems~\cite{goldreich2002zero, 10.1007/978-3-662-53890-6_30}, and incentive mechanism design. As pointed out in~\cite{Vukolic2016}, permissioned consensus protocols rely on a semi-centralized consensus framework and a higher messaging overhead to provide immediate consensus finality and thus high transaction processing throughput. In contrast, permissionless consensus protocols are more appropriate for a blockchain network with loose control on the synchronization and behaviors of the nodes, but may only guarantee probabilistic finality. In the condition of bounded delay and honest majority, permissionless consensus protocols provide significantly better support for network scalability at the cost of a lower processing efficiency.

Provided that the robustness of the consensus protocols is guaranteed, smart contracts are deployed on the distributed virtual computer layer. In brief, this layer abstracts away the details of data organization, information propagation and consensus formation in blockchain networks. As the interoperation layer between the lower-layer protocols and the applications, the virtual computer layer defines the high-level programming language implementation (e.g., Solidity in Ethereum~\cite{Atzei2017}) for encoding smart contracts. It also provides the sandboxed runtime environment (e.g., Ethreum VMs) to ensure the correct execution of
the replicated smart contracts on the network level. The virtual computer layer may adopt different levels of Turing-completeness for smart contract implementation, ranging from stateless circuits in Bitcoin~\cite{nakamoto2008bitcoin} to fully Turing-complete state machines in Ethereum~\cite{buterin2014ethereum} and HyperLedger Fabric~\cite{cachin2016architecture}. Full Turing-completeness enables blockchain networks to perform general-purpose computation in a replicated manner. For this reason, a blockchain network is able to not only provide the services of trusted data recording and timestamping, but also facilitate the functionalities of general-purpose autonomous organization. Therefore, blockchain networks are able to work as the backbone of autonomous organization systems for managing data or transaction-driven interactions among the decentralized entities in the network. On top of the virtual computer layer, the application layer provides the end-user-visible interfaces such as Distributed Applications (DApps)~\cite{8024034, 8260929} and cryptocurrencies.

\subsection{Cryptographic Data Organization}
\label{subsec_data_org}
When viewed as a data structure, a blockchain can be abstracted as an infinitely-growing, append-only string that is canonically agreed upon by the nodes in the blockchain network~\cite{Garay2015}. For data organization, the local blockchain replica of each node is organized in a hierarchical data structure of three levels, namely, the transactions, the blocks and the chain. Each level requires a different set of cryptographic functionalities for the protection of data integrity and authenticity.

\subsubsection{Transactions, Addresses and Signatures}
Transactions are the atomic data structure of a blockchain. Generally, a transaction is created by a set of users or autonomous objects (i.e., smart contracts) to indicate the transfer of tokens from the senders to the specified receivers. A transaction specifies a possibly empty list of inputs associating the token values with the identities (i.e., addresses) of the sending users/objects. It also specifies a nonempty list of outputs designating the redistribution result of the input tokens among the associated identities of the receivers. A transaction can be considered as a static record showing the identities of the senders and the receivers, the token value to be redistributed and the state of token reception. To protect the authenticity of a transaction record, the functionalities of cryptographic hashing and asymmetric encryption are activated:
\begin{itemize}
  \item \emph{Hash Function}: A cryptographic hash function maps at random an arbitrary-length binary input to a unique, fixed-length binary output (i.e., image). With a secure hash function (e.g., SHA-256), it is computationally infeasible to recover the input from the output image. Also, the probability to generate the same output for any two different inputs is negligible.
  \item \emph{Asymmetric Key}: Each node in the blockchain network generates a pair of private and public keys. The private key is associated with a digital signature function, which outputs a fixed-length signature string for any arbitrary-length input message. The public key is associated with a verification function, which takes as input the same message and the acclaimed signature for that message. The verification function only returns \emph{true} when the signature is generated by the signature function with the corresponding private key and the input message.
\end{itemize}
The nodes in the network or the autonomous objects identify themselves by revealing their public keys, namely, the hashcode of their public keys, as their permanent addresses (also known as their pseudo-identities) on the blockchain\footnote{Some cryptocurrency systems (e.g., Monero~\cite{mackenzie2015improving} and ZCash~\cite{hornby2016zcash}) incorporate cryptographic techniques such as one-time signature and group signature to create ephemeral addresses for enhancing anonymity.}. Since each input tuple in a transaction is signed by the associated sending account, the network is able to publicly validate the authenticity of the input through verifying the signature based on the sender's public address.

\subsubsection{Block Organization, Hash Pointer and Merkle Tree}
\begin{figure}[t]
\centering     
\includegraphics[width=.40\textwidth]{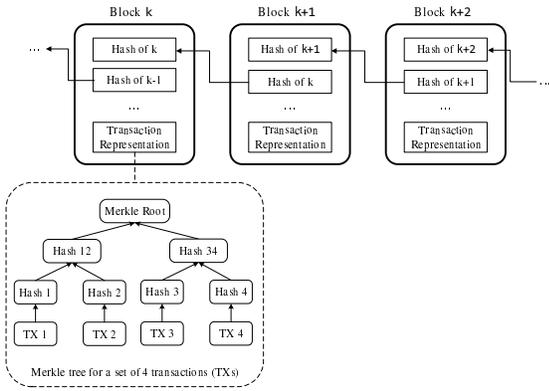}
\caption{Illustration of a chain of blocks, where the transactions in a single block is represented by a Merkle root.}
\label{fig_blockchain}
\end{figure}

A block is a container of an arbitrary subset of transaction records and can only be created by a node participating in the consensus process. To protect the integrity of the transaction records and to specify the ordering of adjacent blocks in a consensus node's local view, a data field known as the hash pointer is kept in the block's data structure. In addition, to reduce the on-chain storage, the cryptographic data structure of Merkle tree is also enabled to generate the tamper-evident digest in the transaction set of a block (see Figure~\ref{fig_blockchain}):
\begin{itemize}
  \item \emph{Hash pointer}: A hash pointer to a block is the hashcode of the concatenated data fields in that block. The hashcode of the current block is stored as the header of that block. The hashcodes of the reference blocks are stored as the hash pointers of a block to indicate that at the local view, the block recognizes that the transactions in the reference blocks are created earlier than those in the current block.
  \item \emph{Merkle Tree}~\cite{Merkle1988}: A Merkle tree represents a transaction set in the form of a binary tree. Therein, each leaf is labeled with the hashcode of a transaction and a non-leaf nodes is labeled with the hashcode of the concatenated labels of its two child nodes. The root node of the Merkle tree is known as the Merkle digest/root. A block storing only the Merkle root of the selected transactions is known to be in a lightweight form, which is sufficient for quick validation and synchronization. When using the lightweight-form storage, the node has to query its peers to retrieve the complete transaction records in the blocks.
\end{itemize}

In addition to the Merkle digest, block header and the hash pointers, a block may also contain auxiliary data fields, whose definition varies with the adopted protocol of block generation based on different consensus schemes. At a local view of the blockchain, the blocks are organized based on the hash pointers to their references/predecessors.  Every blockchain admits a unique block with no reference as the ``genesis block'', namely, the common ancestor block of all valid blocks in the chain. According to the number of hash pointers to the predecessors that are allowed to be kept by a block, the block organization can vary from a linear linked list to a tree of blocks (e.g., GHOST~\cite{Sompolinsky2015}) or a Directed Acyclic Graph (DAG) (e.g., SPECTRE~\cite{sompolinsky2016spectre}). Without specification, we limit most of our discussion on blockchains to the linear-list case, where the total order of the blocks is guaranteed (see Figure~\ref{fig_blockchain}).

\subsection{Blockchain Networks}\label{sub_sec_bc_system}
In a Byzantine environment, the identity management mechanism plays a key role in determining how the nodes in a blockchain network are organized. In an open-access (i.e., public/permissionless) blockchain network, a node can freely join the network and activate any available network functionalities.
Notice that the term ``node'' refers to a logical entity (i.e., the identity of a blockchain user) rather than to a physical device. For example, multiple ``nodes'' associated with different network functionalities can be hosted on the same physical machine. In alternative words, a physical device may appear in multiple identities in the network. Without any authentication scheme, the nodes are organized as overlay P2P networks. Comparatively, in a consortium (i.e., permissioned) blockchain network, only the authorized nodes are allowed to enable the core functionalities such as consensus participation or data propagation. The authorized nodes may be organized in different topologies, e.g., fully connected networks or P2P networks, according to the consensus protocols that the networks adopt. In this paper, we mainly focus on the network protocols in the permissionless cases.

In permissionless blockchain networks, the main goal of the network protocol is to induce a random topology among the nodes and propagate information efficiently for blockchain replica synchronization. Most of the existing blockchain networks employ the ready-to-use P2P protocols with slight modification for topology formation and data communication. For peer discovery and topology maintenance, the nodes in Bitcoin-like blockchain networks rely on querying a hard-coded set of volunteer DNS servers, which return a random set of bootstrapping nodes' IP addresses for the new nodes to initialize their peer lists~\cite{6688704, Biryukov:2014:DCB:2660267.2660379}. Nodes then request or advertise addresses based on these lists. In contrast, the Ethereum-like networks adopt a Kademlia-inspired protocol based on Distributed Hash Tables (DHTs)~\cite{Maymounkov2002} for peer/route discovery\footnote{Kademlia measures the node distance using XOR distance of the node addresses (hash values). The $k$-closest nodes are selected as neighbors.} through UDP connections. In blockchain networks, the connection of a node to a peer is managed based on reputation using a penalty score. A node will increase the penalty score of the peer sending malformed messages until the IP address of the faulty node is locally banned~\cite{Biryukov:2014:DCB:2660267.2660379, ehereumwire}.

To replicate the blockchain over all nodes in the network, the messages of transactions and blocks are ``broadcast'' through flooding the P2P links in a gossip-like manner. Typically, a P2P link in blockchain networks is built upon a persistent TCP connection after a protocol-level three-way handshake, which exchanges the replica state and the protocol/software version of each node~\cite{ehereumwire, gencer2018decentralization}. After the connections to the peer nodes are established, another three-way handshake occurs for a node to exchange new transactions/blocks with its neighbors. The node first notifies its peers with the hashcode of the new transactions/blocks that it receives or generates. Then, the peers reply with the data-transfer request specifying the hashcode of the information that they need. Upon request, the transfer of transactions/blocks is done via individual transfer messages\footnote{For example, the details of handshake and synchronization in the Ethereum network are defined in the DEVp2p Wire Protocol~\cite{ehereumwire}.}. The data transfer in blockchain networks is typically implemented based on the HTTP(s)-based Remote Procedure Call (RPC) protocol, where the messages are serialized following the JSON protocol~\cite{ehereumwire}.

\begin{figure}[t]
\centering     
\includegraphics[width=.38\textwidth]{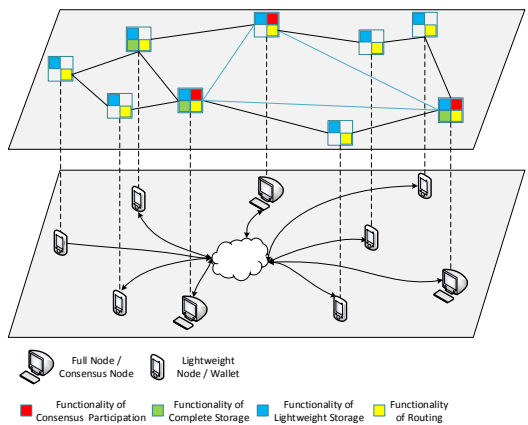}
\caption{Illustration of the nodes' roles in a permissionless blockchain network. The P2P links between consensus nodes are shown in blue.}
\label{fig_network}
\end{figure}

\begin{figure}[t]
\centering     
\includegraphics[width=.38\textwidth]{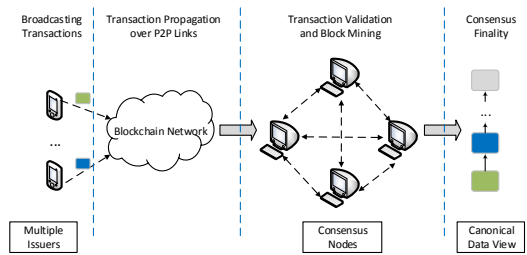}
\caption{The life cycle of blockchain transactions. Note that transaction validation and blockchain mining may happen at the same time with transaction propagation, depending on the consensus protocol adopted by the blockchain.}
\label{fig_transaction_life_cycle}
\end{figure}

An open-access blockchain network does not explicitly specify the role of each node. Nevertheless, according to the enabled functionalities, the nodes in the network can be categorized as the lightweight nodes, the full nodes and the consensus nodes~\cite{antonopoulos2014mastering}. Basically, all nodes are required to enable the routing functionality for message verification/propagation and connection maintenance. A lightweight node (e.g., wallets) only keeps the header of each block in its local storage. A full node stores locally a complete and up-to-date replica of the canonical blockchain. Compared with the lightweight nodes, a full node is able to autonomously verify the transactions without external reference. A consensus node enables the functionality of consensus participation. Therefore, it is able to publish new blocks and has a chance to influence the state of the canonical blockchain. A consensus node can adopt either complete storage or lightweight storage. In Figure~\ref{fig_network}, we present an example of different node types in a public blockchain network. Meanwhile, the lifecycle of a new transaction is shown in Figure~\ref{fig_transaction_life_cycle}. {It is worth noting that the consensus nodes are often referred to as the ``miners'' or ``mining nodes'' of blocks in the context of blockchain consensus formation, especially when token rewards of block proposal are involved.} Meanwhile, different roles of nodes lead to the inconsistency in their interests.  Namely, the transaction-issuing nodes (e.g., lightweight nodes) may not be the transaction-approving nodes (i.e., consensus nodes). For this reason, caution needs to be taken in protocol design to ensure that the consensus nodes act on behalf of the others in a trustless environment, especially on the consensus layer.

\subsection{Consensus in Blockchain Networks}\label{sub_sec_p2p_integritgy}
In the context of distributed system, the issue of maintaining the canonical blockchain state across the P2P network can be mapped as a fault-tolerant state-machine replication problem~\cite{raynal2010communication}. In other words, each consensus node maintains a local replicate (i.e., view) of the blockchain. An agreement (i.e., consensus) on the unique common view of the blockchain is expected to be achieved by the consensus nodes in the condition of Byzantine/arbitrary failures\footnote{See~\cite{Schneider:1990:IFS:98163.98167,Castro:2002:PBF:571637.571640} for the formal definition of Byzantine failures.}. In blockchain networks, Byzantine failures cause faulty nodes to exhibit arbitrary behaviors including malicious attacks/collusions (e.g., Sybil attacks~\cite{Douceur2002} and double-spending attacks~\cite{conti2017survey}), node mistakes (e.g., unexpected blockchain fork due to software inconsistency~\cite{buterin2013bitcoin}) and connection errors. We can roughly consider that the sequence of blocks represents the blockchain state, and the confirmation of a transaction incurs a blockchain state transition. According to~\cite{raynal2010communication, Cachin2001}, a blockchain updating protocol is said to achieve the (probabilistic) consensus (a.k.a. atomic broadcast\footnote{Here, the semantic of ``broadcast'' is consistent with that in the context of distributed system/database. Namely, a message is atomically broadcast when it is either received by every nonfaulty node, or by none at all.}~\cite{raynal2010communication,doi:10.1093/comjnl/bxh145,cachin2017blockchains}) in a Byzantine environment if the following properties are (probabilistically) satisfied~\cite{bano2017consensus}:
\begin{itemize}
  \item \emph{Validity (Correctness)}: If all the honest nodes activated on a common state propose to expand the blockchain by the same block, any honest node transiting to a new local replica state adopts the blockchain headed by that block.
  \item \emph{Agreement (Consistency)}: If an honest node confirms a new block header, then any honest node that updates its local blockchain view will update with that block header.
  \item \emph{Liveness (Termination)}: All transactions originated from the honest nodes will be eventually confirmed.
  \item \emph{Total order}: All honest nodes accept the same order of transactions as long as they are confirmed in their local blockchain views.
\end{itemize}

The consensus protocols vary with different blockchain networks. Since the permissioned blockchain networks admit tighter control on the synchronization among consensus nodes, they may adopt the conventional Byzantine Fault-Tolerant (BFT) protocols (c.f., the primitive algorithms described in~\cite{miller2014anonymous,sun2014solving}) to provide the required consensus properties. A typical implementation of such protocols can be found in the Ripple network~\cite{schwartz2014ripple}, where a group of synchronized Ripple servers perform blockchain expansion through a voting mechanism. Further, if an external oracle is introduced to designate the primary node for block generation (e.g., with HyperLedger Fabric v0.5~\cite{cachin2016architecture}), Practical BFT (PBFT)~\cite{Castro:2002:PBF:571637.571640} can be adopted to implement a three-phase commit scheme for blockchain expansion. In a network of $N$ consensus nodes, the BFT-based protocols are able to conditionally tolerate $\lfloor\frac{N-1}{5}\rfloor$ (e.g.,~\cite{schwartz2014ripple}) to $\lfloor\frac{N-1}{2}\rfloor$ (e.g.,~\cite{liu2016xft}) faulty nodes.

On the contrary, permissionless blockchain networks admit no identity authentication or explicit synchronization schemes. Therefore, the consensus protocol therein is expected to be well scalable and tolerant to pseudo identities and poor synchronization. Since any node is able to propose the state transition with its own candidate block for the blockchain header, the primary goal of the consensus protocol in permissionless networks is to ensure that every consensus node adheres to the ``longest chain rule''~\cite{7423672}. Namely, when the blocks are organized in a linked list, at any time instance, only the longest chain can be accepted as the canonical state of the blockchain. Due to the lack of identity authentication, the direct voting-based BFT protocols do not fit in permissionless blockchain networks. Instead, the incentive-based consensus schemes such as the Nakamoto consensus protocol~\cite{nakamoto2008bitcoin} are widely adopted.

\subsection{Nakamoto Consensus Protocol and Incentive Compatibility}\label{sub_sec_mining_intro}
To jointly address the problems of pseudonymity, scalability and poor synchronization, Nakamoto proposed in~\cite{nakamoto2008bitcoin} a permissionless consensus protocol based on a framework of cryptographic block-discovery racing game. This is also known as the Proof of Work (PoW) scheme~\cite{7423672, dinh2017untangling}. From a single node's perspective, the Nakamoto consensus protocol defines three major procedures, namely, the procedure of chain validation, the procedure of chain comparison and extension and the procedure of PoW solution searching~\cite{Garay2015}.
The chain validation predicate provides a Boolean judgment on whether a given chain of blocks has the valid structural properties. It checks if each block in the chain provides valid PoW solution and no conflict between transactions as well as the historical records exists. The function of chain comparison and extension compares the length of a set of chains, which may be either received from peer nodes or locally proposed. It guarantees that an honest node only adopts the longest proposal among the candidate views of the blockchain. The function of PoW solution searching is the main ``workhorse'' of the protocol and defines a cryptographic puzzle-solving procedure in a computation-intensive manner.

In brief, PoW solution requires exhaustively querying a cryptographic hash function for a partial preimage generated from a candidate block, whose hashcode satisfies a pre-defined condition. For simplicity of exposition, let $\mathcal{H}(\cdot)$ denote the hash function and $x$ denote the binary string assembled based on the candidate block data including the set of transactions (e.g., Merkle root), the reference hash pointers, etc. Then, we can formally define the PoW puzzle and solution as follows:
\begin{Definition}
  \label{def_puzzle_pow}
  Given an adjustable hardness condition parameter $h$, the process of PoW puzzle solution aims to search for a solution string, $nonce$, such that for a given string $x$ assembled based on the candidate block data, the hashcode (i.e, the target block header $bh$) of the concatenation of $x$ and $nonce$ is smaller than a target value $D(h)$:
  \begin{equation}
    \label{eq_puzzle_pow}
    bh = \mathcal{H}(x\Vert nonce)\le D(h),
  \end{equation}
  where for some fixed length of bits $L$, $D(h)=2^{L-h}$.
\end{Definition}

The Nakamoto protocol is computation-intensive since to win the puzzle solving race, a node needs to achieve a hash querying rate as high as possible. This property financially prevents the Sybil attacks of malicious nodes by merely creating multiple pseudo identities. On the other hand,  the economic cost (mainly electricity consumption) also renders it impractical for any node to voluntarily participate the consensus process at a consistent economic loss. To ensure proper functioning of a permissionless blockchain network, the Nakamoto protocol introduces incentives to probabilistically award the consensus participants based on an embedded mechanism of token supply and transaction tipping~\cite{nakamoto2008bitcoin}. From a game theoretic point of view, an implicit assumption adopted by the Nakamoto consensus protocol is that all the participant nodes are individually rational~\cite{nisan2007algorithmic}. In return, the consensus mechanism is expected to be \emph{incentive compatible}. In other words, the consensus protocol should ensure that any consensus node will suffer from finical loss whenever it deviates from truthfully following the protocol.

\begin{figure}[t]
\centering     
\includegraphics[width=.28\textwidth]{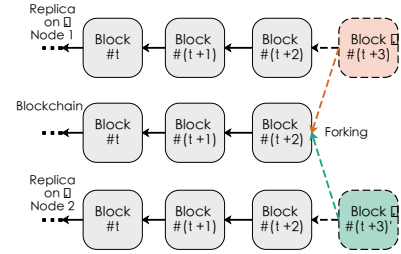}
\caption{A (temporary) fork happens at nodes 1 and 2 when their local PoW processes lead to different proposals of the new blockchain header, i.e., $(t+3)$ and $(t+3)'$ at the same time. Both $(t+3)$ and $(t+3)'$ satisfy (\ref{eq_puzzle_pow}).}
\label{fig_blockchain_forking}
\end{figure}
However, the incentive compatibility of the Nakamoto protocol has been openly questioned~\cite{Babaioff:2012:BRB:2229012.2229022, athey2016bitcoin, 6824541, Eyal2014}. Since the Nakamoto protocol allows nodes to propose arbitrary blocks from their local pending transaction set, it is inevitable for the network to experience blockchain expansion race with a (temporary) split, i.e., fork, in the local views of the blockchain state~\cite{7423672, conti2017survey} (see Figure~\ref{fig_blockchain_forking}). To guarantee the consensus properties and thus convergence to one canonical blockchain state, the Nakamoto protocol relies on the assumption that the majority of the consensus nodes follow the longest chain rule and are altruistic in information forwarding. It has been found in~\cite{Babaioff:2012:BRB:2229012.2229022, ersoy2017information} that rational consensus nodes may not have incentive for transaction/block propagation. As a result, the problem of blockchain forking may not be easily resolved in the current framework of the Nakamoto protocol. Special measures should be further taken in the protocol design, and a set of folklore principles has been suggested to gear the consensus mechanism towards a protocol for secured and sustainable blockchain networks~\cite{8306870, debus2017consensus,kroll2013economics,7163021}:
\begin{itemize}
  \item The consensus mechanism should enforce that propagating information and extending the longest chain of block are the monotonic strategies of the consensus nodes~\cite{kroll2013economics}. In other words, all the sub-stages in the consensus process should be incentive-compatible in an open environment with the tolerance to Byzantine and unfaithful faults.
  \item The consensus mechanism should encourage decentralization and fairness. Namely, it should not only discourage coalition, e.g., botnets and mining pools~\cite{rosenfeld2011analysis, Garay2015}, but also make the consensus process an uneasy prey of the adversaries with cumulated computation power.
  \item The consensus mechanism should strike a proper balance between processing throughput and network scalability~\cite{cachin2017blockchains, 8332496}.
\end{itemize}

\section{Distributed Consensus Mechanisms Based on Proof of Concepts}
\label{sec_consensus}
Based on the technical components of permissionless blockchain networks introduced in Section~\ref{sec_preliminary}, now we are ready to review the details about the designing methodologies of the consensus protocol for permissionless blockchains. In this section, we start by presenting the consensus protocols in the most prevalent blockchain networks in a uniform framework. Then, we explore the different approaches of extending/modifying the protocol to meet a series of specific performance requirement.

\subsection{Permissionless Consensus via Zero-Knowledge Proofs}
\label{sec_zkp_consensus}
\begin{figure}[t]
\centering     
\includegraphics[width=.38\textwidth]{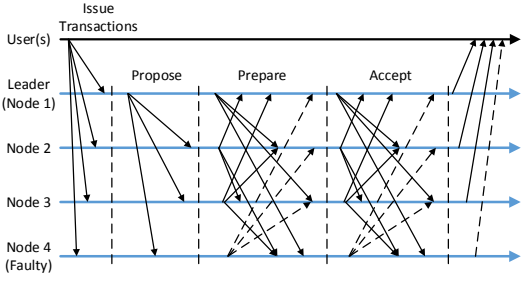}
\caption{BFT-based message pattern of three-way handshake in permissioned blockchains, e.g., Hyperledger Fabric using BFT-SMaRt~\cite{sousa2017byzantine}. The message is formed based on the granularity level of blocks, i.e., a batch of transactions.}
\label{fig_blockchain_PBFT}
\end{figure}

For traditional BFT consensus protocols, e.g., Byzantine Paxos~\cite{cachin2009yet} and PBFT~\cite{Castro:2002:PBF:571637.571640}, it is generally necessary to assume a fully connected topology among the consensus nodes as well as a leader-peer hierarchy for block proposal. The BFT consensus process is organized explicitly in rounds of three-way handshakes, thus synchronization between nodes with bounded execution time and message latency is also required. As illustrated in Figure~\ref{fig_blockchain_PBFT}, only the leader is responsible for proposing new blocks to a consortium of peer nodes at the proposal (pre-prepare) phase. This is followed by two all-to-all messaging phases, where a peer node only accepts the proposal (i.e., commit) when it receives more than a certain number of proposal approvals from the other peers (e.g., $\lfloor\frac{n+f+1}{3}\rfloor$ with PBFT for a network of $n$ honest nodes and $f$ Byzantine nodes). These classical state-machine replication approaches guarantee the properties of deterministic agreement and liveness in Byzantine environment, and are well-known for their low processing latency~\cite{Vukolic2016}. However, the characteristics of leader-peer hierarchy and high communication complexity in $\Theta(n^2)$~\cite{cachin2009yet} naturally require the BFT-based blockchain consensus protocols to be implemented in a small-scale permissioned network with centralized admission control. In order to achieve full decentralization and high consensus scalability, alternative approaches such as Nakamoto protocols become critical in the design of blockchain's consensus layer.

According to our discussion in Section~\ref{sub_sec_mining_intro}, the primary functionality of PoW in the Nakamoto protocol is to simulate the leader election in the traditional BFT protocols. The PoW process abstracted by Definition~\ref{def_puzzle_pow} is essentially a verifiable process of weighted random coin-tossing, where the probability of winning is no longer uniformly associated with the nodes' identities but in proportion to the resources, e.g., hashrate casted by the nodes. Then, we can consider that each new block is generated by a time-independent ``lottery'', where the probability of being elected as the leader for block proposal depends on the ratio between the casted resource of a node (or a node coalition) and the total resources presented in the entire network. Let $w_i$ denote the resource held by node $i$ in a network of node set $\mathcal{N}$, then, the probability of node $i$ winning the leader-election in a PoW-like process should follow:
\begin{equation}
    \label{eq_winning_prob}
    \Pr\nolimits_i^{\textrm{win}}=\frac{w_i}{\sum_{j\in\mathcal{N}}w_j},
\end{equation}
where $w_i$ generalizes the share of any verifiable resource such as computational power~\cite{nakamoto2008bitcoin}, memory~\cite{hornby2016zcash}, storage~\cite{Kopp2016}, etc. In contrast to the BFT protocols, the peer nodes accept the received block proposal following the longest-chain-rule after they verify the validity of the block and the transactions therein. Since no all-to-all messaging phase is needed, the Nakamoto protocol may have a much smaller message complexity $\Omega(n)$ when the majority of the peers are honest~\cite{miller2014anonymous}.

As the core component of the Nakamoto protocol, the PoW scheme originates from the idea of indirectly validating nodes' identities in pseudonymous P2P networks through an identity pricing mechanism~\cite{Jakobsson1999, aspnes2005exposing}. More specifically, the PoW scheme described by Definition~\ref{def_puzzle_pow} is originally designed to measure the voting power or the trustworthiness of a node according to the constrained resources presented by the node in the P2P network. Thus, the tolerable fraction of Byzantine nodes in BFT protocols is replaced by a limited fraction of the total computational power of the network~\cite{aspnes2005exposing}. Compared with the original design, the PoW scheme in blockchain networks is no longer used for direct identity verification between peers. Instead, the PoW processes of all the nodes in a blockchain network are expected to collectively simulate a publicly verifiable random function to elect the leader of block proposal following the distribution given by (\ref{eq_winning_prob}). Based on such a design paradigm, PoW can be generalized into the framework of Proof-of-Concepts (PoX) (cf.~\cite{7423672}). With PoX, the nodes in the network are required to non-interactively prove the possession or commitment of certain measurable resources beyond hashrates in PoW. Furthermore, their collective behavior should also yield a stochastic process for leader assignment following the distribution given in (\ref{eq_winning_prob}).

From a network-level perspective, PoX generally relies on a pseudorandom oracle to provide the property of verifiable unpredictability. It also needs to implement a one-way cryptographic puzzle for the proof of resource devoting in the framework of non-interactive ZK Proofs (ZKPs). A conventional ZKP system consists of two parties, namely, the prover executing a computationally unbounded strategy to generate the proof of an assertion without releasing it and the verifier executing a probabilistic polynomial-time strategy to verify it. A party is non-interactive when it can only choose between publishing messages to the network and remaining passive. Otherwise it is interactive. In the context of blockchain consensus protocols, the ZKP framework is extended from proving a private input (i.e., knowledge) to proving possession/consumption of a minimum amount of resource (e.g., computational work). Recent studies haven shown that with specific puzzle design, proof of knowledge and proof of work can be incorporate into a single framework of indistinguishable Proofs of Work or Knowledge (PoWorK)~\cite{10.1007/978-3-662-53890-6_30}, where the prover of work makes calls to a certain puzzle solving algorithm instead of sampling from a non-polynomial language witness relation distribution. In general, the adopted puzzle has to satisfy the basic soundness and completeness properties~\cite{mohr2007survey, goldreich2002zero}. Namely, an invalid proof should always be rejected by nonfaulty verifying nodes while a valid proof should always be accepted by nonfaulty verifiers. A complexity gap is expected such that the puzzle is easy to verify (in polynomial-time) but (moderately) hard for adversaries to invert/solve~\cite{Alwen2017}. Furthermore, in permissionless blockchain networks, any node is able to publish arbitrary block proposals. In this situation, a 3-step interactive prover-verifier ZK scheme with verifier-designated challenges will lead to excessive message overhead. This is the critical reason for requiring a non-interactive puzzle design. Following the generation-computation-verification paradigm of non-interactive puzzles (cf. the verifiable random function defined in~\cite{814584}), we can abstract a PoX process into the three stages described in Table~\ref{table_pox_stage}.
\begin{table}[ht]
  \centering
  \scriptsize
  \caption{Three-stage Abstraction of a PoX Process}
  \renewcommand{\arraystretch}{1.3}
 \begin{tabular} {|p{2.4cm} |p{5.4cm} |}
 \hline
 Initialization (generator of random seed or keys) & The \emph{initialization stage} provides the prover and the verifier the necessary information to run in subsequent stages according to the PoX specifications. Typical non-interactive ZKP systems, e.g., zk-SNARK~\cite{Ben-Sasson2013} have to query a trusted third-party key/random seed generation protocol to produce a common reference string for both the prover and the verifier.\\
 \hline
 Execution (challenge and proof generator) & For non-interactive ZKP, the \emph{execution stage} requires the prover to generate according to the common reference string a random challenge that constitutes a self-contained, uncompromisable computational problem, namely, the puzzle. Meanwhile, a corresponding proof (a.k.a. witness or puzzle solution) is also generated.\\
 \hline
 Verification & In the \emph{verification stage}, a verifier checks about the proof's correctness, which is determined solely based on the information issued by the prover. \\
 \hline
\end{tabular}
\label{table_pox_stage}
\end{table}

With the paradigm of PoX described above, we are now ready to investigate the puzzle design problem for different PoX schemes, which can be seen as modification or extension to the existing PoW-based Nakamoto protocol (see~\cite{king2012ppcoin, Kiayias2017, Filecoin, Bentov:2014:PAE:2695533.2695545, Milutinovic:2016:PLE:3007788.3007790} for examples). Since a trusted third party does not exist in a permissionless blockchain network, special caution should be taken in the puzzle design such that the freshness of the puzzle is guaranteed at the execution stage. Namely, the puzzle solution is unpredictable and the proof is non-reusable. Theoretical analyses of blockchain networks, e.g.,~\cite{Kiayias2017} may assume such a property on the condition that the network has access to a universal random sampler (a.k.a., random oracle) or an ideal randomness beacon\footnote{The concept of random beacon service is first proposed in~\cite{RABIN1983256}, where a trusted third party periodically emits random integers to the public.}.
Nevertheless, due to full decentralization of the permissionless blockchain networks, a case-by-case study for different PoX schemes is usually needed for practical implementation of the random oracle in order to prevent puzzle grinding and leader election manipulation. Apart from the aforementioned properties of non-invertibility, completeness, soundness and freshness, the other requirements for puzzle design in PoX may include but are not limited to the following:
\begin{itemize}
  \item The puzzle should be resistant to the aggregation~\cite{biryukov2017equihash} or outsourcing~\cite{Miller:2015:NSP:2810103.2813621} of the computational resources.
  \item The puzzle-solving process should be eco-friendly~\cite{Bentov2016,king2012ppcoin,moran2017rational,Milutinovic:2016:PLE:3007788.3007790,Bentov:2014:PAE:2695533.2695545}.
  \item In addition to providing incentive based on resource pricing mechanism, the puzzle-solving process should provide useful services in the meanwhile~\cite{Filecoin,ball2017proofs}.
\end{itemize}

\subsection{Nakamoto Protocol Based on Primitive Proof of Work}\label{sub_sec_POW}
As we have reviewed in the previous discussion, the primitive PoW scheme proposed in~\cite{nakamoto2008bitcoin} works to financially disincentivize the Sybil attacks on block proposal and maintains a biased random leader election process in proportion to the hashrate casted by each node. Recall that the input string $x$ to the PoW puzzle is a concatenation of the previous block's hash pointer and the payload data of the proposed block. For the puzzle design of PoW, the reason of choosing the hash function $\mathcal{H}(\cdot)$ in (\ref{eq_puzzle_pow}), e.g., SHA-256 in practice lies in the fact that a hash function is computationally indistinguishable from a pseudorandom function, if it preserves the properties of collision resistance\footnote{The collision probability of $\mathcal{H}(\cdot)$ is $e^{-\Omega(L)}$ and thus negligible~\cite{Garay2015}.} and pre-image resistance~\cite{al2010cryptographic}. Since the random output of $\mathcal{H}(\cdot)$ is time-independent and only determined by the input string, it plays the role of an uncompromisable random oracle and outputs a unique, unpredictable result every time when it is queried with a different $x$~\cite{garay2017proofs}. This means that a node in the blockchain network is able to construct a fresh random challenge solely based on its block proposal without referring to any designated verifier or third-party initializer. Meanwhile, it is well-known that with a proper cryptographic hash function, the search for a preimage $(x, nonce)$ satisfying the condition $\mathcal{H}(x\Vert nonce)\!\le\!2^{L-h}$ in (\ref{eq_puzzle_pow}) cannot be more efficient than exhaustively querying the random oracle for all $nonce\!\in\![0, 2^L]$. This leads to a puzzle time complexity of $\mathcal{O}(2^h)$~\cite{debus2017consensus}. On the other hand, verifying the puzzle only requires a single hash query. Therefore, the properties of non-invertibility, completeness, soundness and freshness are all satisfied by the PoW puzzle given by Definition~\ref{def_puzzle_pow}.

For a given difficulty level $D(h)$ in (\ref{eq_puzzle_pow}), each single query to $\mathcal{H}(\cdot)$ is an i.i.d. Bernoulli trial with a success probability
\begin{equation}
  \label{eq_prob_success}
  \Pr\left(y:\mathcal{H}(x\Vert y)\le D(h)\right)=2^{-h}.
\end{equation}
We adopt the typical assumption of loosely network synchronization for analyzing PoW-based blockchains~\cite{garay2017proofs, Garay2015}. Namely, all messages are delivered with bounded delay in one round. Then, (\ref{eq_prob_success}) indicates that the frequency for a node to obtain the puzzle solutions during a certain number of loosely synchronized rounds is a Bernoulli process. Since the probability given in (\ref{eq_prob_success}) is negligible for a sufficiently large $h$ with cryptographic hash functions $\mathcal{H}(\cdot)$, the Bernoulli process of node $i$ converges to a Poisson process as the time interval between queries/trails shrinks~\cite{miller2014anonymous}.

To analyze the PoW scheme, let $w_i$ in (\ref{eq_winning_prob}) refer to the number of queries that node $i$ can make to $\mathcal{H}(\cdot)$ in a single round. Then, we can approximate the rate of the Poisson process for node $i$'s puzzle solution by $\lambda_i\!=\!w_i/2^h$~\cite{Kraft2016}. Note that every node in the network is running an independent puzzle-solving process. Since a combination of $N$ independent Poisson processes is still a Poisson process, then, the collective PoW process of a network with $N$ nodes has a rate
\begin{equation}
    \label{eq_poisson_rate}
    \lambda=\sum_{i=1}^{N}\lambda_i=\frac{\sum_{i=1}^{N}w_i}{2^h}.
\end{equation}
The property of the combined Poisson processes in (\ref{eq_poisson_rate}) leads to the probability distribution for leader election in (\ref{eq_winning_prob}). From a single node's perspective, the repeated PoW puzzle-solving processes take the form of a block-proposal competition across the network. From the perspective of the network, for a given difficulty level $D(h)$, this puzzle-solving race simulates a verifiable random function for leader election and guarantees to follow the distribution in (\ref{eq_winning_prob}). Most importantly, it tolerates any fraction of the Byzantine nodes
in the network.

Nevertheless, the PoW by itself cannot guarantee any of the principle Byzantine consensus properties as described in Section~\ref{sub_sec_p2p_integritgy}. On top of the designed PoW puzzle and the P2P information diffusion functionality, three external functions are abstracted in~\cite{Garay2015} to describe the Nakamoto consensus protocol from a single node's perspective. These functions are
\begin{enumerate}
\item the \emph{chain reading function} that receives as input a blockchain and outputs an interpretation for later use;
\item the \emph{content validation function} that validates a blockchain replica and checks the data consistency with the applications (e.g., Bitcoin) on top of the blockchain;
\item the \emph{input contribution function} that compares the local and the received views of the blockchain and adopts the ``best'' one following the rule of longest chain.
\end{enumerate}
The input contribution function realizes the puzzle execution stage and the content validation function realizes the puzzle verification stage in Table~\ref{table_pox_stage}. Due to the independent Poisson processes in the block-proposal competition, more than one node may propose to extend the blockchain using different blocks with corresponding valid PoW solutions at the same time. As a result, the nodes may read from the network multiple valid views of the blockchain and choose different forks as their ``best'' local views (see also Figure~\ref{fig_blockchain_forking}). Theoretically, it has been shown in~\cite{7756226} that deterministic consensus in permissionless blockchain networks cannot be guaranteed unless all non-faulty nodes are reachable from one to another and the number of consensus nodes is known. For this reason, in~\cite{Garay2015,garay2017proofs, 10.1007/978-3-319-63688-7_10}, Garay et al. propose to capture the properties of validity, agreement and liveness of the Nakamoto consensus protocol by the three chain-based properties in Table~\ref{table_pox_properties}. Then, the PoW-based Nakamoto protocol can be modeled as a probabilistic Byzantine agreement protocol.
\begin{table}[ht]
  \centering
  \scriptsize
  \caption{Three Properties of Nakamoto Protocols for Blockchains}
  \renewcommand{\arraystretch}{1.3}
 \begin{tabular} {|p{1.2cm} |p{1.3cm}|p{5.0cm} |}
 \hline
  Nakamoto Protocol-Specified Properties & Corresponding Properties of Byzantine Agreement & Explanation in Details\\
 \hline
 Common-prefix property & Agreement (and permanent order) & In the condition of multiple local blockchain views due to forking, the \emph{common-prefix property} indicates that after cutting off (pruning) a certain number of block from the end (header) of the local chain, an honest node will always obtain a sub-chain that is a prefix of another honest node's local view of the blockchain.\\
 \hline
 Chain-quality property & Validity & Among a given length of consequent blocks in the local blockchain view of an honest node, the number of blocks that is proposed by Byzantine nodes (adversaries) is upper-bounded.\\
 \hline
 Chain-growth property & Liveness & For any given rounds of block proposals, the number of blocks appended to the local view of any honest node is lower-bounded.\\
 \hline
\end{tabular}
\label{table_pox_properties}
\end{table}

In order to quantify the Byzantine agreement properties for blockchains, three conditions, i.e., the upper-bounded information diffusion delay, a ``flat network'' with equal and limited hashrates and the upper-bounded number of Byzantine nodes are assumed in~\cite{garay2017proofs, Garay2015, 10.1007/978-3-319-63688-7_10}. It is shown in~\cite{Garay2015} that the three properties in Table~\ref{table_pox_properties} are quantified by three parameters, namely, the collective hashrates of the honest nodes, the hashrate controlled by the adversaries and the expected block arrival rate of the network-level Poisson process given in (\ref{eq_poisson_rate}). It has been further proved in~\cite{Garay2015} that under the condition of honest majority, the basic properties of validity and agreement are satisfied by the Nakamoto protocol with overwhelming probability. Furthermore, the common-prefix property and the chain-growth property formalize the presumption in~\cite{nakamoto2008bitcoin} that a transaction is secured when a sufficient length of subsequent blocks is appended to the chain. In other words, when a block is a certain number of blocks deep from the end of the chain, or equivalently, the repeated block-proposal competition has passed sufficiently many rounds, the transaction data in that block is non-reversible/persistent and thus guaranteed to be double-spending proof. It is worth noting that the studies in~\cite{Garay2015, 10.1007/978-3-319-63688-7_10} provide a generalizable approach for evaluating the security and the efficiency of the PoX-based Nakamoto protocols in permissionless blockchains. Based on the quantitative analysis of the properties in Table~\ref{table_pox_properties}, the same framework of security evaluation has been adopted by the studies in consensus protocols using other types of puzzle design such as Proof of Stakes (PoS)~\cite{Kiayias2017, cryptoeprint:2017:656}.

\begin{figure}[t]
\centering     
\subfigure[]{\label{fig_hashrate}\includegraphics[width=.33\textwidth]{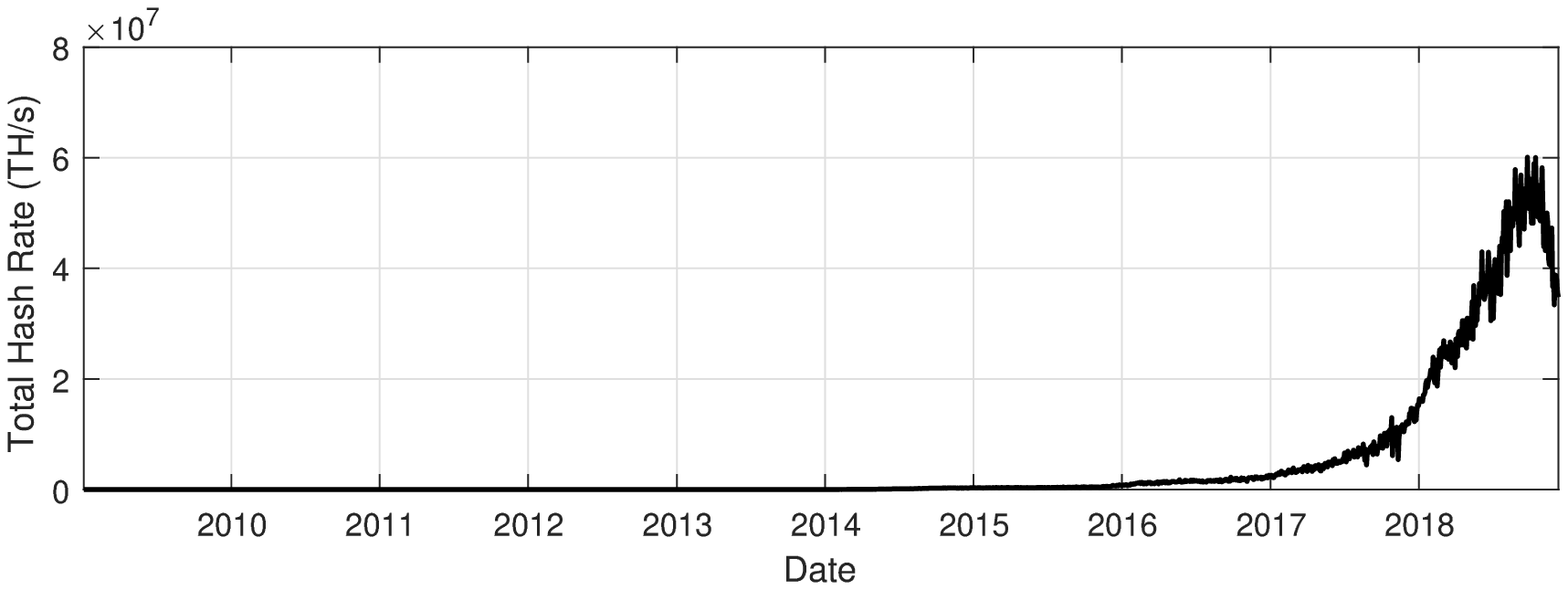}}
\subfigure[]{\label{fig_difficulty}\includegraphics[width=.33\textwidth]{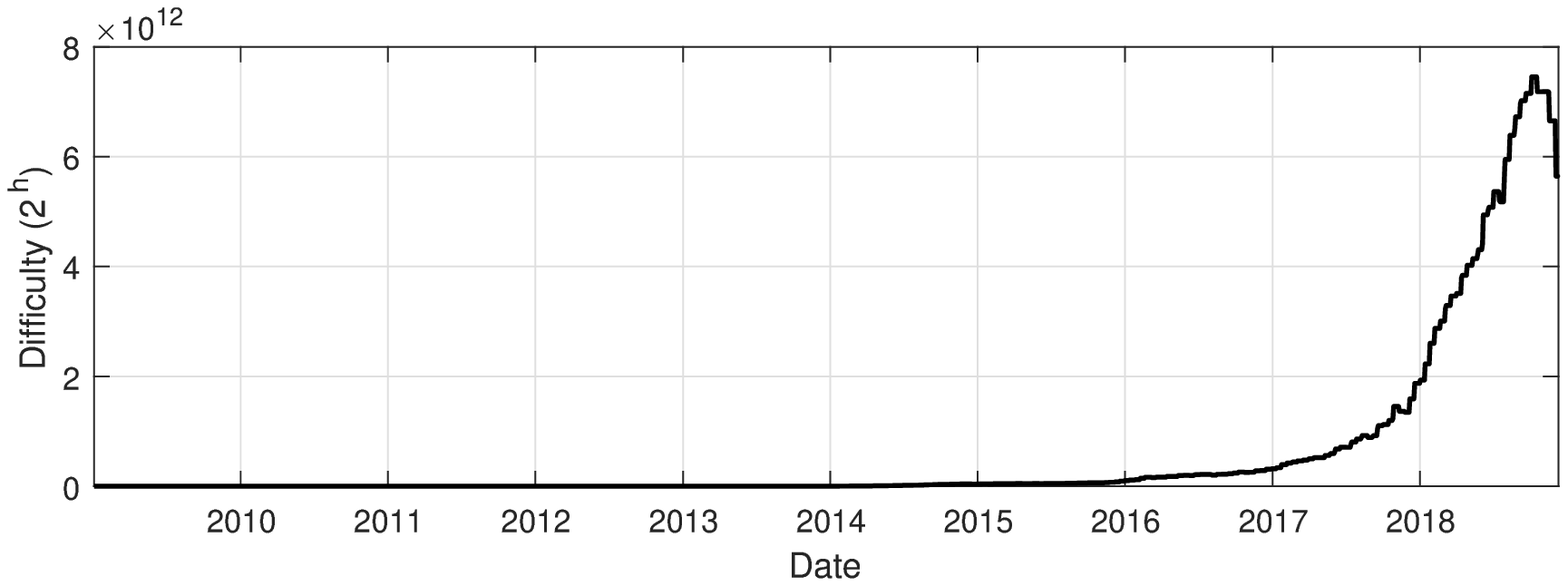}}
\caption{Evolution of (a) the total hash rate and (b) the PoW puzzle difficulty in the Bitcoin network over time. Data source: {https://www.blockchain.com}.}
\label{fig_Bitcoin_statics}
\end{figure}

Due to the open access nature of permissionless blockchains, the hashrate presented in a practical blockchain network is generally unstable. As indicated by Figure~\ref{fig_Bitcoin_statics}, since the introduction of the Application Specific Integrated Circuit (ASIC) for hash acceleration in 2013, the practical PoW-based blockchain networks, e.g., Bitcoin, have experienced an explosive increase of the total hashrate with huge fluctuation~\cite{8048662}. Practically, blockchain networks adopt a heuristic, periodic difficulty-adjustment policy to maintain a roughly fixed time interval, i.e., $\lambda^{-1}$ in (\ref{eq_poisson_rate}), between two neighbor blocks. However, the expected value of $\lambda^{-1}$ is usually chosen in an arbitrary manner and is frequently reduced in favor of a higher transaction throughput (see Litecoin~\cite{coinmarketcap} and ZCash~\cite{hornby2016zcash} for example). Following the assumption of partial synchronization~\cite{Garay2015}, the roughly fixed time interval indeed implies an upper bound for the information dissemination latency in the P2P network~\cite{kiayias2015speed}.

With such a consideration in mind, a theoretical study is provided in~\cite{pass2017analysis} between the upper bound of the information latency and the persistence of the block data in a node's local view of the blockchain. Consider a flat network of $N$ nodes with a maximum block propagation delay of $T$. It is found in~\cite{pass2017analysis} that for a given fraction of adversary node $\rho$ ($0\!\le\!\rho\!<\!0.5 $), the block generation probability for each node should satisfy the following condition in order to ensure the property of data persistence (Theorem 1.1 in~\cite{pass2017analysis}):
\begin{equation}
  \label{eq_persistence}
  \Pr\nolimits^g_i\le\frac{1}{T\rho\sum_{i=1}^Nw_i},
\end{equation}
where $\Pr\nolimits^g_i$ can be calculated based on (\ref{eq_prob_success}) and a given hashrate.

Furthermore, the block interval rules the trade-off between security and efficiency. The formal refers to the degree of fulfillment (i.e., the probabilistic consistency) of the Byzantine agreement properties, whereas the latter refers to the transaction throughput, which can be measured in the number of confirmed transactions per second. In~\cite{kiayias2015speed, 6688704}, examination on the block propagation delay $T$ in (\ref{eq_persistence}) shows that a safe upper bound on $T$ is jointly determined by the block size, the network scale measured in hop counts, and the average round-trip time of the links. The empirical study in~\cite{6688704} reveals that for small-size blocks, e.g., less than 20kB for Bitcoin, the round-trip delay is the dominant factor of the block propagation delay. Otherwise, transaction validation time becomes the major factor of the block propagation delay, which grows linearly with respect to the size of a block, e.g., 80ms/kB for Bitcoin. In~\cite{Croman2016}, an implicit metric to capture the impact of network scale on the block propagation delay is adopted. Therein, the ratio between the block size and the propagation time required to reach a certain percentage of the nodes in the network is measured for the Bitcoin network. The experiments show that in the Bitcoin network with 55kb/s propagation rate for 90\% of the nodes, the block interval should not be smaller than 12s, which leads to a peak transaction throughput of 26TX/s for 250Byte transactions.

Furthermore, the studies in~\cite{rizun2016subchains, 8326513} also consider the impact of the propagation delay on the incidence of abandoning a proposed block with valid PoW solution. More specifically, finding a valid puzzle solution does not necessarily mean that the proposed block will be finally accepted by the network. Due to the propagation delay, a blockchain fork (see Figure~\ref{fig_blockchain_forking}) can only be adopted as the canonical blockchain state when it is first disseminated across the network. By considering both the round-trip delay and the block verification delay, the average block propagation delay across a P2P network is modeled as a function of the block size $s$ in~\cite{8326513}:
\begin{equation}
  \label{eq_avg_block_delays}
  T(s) = T_p(s) + T_v(s) = \frac{s}{a C}+b s,
\end{equation}
where $a$ is a network scale-related parameter, $C$ is the average effective channel capacity of each link~\cite{rizun2015transaction} and $b $ is a coefficient determined by both the network scale and the average verification speed of each node (cf.~\cite{6688704}). Based on (\ref{eq_avg_block_delays}), the probability for the network to abandon/orphan a valid block proposal of size $s$ due to the delay of block diffusion is modeled as follows~\cite{rizun2016subchains, 8326513}:
\begin{equation}
  \label{eq_orphan_prob}
  \Pr\nolimits^{\textrm{Orphan}}(s)=1-e^{-\lambda T(s)},
\end{equation}
where $\lambda$ is the expected block arrival rate.

From a user's perspective, it is insufficient to know only the network-level probability of block orphaning due to the latency. Alternatively, it is of more interest to determine the safe time interval between locally observing on the chain a transaction and confirming it. With this in mind, the study in~\cite{pass2017analysis} considers a scenario where the adversary gets additional computation time by delaying the block propagation with a certain number of rounds $\Delta$. Based on the analysis of the common-prefix property~\cite{Garay2015}, a new metric, i.e., $K$-consistency is proposed in~\cite{pass2017analysis} to examine whether any two honest nodes are able to agree on the blockchain state that is at least $K$ blocks deep from the end of the chain. Let $\alpha$ and $\beta$ denote the probabilities that an honest node and the attackers can propose a valid block within a round, respectively. The analytical study in~\cite{pass2017analysis} (cf.~\cite[Lemma 8]{kiayias2015speed}) shows that the required waiting time $T$ is jointly determined by $\alpha$, $\beta$, $\Delta$ and the parameter determining the searching space of the hash function, i.e., $L$ in Definition~\ref{def_puzzle_pow}. More specifically, as long as the following condition is satisfied with an arbitrarily small constant $\delta>0$ (see~\cite[Theorem 1.2]{pass2017analysis})
\begin{equation}
  \label{eq_consistency_condition}
  \alpha(1-(2\Delta+2)\alpha)\ge(1+\delta)\beta,
\end{equation}
and $K\!>\!K_0(L)\!=\!c\log(L)$ for some constant $c$, the Nakamoto protocol satisfies the property of $K$-consistency (except with negligible probability in $K$). However, the closed-form threshold $K_0(L)$ for $K$-consistency is not provided in~\cite{pass2017analysis}.

\subsection{Proof of Concepts Attached to Useful Resources}\label{sub_sec_pour}
Under the framework of Nakamoto protocol, a number of alternative PoX schemes have been proposed to replace the original PoW scheme in permissionless blockchain networks. Generally, these PoX schemes aim at two major designing goals, i.e., to incentivize useful resource provision, e.g.,~\cite{Filecoin, ghosh2014torpath, Kopp2016, 203890, ball2017proofs} and to improve the performance, e.g., in terms of security, fairness and eco-friendliness~\cite{park2015spacecoin, Blocki2016, moran2017rational} of the blockchain networks. Starting from this subsection, we will focus on the principles of puzzle design discussed in Section~\ref{sec_zkp_consensus} and provide a close examination on different PoX schemes in the literature.

With the purpose of useful resource provision, the idea of ``Proof of Useful Resources'' (PoUS) has been proposed to tackle the resource wasting problem of PoW. Instead of enforcing the consumption of computational cycles for merely hash queries, a number of studies are devoted to the design of puzzles that are attached to useful work. An early attempt, i.e., Primecoin~\cite{king2013primecoin}, proposed to replace the PoW puzzle in (\ref{eq_puzzle_pow}) by the puzzle of searching three types of prime number chains, i.e., the Cunningham chain of the first/second kind or the bi-twin chain~\cite{andersen2005cunningham}. However, the verification stage of Primecoin puzzle is based on classical Fermat test of base two (pseudoprime)~\cite{king2013primecoin}, hence violates the principle of soundness in non-interactive ZKP. Meanwhile, since the induced solution arrival does not follow the i.i.d. Bernoulli model in (\ref{eq_prob_success}), the Primecoin puzzle does not simulate the random distribution for leader selection as required by (\ref{eq_winning_prob}).

In~\cite{8171383}, a similar scheme, i.e., the proof of exercise is proposed to replace the preimage searching problem in PoW with the useful ``exercise'' of matrix product problems. The scheme uses a pool of task proposals to replace the PoW-based puzzle solving processes by the computation tasks offered by non-authenticated clients. Each consensus node needs to bid for a specific task to determine its puzzle. For this reason, the puzzle solution-generating scheme behaves more like a Computation as a Service (CaaS) platform. Since the matrix problems in the task pool may present different complexity levels, the puzzle competition does not fully simulate on the network level the random distribution in (\ref{eq_winning_prob}). Also, the solution verification can only be done probabilistically due to the lack of $O(n)$ verification schemes. Therefore, the proposed scheme in~\cite{8171383} suffers  from the same problems as in the Primecoin~\cite{king2013primecoin}.

In~\cite{ball2017proofs}, a new puzzle framework, i.e., useful Proof of Work (uPoW) is designed to replace the primitive PoW puzzle in (\ref{eq_puzzle_pow}) with a specific set of problems satisfying not only the properties of completeness, soundness and non-invertibility (hardness), but also the additional requirement of usefulness. Here, the usefulness is implied in the execution stage of the puzzle (cf. Table~\ref{table_pox_stage}). Formally, by assuming completeness and soundness, the properties of usefulness can be defined as follows (cf.~\cite[Definition 1]{ball2017proofs}):
\begin{Definition}[Usefulness]
  \label{def_usefulness}
  Suppose that a challenge $c_x$ and an accompanying puzzle solution (proof) $s$ are generated from an input string $x$. If there exists an algorithm $\textrm{Recon}(c_x,s)$ such that for a target function $F(\cdot)$ its output satisfies $\textrm{Recon}(c_x,s)\!=\!F(x)$, the challenge is known to be useful for delegating the computation of $F(x)$.
\end{Definition}
The study in~\cite{ball2017proofs} proposes to replace preimage searching in (\ref{eq_puzzle_pow}) with a family of one-way functions satisfying the property of fine-grained hardness~\cite{Ball:2017:AFH:3055399.3055466} for uPoW puzzle design. Namely, the PoW puzzle is proposed to be replaced by the problem of known worst-case-to-average-case complexity reduction. A special case of uPoW puzzles based on the problem of $k$-Orthogonal Vectors ($k$-OV) is discussed. In brief, the solution to $k$-OV performs an exhaustive search over $k$ sets of identical-dimension vectors and determines whether for each set there exists a vector such that these $k$ vectors are $k$-orthogonal. In order to construct non-interactive proofs, uPoW in~\cite{ball2017proofs} employs the hash function $\mathcal{H}(\cdot)$ as a random oracle. Simply put, given the number of vectors in each set, non-interactive uPoW treats the elements of each vector as the random coefficients of polynomials with the identical order. uPoW initializes the first element of each vector, i.e., the lowest order coefficient with a publicly known input string $x$ and then uses it as the input to $\mathcal{H}(\cdot)$ for generating the next-order coefficient. The output of $\mathcal{H}(x)$ will then be iteratively used as the input for generating the next-order coefficient. This can be considered as a typical example of applying the Fiat-Shamir scheme\footnote{The Fiat-Shamir scheme takes a similar form to the process of digital signature verification, see~\cite{Fiat:1987:PYP:36664.36676} for the definition.} to construct non-interactive PoW out of interactive ZKP schemes. With such an approach, uPoW does not need to explicitly define the vector sets. It also guarantees that the solutions of $k$-OV found by each prover follow a Bernoulli distribution. Therefore, the uPoW scheme fits well in the existing Nakamoto protocols by simulating a provable random function. As stated in~\cite{ball2017proofs}, besides $k$-OV, uPoW is compatible with computation delegation for other problems such as 3SUM~\cite{Ball:2017:AFH:3055399.3055466}, all-pairs shortest path~\cite{Ball:2017:AFH:3055399.3055466}, and any problem that reduces to them\footnote{These problems should be worst-case hard for some time bound and can be represented by low-degree polynomials.}.

Schemes that are similar to uPoW can also be found in~\cite{203890}. In~\cite{203890}, the problem of untrusted computational work assignment is addressed in a Trusted Execution Environment (TEE). The TEE can be constructed using Intel Software Guard Extensions (SGX), which is a set of new instructions available on certain Intel CPUs to protect user-level codes from attacks by hardware and other processes on the same host machine. In the permissionless network, the clients supply their workloads in the form of tasks that can be run in an SGX-protected enclave (i.e., protected address space). The study in~\cite{203890} exploits the truthfulness-guaranteeing feature of the Intel attestation service~\cite{johnson2016intel} in the SGX-protected platform to verify and measure the software running in an enclave. With the designed puzzle, the work of each consensus node is metered on a per-instruction basis, and the SGX enclave randomly determines whether the work results in a valid block proof by treating each instruction as a Bernoulli trial. Based on the TEE, each executed useful-work instruction is analogous to one hash query in the primitive PoW, and the enclave module works as a trusted random oracle.

Apart from delegation of useful computation, PoX can also be designed to incentivize distributed storage provision. For example, Permacoin~\cite{6956582} proposes a scheme of Proof of Retrievability (PoR) in order to distributively store an extremely large size of data provided by an authoritative file dealer. The file dealer divides the data into a number of sequential segments and publishes the corresponding Merkle root using the segments as the leaves. A consensus node uses its public key and the hash function to select a random group of segment indices for local storage. For each locally stored segment, the node also stores the corresponding Merkle proof derived from querying the Merkle tree. The challenge-proof pair is generated based on a subset of the locally stored segments and the corresponding Merkle proof.  To ensure the non-interactiveness and freshness of the puzzle (cf. interactive PoR in~\cite{Juels:2007:PPR:1315245.1315317}), the node needs a publicly known and non-precomputable puzzle ID to seed the process of segment selection called ``scratch-off''. To help the readers understand the puzzle generation process, we present a simplified execution stage of PoR as follows (see also~\cite[Figure 1]{6956582}):
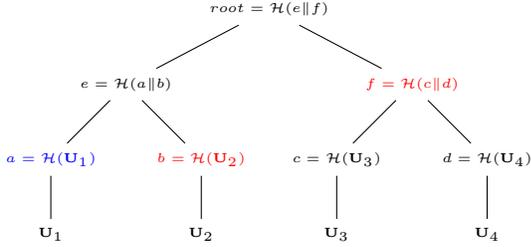
\begin{figure}
  \centering     
  \begin{tikzpicture}[level distance=1.0cm,
    level 1/.style={sibling distance=3.8cm},
    level 2/.style={sibling distance=2.0cm}]
    \tikzstyle{every node}=[font=\tiny]
    \node {$root=\mathcal{H}(e\Vert f)$}
      child
      {
        node {$e=\mathcal{H}(a\Vert b)$}
        child
        {
          node {\color{blue}$a=\mathcal{H}(\mathbf{U}_1)$}
          child
          {
            node {$\mathbf{U}_1$}
          }
        }
        child
        {
          node {\color{red}$b=\mathcal{H}(\mathbf{U}_2)$}
          child
          {
            node {$\mathbf{U}_2$}
          }
        }
      }
      child
      {
        node {\color{red}$f=\mathcal{H}(c\Vert d)$}
        child
        {
          node {$c=\mathcal{H}(\mathbf{U}_3)$}
          child
          {
            node {$\mathbf{U}_3$}
          }
        }
        child
        {
          node {$d=\mathcal{H}(\mathbf{U}_4)$}
          child
          {
            node {$\mathbf{U}_4$}
          }
        }
      };
  \end{tikzpicture}
  \caption{Illustration of Merkle proof: for segment $\mathbf{U}_1$, the Merkle proof is obtained by climbing up the tree until the root (as the nodes in red).}
  \label{fig_Merkle_tree}
\end{figure}

\begin{itemize}
  \item {The execution stage of PoR}: suppose a node is given the key pair $(sk, pk)$, the puzzle ID $id_{puz}$, the vector of locally stored segment indices $\mathbf{v}$, the required number of Merkle proofs $k$, the vectors of all the file segments $\mathbf{U}$ and the corresponding Merkle proof vector $\pi$. The random IDs of the local segments for challenge generation can be determined by:
  \begin{eqnarray}
    \label{eq_puzzle_generation}
    \forall 1\!\le\! j\!\le\! k: r_j\!=\!\mathbf{v}\left(\mathcal{H}(id_{puz}\Vert pk\Vert j\Vert nonce) \!\!\!\mod \vert \mathbf{v}\vert\right),
  \end{eqnarray}
  where $nonce$ is a random value chosen by the node. For each segment $\mathbf{U}(\mathbf{v}(r_j))$ in the challenge, the proof is in the form of $\left(pk_i, nonce, \mathbf{U}(\mathbf{v}(r_j)), \pi(\mathbf{v}(r_j))\right)$.
\end{itemize}

The execution stage of PoR in~\cite{6956582} is composed of a fixed number of queries to the random oracle $\mathcal{H}$. Thereby, although PoR satisfies the principle properties of non-interactive ZKP, it does not simulate the random leader election process. In this sense, the proposed PoR scheme may not be able to achieve the claimed goal of ``repurposing PoW'' in~\cite{6956582}. Instead, it is more similar to the existing systems such as Stoj~\cite{Storj2016}, Sia~\cite{vorick2014sia} and TorCoin~\cite{ghosh2014torpath}, where PoX is only used to audit the execution of the smart contracts or script-based transactions instead of facilitating the consensus mechanism.

Further improvement to PoR can be found in the proposals of KopperCoin~\cite{Kopp2016} and Filecoin~\cite{Filecoin}. In~\cite{Kopp2016}, KopperCoin adopts the same framework of distributed storage for a single file as in Permacoin~\cite{6956582}. Compared with Permacoin, the main improvement of the puzzle design in KopperCoin is to simulate the random leader election process for block proposal. KopperCoin introduces a bitwise XOR-based distance metric between the index of a locally stored data segment and a random, publicly known challenge $c$. A node needs to provide the valid Merkle proof (PoR) of a segment, of which the index (denoted by $j$) should satisfy the following condition:
\begin{equation}
  \label{eq_por_difficulty_segment}
  \mathcal{H}(x)\cdot 2^{|j\oplus c|}\le D(h),
\end{equation}
where the block payload $x$ and the difficulty threshold $D(h)$ are defined in the same way as in Definition~\ref{def_puzzle_pow}. Compared with (\ref{eq_puzzle_pow}), the solution searching for (\ref{eq_por_difficulty_segment}) is now performed within the range of the locally-stored segment indices. The more segments a node offers to store, the better chance the node has to find a solution to (\ref{eq_por_difficulty_segment}). Again, the generation of the public, unpredictable random challenge $c$ can be derived based on hashing the header of the most recent block. This approach presents another example of applying the Fiat-Shamir transformation to realize non-interactiveness~\cite{Fiat:1987:PYP:36664.36676}.

\begin{figure}[t]
\centering     
\includegraphics[width=.40\textwidth]{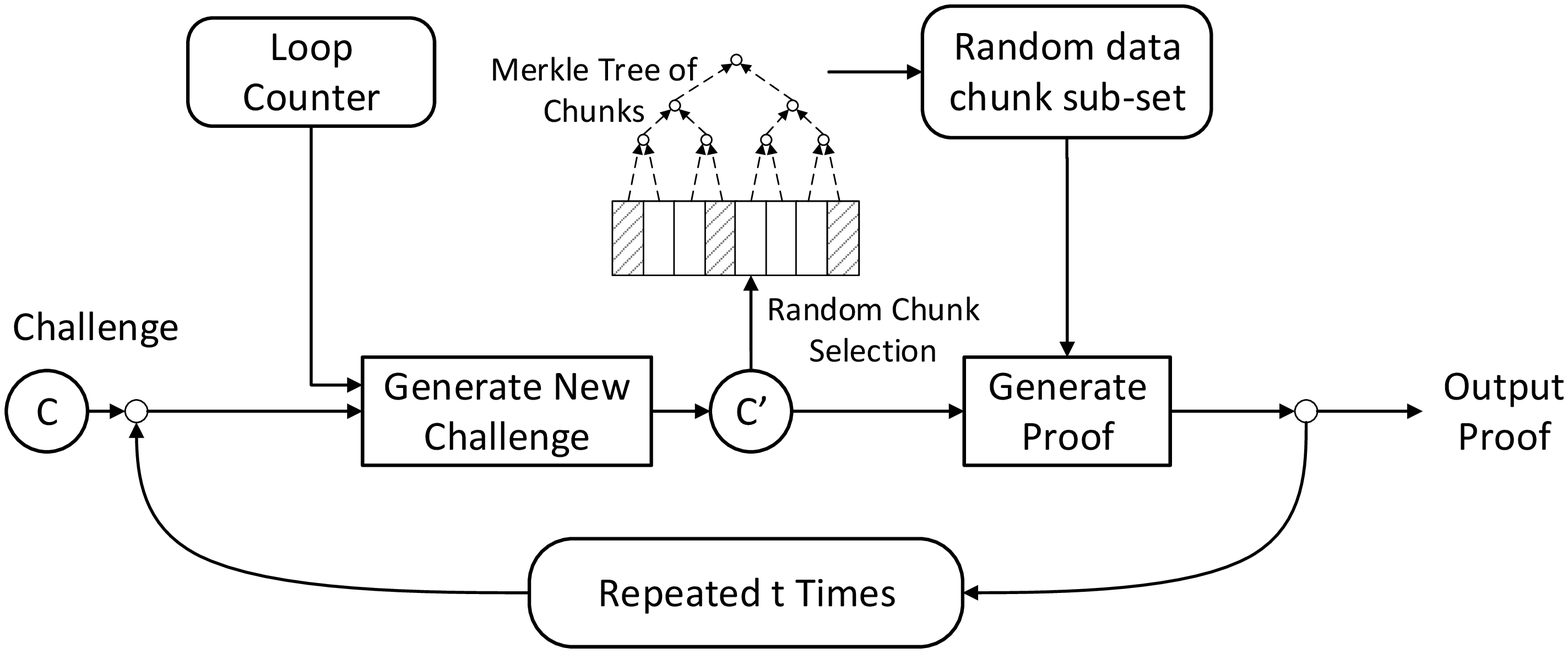}
\caption{Illustration of the PoST scheme based on iterative PoR over time.}
\label{fig_post}
\end{figure}
In the Filecoin network~\cite{Filecoin}, the concept of ``spacetime'' is introduced to allow metering the data stored in the network with an expiry time. Filecoin aims to provide the functionality of recycling and re-allocating the storage on the provider (miner) side as well as easing the files retrieval process on the client side. Like in the proof-of-exercise scheme, Filecoin designs the market for storage and retrieval of multiple files based on smart contracts. A new puzzle, i.e., Proof of SpaceTime (PoST)~\cite{moran2017rational}, is adopted based on the intuition of generating a PoR sequence during a certain period to prove the holding time of useful storage. As illustrated by Figure~\ref{fig_post}, the major difference of PoST from PoR lies in the repeated execution phases for challenge updating without rerunning the initialization stage. Namely, a consensus node is required by the Filecoin network to submit PoR (e.g., in a similar way to Permacoin~\cite{6956582}) every time when the blockchain is extended by a certain number of blocks. Instead of simulating random leader election based on adjustable difficulty~\cite{Kopp2016}, the Filecoin network uses the following mechanism to determine whether a node $i$ is elected for block proposal:
\begin{equation}
  \label{eq_filecoin}
  \frac{1}{2^L}\mathcal{H}(t\vert \textrm{rand}(t))\le\frac{w_i}{\sum_{j\in\mathcal{N}}w_j},
\end{equation}
where $t$ is the index of consensus round (i.e., block index), $L$ is the output string length of the hash function (see (\ref{eq_puzzle_pow})), $\textrm{rand}(\cdot)$ is an assumed random oracle, and $w_i$ represents the storage power of node $i$ (see also (\ref{eq_winning_prob})). It is worth noting that the evaluation of $w_i$ in (\ref{eq_filecoin}) can only be done through PoST. Thus, the Filecoin network admits a double-challenge scheme, where the leader election is performed based on a second challenge, i.e., (\ref{eq_filecoin}). The nodes with the better quality of PoST proofs (storage power) are more likely to win the second challenge. Under the framework of double challenges, a similar approach of puzzle design can also be found in the proof of space-based cryptocurrency proposal known as SpaceMint~\cite{moran2017rational, park2015spacecoin}.

\subsection{Proof of Concepts for Performance Improvement}
\label{susub_sec_pox_pi}
Alternative PoX schemes have also been designed with the emphasis on improving the performance of PoW in the aspects such as security, fairness and sustainability. To alleviate the problem of computation power centralization due to the massive adoption of ASICs, memory-hard PoW, also known as the Proof of Memory (PoM), is adopted by ZCash~\cite{hornby2016zcash} and Ethereum~\cite{buterin2014ethereum} networks. In the ZCash network, the Equihash scheme~\cite{biryukov2017equihash} is adopted based on the generalized birthday problem~\cite{10.1007/3-540-45708-9_19}. The study in~\cite{biryukov2017equihash} has pointed out that any identified NP-complete problem can be the natural candidate for the PoX puzzle due to their proved hardness, as long as the solution verification can be completed in polynomial time. However, a puzzle design only satisfying the hardness requirement may not be able to combat the botnet or ASIC-based manipulation of hashrate. Thus, a suitable PoX is expected to be ``optimization-free'' and ``parallelism-constraint''. Namely, the solution searching process cannot be sped up by using alternative algorithms or through parallelization.

An ideal approach of imposing parallelism constraint is to ensure that the PoW scheme is inherently sequential. However, an inherently sequential NP problem that is known to be verified in short time is yet to be found~\cite{biryukov2017equihash}. Therefore, the study in~\cite{biryukov2017equihash} adopts an alternative approach by imposing enormous memory bandwidth to the parallel solution of the puzzle. According to~\cite{10.1007/3-540-45708-9_19}, the generalized $k$-dimensional birthday problem is to find $k$ strings of $n$ bits from $k$ sets of strings, such that their XOR operation leads to zero. Equihash employs the hash function $\mathcal{H}(\cdot)$ to randomly generate the $k$ strings using the block payload data $x$ and a nonce (as in (\ref{eq_puzzle_generation})), such that both the XOR-based birthday problem solution and a PoW preimage of a given difficulty are found. It is shown in~\cite{10.1007/3-540-45708-9_19} that the best solution algorithm to this problem presents $O(2^{n/k})$ complexity in both time and space and thus is memory-intensive. More importantly, for a $k$-dimensional problem, a discounting factor $1/q$ in memory usage leads to $O(q^{k/2})$ times more queries to the hash function. Due to the physical memory bandwidth limit, the computation advantage of parallelization is limited. These properties guarantee the ASIC-resistance of Equihash.

\begin{figure}[t]
\centering     
\includegraphics[width=.48\textwidth]{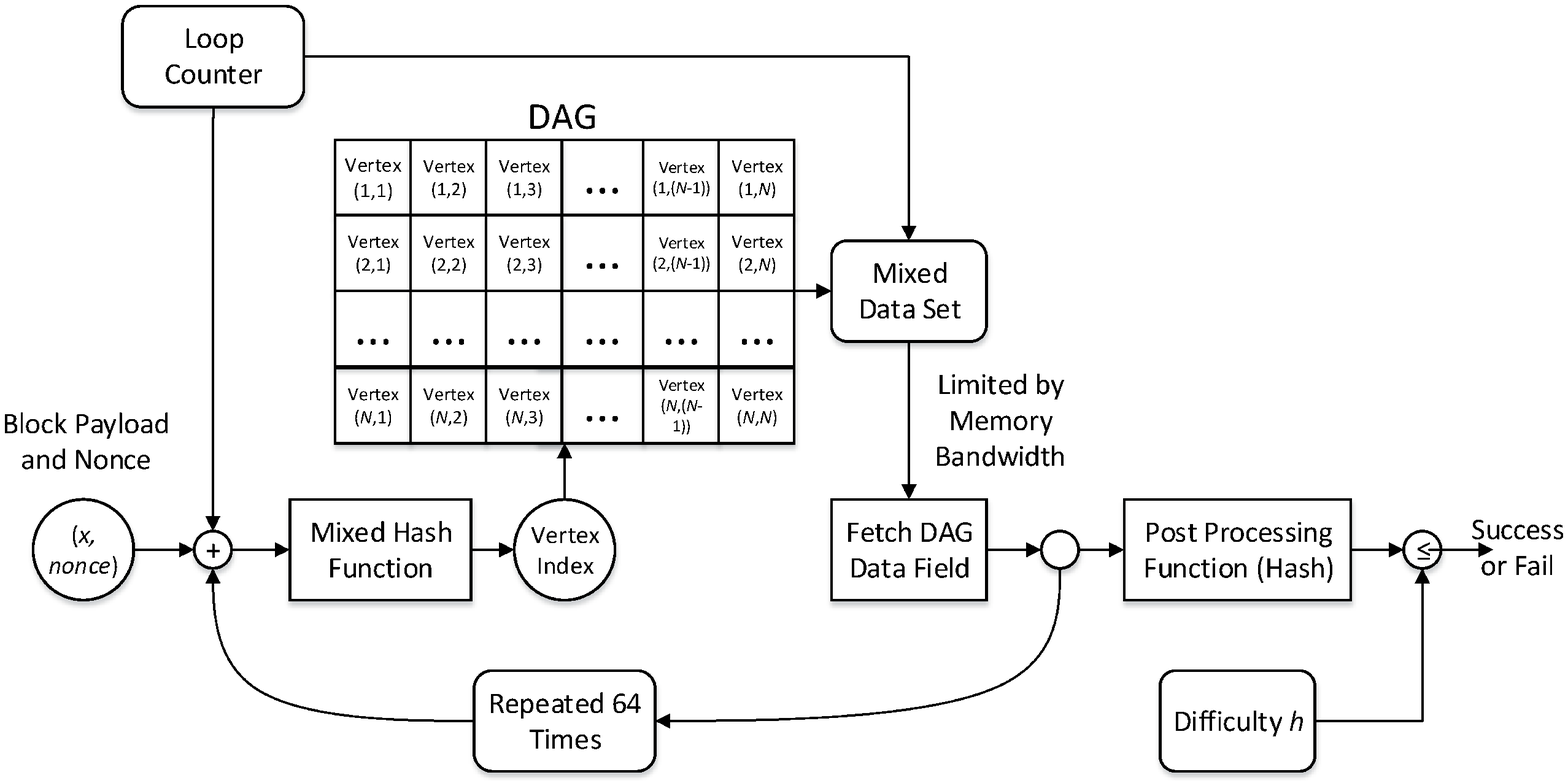}
\caption{One query to the random oracle in Ethash for a given $nonce$ based on the iterative mixed hash operation for vertex searching.}
\label{fig_ethash}
\end{figure}

With the same purpose of preventing the ``super-linear'' profit through hashrate accumulation, Ethereum currently adopts a different puzzle design known as Ethash for ASIC resistance~\cite{wood2014ethereum}. Ethash requires the consensus nodes to search for the PoW puzzle solution based on a big pseudorandom dataset, which increases linearly over time. The dataset is organized as the adjacency matrix of a DAG, where each vertex represents a randomly generated data field of 128 bits. In the execution stage of Ethash, the node starts a one-time search of the solution with a hash query, and uses the concatenation of the block payload and a nonce to seed the hash function for locating a random vertex in the DAG. Then, the search is completed in a fixed-iteration loop of queries to the hash function, for which the output of the last iteration, i.e., the data field of the last vertex in the path is used as the input to determine the position of the next vertex in the DAG. The final output of the loop is used to check against the preimage condition as in (\ref{eq_puzzle_pow}). As illustrated in Figure~\ref{fig_ethash}, the designed puzzle of Ethash makes the searching algorithm inherently sequential. With Ethash, the rate of data field fetching from the DAG is limited by the memory bandwidth. Then, paralleling the hash queries with ASICs cannot lead to much performance improvement in a single search of the puzzle solution.

Ethash~\cite{wood2014ethereum} only makes the puzzle solution partially sequential within a single attempt of preimage search. Therefore, Ethash still faces the problem of PoW outsourcing since a consensus node can divide the puzzle solution search into multiple sub-problems and outsource them to different ``mining workers'' (i.e., puzzle solvers). Such a problem is also known as the formation of mining coalition (pool)~\cite{Eyal2014} and may result in a serious problem of consensus manipulation by a handful of full nodes~\cite{7163021}. In~\cite{Miller:2015:NSP:2810103.2813621}, a nonoutsourceable ``scratch-off puzzle'' is proposed to disincentivize the tendency of mining task outsourcing. Intuitively, when a node effectively outsources its puzzle-solving work to some mining machines, we call the puzzle nonoutsourceable if these miners can steal the block proposal reward of that node without producing any evidence to implicate themselves.
The study in~\cite{Miller:2015:NSP:2810103.2813621} employs Merkle proofs for puzzle design, which can be considered as a generalization of the PoR~\cite{6956582}. In~\cite{Miller:2015:NSP:2810103.2813621}, a Merkle tree is created based on a number of random strings. To generate a fresh puzzle, a node queries the hash function for the first time with a random nonce and the constructed Merkle root. The output of this query is used to select a random subset of distinct leaves on the Merkle tree. Then, the concatenation of the Merkle proofs for each leaf in subset and the same nonce is used as the input to the second query of the hash function. The output is used to compare with the preimage condition as given in (\ref{eq_puzzle_pow}). If a solution (nonce) is found, the payload of the proposed block is used as the input of the third query to the hash function, and the output is used to select another subset of random leaves on the Merkle tree. The corresponding Merkle proofs are treated as the ``signature'' of the payload of the proposed block. With such puzzle design, mining workers only need to know a sufficiently large fraction of the Merkle tree leaves to ``steal'' the reward by replacing the Merkle proof-based signature with their own proofs.

It is worth noting that the nonoutsourceable puzzle in~\cite{Miller:2015:NSP:2810103.2813621} is generated in such a way to make the preimage search for (\ref{eq_puzzle_pow}) independent of the payload of the proposed block, i.e., using the randomly generated Merkle tree. Then, a mining worker is able to replace the original payload including the public keys from the outsourcer by its own payload without being detected. A similar proposal of nonoutsourceable puzzle can be found in~\cite{Daian2017}, where a nonoutsourceable puzzle is designed based on two-fold puzzle. Namely, an inner puzzle is solved as a typical PoW puzzle, whose solution is used as the input of an additional PoW puzzle known as the outer puzzle. To prevent outsourcing the work load, a mining worker's signature is required for the inner puzzle solution to be used by the outer puzzle. However, it is pointed out in~\cite{Daian2017} that such design can only be considered heuristic and is not guaranteed to have the formal properties of “weak outsourceability”~\cite{Miller:2015:NSP:2810103.2813621}.

\begin{figure}[t]
\centering     
\includegraphics[width=.35\textwidth]{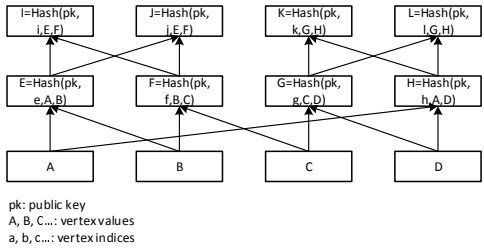}
\caption{An example of DAG formation based on the hash of the parent vertices: for miner $i$ adopting a public key $pk_i$, the value $v_j$ of the $j$'s vertex in its DAG with $m$ parent vertices $\{p1,\ldots,pm\}$ is obtained as $v_j=H(pk_i, j, v_{p1},\ldots,v_{pm})$.}
\label{fig_dag_formation}
\end{figure}

Apart from the manipulation-resistant puzzles, other puzzles are proposed in~\cite{park2015spacecoin, Blocki2016} with the emphasis on eco-friendliness. Therein, the major goal is to reduce/remove the repeated hash queries to curb energy consumption due to hash queries. In~\cite{park2015spacecoin}, the SpaceMint network is proposed based on Proof of SPace (PoSP)~\cite{Dziembowski2015}. Similar to PoR~\cite{6956582}, PoSP requires the consensus nodes to provide non-interactive proofs of storage dedication during puzzle solution searching. The major difference from PoR lies in that PoSP does not need the prover to store useful data (from the verifiers), and the proof is based on a large volume of random data stored on the provers' hard drive. As in Ethash~\cite{wood2014ethereum}, the committed space is also organized as a DAG, where the value of each vertex is determined based on the hash of its parent vertices (see Figure~\ref{fig_dag_formation}). A consensus node is required to use the hash of an earlier block as the seed to sample a random set of vertex values. The set of the vertex values forms the challenge of the node's local PoSP puzzle. If the node is able to provide the Merkle proofs for all the vertices in the challenge set, namely, the sibling vertices that lie on the path between each challenge vertex and the end vertex in the DAG with no outgoing edge, the proposed block is considered a valid block candidate. SpaceMint also proposes to measure the quality of a set of Merkle proofs based on the hash value of the concatenated vertex in a Merkle tree. Then, the blockchain network is able to select the block with the best quality of proof from the candidate blocks when a fork occurs.

The study in~\cite{Blocki2016} proposes to introduce a human-in-the-loop puzzle, i.e., the Proof of Human-work (PoH) into the Nakamoto protocol. The designing goal of PoH is to guarantee the properties of eco-friendliness, usefulness and centralization-resistance at the same time. It is proposed in~\cite{Blocki2016} that PoH should be able to provide non-interactive, computer-generated puzzles which are moderately hard for a human but hard for a computer to solve, even for the computer that generates the puzzles. PoH is inspired by the widely-adopted systems of Completely Automated Public Turing-Test to tell Computers and Humans Apart (CAPTCHA)~\cite{vonAhn2003}. Traditional CAPTCHA systems usually take human-efficient input (e.g., images) with a known solution and generate the puzzle based on distortion to the solution. For PoH, a universal sampler~\cite{Hofheinz2016} is assumed to be available to generate a random CAPTCHA instance for the consensus node such that the puzzle-generating machine is not able to directly obtain the puzzle solution. Then, the node (i.e., miner) needs human work to obtain the corresponding solution of the CAPTCHA puzzle. A two-challenge puzzle design is adopted and the solution of the CAPTCHA puzzle is used as the input of a small PoW puzzle as defined in (\ref{eq_puzzle_pow}). A complete PoH solution includes a CAPTCHA solution and a nonce such that they together satisfy the preimage condition in (\ref{eq_puzzle_pow}). PoH implicitly assumes that some Artificial Intelligence (AI) problems (e.g., recognition of distorted audios or images) are human-efficient but difficult for machines. Then, by selecting a proper underlying CAPTCHA scheme, it is possible to extend the PoH with a variety of meaningful human activities ranging from that educational purposes to a number of socially beneficial programs~\cite{Hofheinz2016}.

For a progressive summary, we summarize in Table~\ref{table_pox_comparison}  the major properties of the PoX schemes reviewed in this section.
\begin{table*}[t]
  \centering
  \scriptsize
  \caption{Comparison of Different PoX Schemes for Permissionless Blockchains}
  \renewcommand{\arraystretch}{1.3}
 \begin{tabular}{|p{1.75 cm} | p{2.5 cm}| p{1.5cm}| p{2.2cm} |p{1.2cm} | p{1.2cm}| p{2.3cm}| l|}
 \hline
 \makecell{Puzzle Name} & \makecell{Origin of Hardness \\(One-way Function)} & \makecell{Designing \\ Goal} & \makecell{Implementation \\Description} &\makecell{ZKP \\ Properties} & \makecell{Simulation \\of Random \\ Function}& \makecell{Features of \\Puzzle Design} &\makecell{Network \\ Realization} \\
 \hline
 \makecell[l]{Primitive proof of \\ work \cite{Garay2015, garay2017proofs}} & {Partial preimage search via exhaustive queries to the random oracle} & Sybil-proof & {Repeated queries to cryptographic hash function}&  Yes & Yes & Single challenge & \makecell[l]{Bitcoin~\cite{nakamoto2008bitcoin},\\ Litecoin~\cite{coinmarketcap}} \\
 \hline
 \makecell[l]{Proof of exercise\\ \cite{8171383}}  & {Matrix product} & Computation delegation &  Probabilistic verification&  N/A & No & Single challenge & N/A\\
 \hline
 \makecell[l]{Useful proof of \\work~\cite{ball2017proofs}}   & {$K$-orthogonal vector, 3SUM, all-pairs shortest path, etc.} & Computation delegation &  {Non-interactiveness via Fiat-Shamir transformation}&  Yes & Yes &Single challenge with sequential hash queries & N/A\\
 \hline
 \makecell[l]{Resource-efficient \\ mining~\cite{203890}} & N/A & Computation delegation & Guaranteed by TEE & Yes & Yes & Trusted random oracle implemented by dedicated hardware & N/A \\
 \hline
 \makecell[l]{Proof of \\retrievability  \cite{Juels:2007:PPR:1315245.1315317}}  & {Merkle proofs of file fragments in the Merkle tree} & Distributed storage & Non-interactiveness via Fiat-Shamir transformation and random Merkle proofs & Yes & Conditional & Two-stage challenge &\makecell[l]{Permacoin~\cite{6956582}, \\KopperCoin~\cite{Kopp2016}}\\
 \hline
 \makecell[l]{Proof of space\\-time~\cite{Filecoin}}  & \makecell[l]{The repeated proof of \\ retrievability over time} & Decentralized storage market &  \makecell[l]{Repeated PoR}& Yes & Conditional & Two-stage challenge and repeated PoR over time& Filecoin~\cite{Filecoin}\\
 \hline
 \makecell[l]{Equihash~\cite{biryukov2017equihash}}  & {The generalized birthday problem} & ASIC resistance &  Time-space complexity trade-off in proof generation~\cite{biryukov2017equihash}&  Yes& Yes & Memory-hard & ZCash~\cite{hornby2016zcash}\\
 \hline
 \makecell[l]{Ethash~\cite{wood2014ethereum}}  & Random path searching a random DAG & ASIC resistance &  Repeated queries to cryptographic hash function& Yes & Yes & Sequential, memory-hard puzzle &  Ethereum~\cite{buterin2014ethereum}\\
 \hline
 \makecell[l]{Nnonoutsourceable \\ scratch-off  puzzle\\ \cite{Miller:2015:NSP:2810103.2813621}} & Generalization of proof of retrievability & Centralization resistance &  Random Merkle proof & Yes & Yes & Two-stage challenge & N/A\\
 \hline
 \makecell[l]{Proof of space\\ \cite{Dziembowski2015}} & Merkle proofs of a vertex subset in a random DAG & Energy efficiency & Random Merkle proof &  Yes & Yes & Two-stage challenge and measurement of proof quality& SpaceMint~\cite{Dziembowski2015}\\
 \hline
 \makecell[l]{Proof of human \\ work~\cite{Blocki2016}} & Radom CAPTCHA puzzle requiring human effort & Useful work and energy efficiency &  {CAPTCHA and PoW} &  Yes & Yes & Human in the loop & N/A \\
 \hline
\end{tabular}
\label{table_pox_comparison}
\end{table*}

\section{Strategies of Rational Nodes in the Framework of Nakamoto Consensus Protocols}
\label{sec:AMS}

%
In this section, we review the studies on the incentive compatibility of the Nakamoto consensus protocols. By adopting the basic assumption on rationality of the consensus nodes (i.e., block miners), we provide a comprehensive survey on the node strategies in the consensus process for block mining. It is worth noting that most of the analysis in the literature about the consensus nodes' mining strategies are presented in the context of the PoW-based Bitcoin network. Nevertheless, they can be readily extended to other PoX schemes under the framework of Nakamoto protocols. In particular, we focus on the game theoretic formulation of resource allocation during the mining process, and then explore how miners can exploit the vulnerability of the incentive mechanism of the Nakamoto protocols in permissionless blockchain networks.

\subsection{Incentive Compatibility of Nakamoto Protocols}
\label{subsub_compatibility}
For Nakamoto protocols, monetary incentive plays the key role to ensure that most of the consensus nodes/miners follow the rules of blockchain state transition during the puzzle solution competition. In permissionless blockchain networks, the incentive mechanism is built upon the embedded digital token issuing and transferring schemes. In a typical PoW-based blockchain network, the leader/winner in the block proposal competition not only collects transaction fees from the approved transactions in the new block, but also gets token issuing reward, e.g., the ``coinbase reward'' in Bitcoin, for expanding the blockchain with the new block. For this reason, the puzzle competition process is compared to the process of ``gold mining'', since by casting resources into the competition, the nodes expect to receive monetary rewards carried by the tokens. As a result, the consensus participant nodes are better known as block ``miners'' to the public.

\begin{figure}[t]
\centering     
\subfigure[]{\label{fig_pool_share_btc}\includegraphics[width=.21\textwidth]{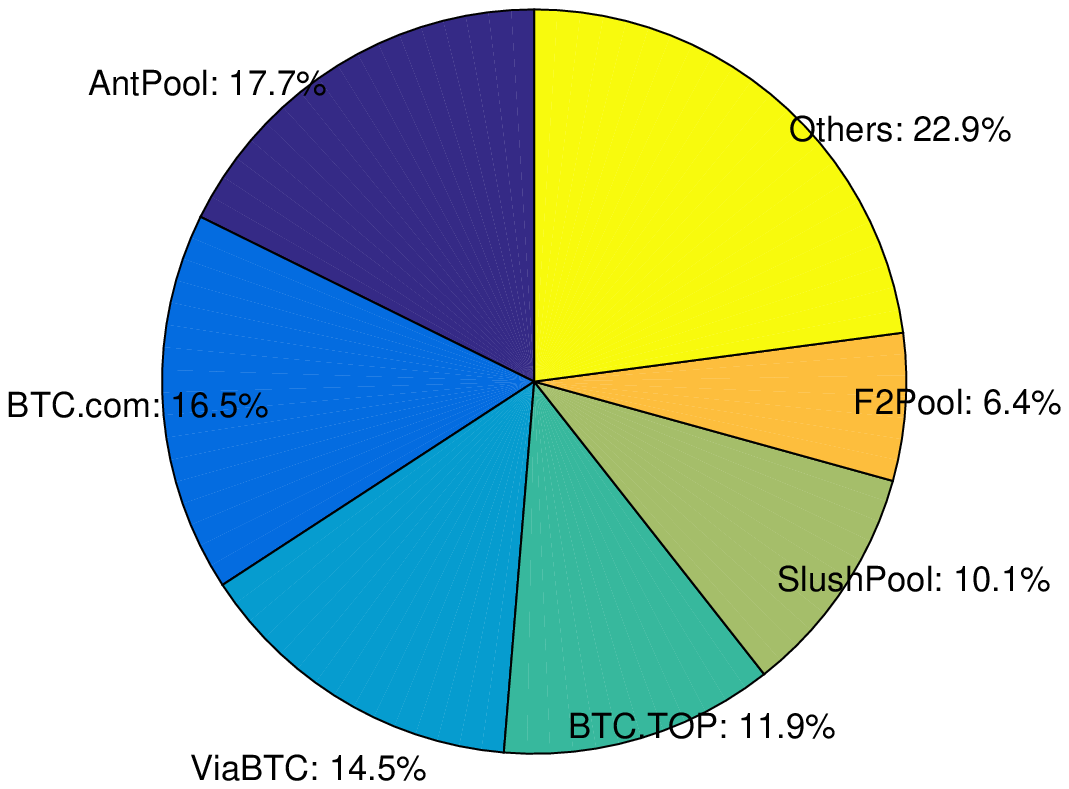}}
\subfigure[]{\label{fig_pool_share_eth}\includegraphics[width=.21\textwidth]{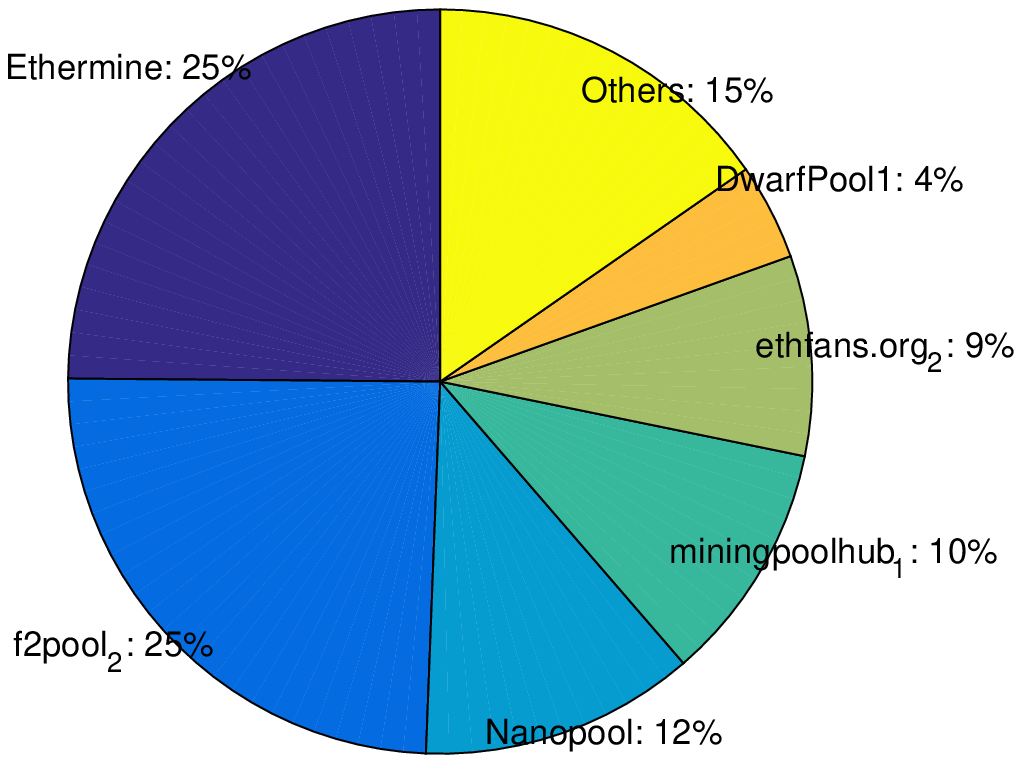}}
\caption{Hash rates controlled by mining pools in (a) Bitcoin (data source: {https://blockchain.info}) and (b) Ethereum (data source: {https://etherscan.io}).}
\label{fig_pool_share}
\end{figure}

In~\cite{kroll2013economics} the consensus in blockchain networks is divided into three folds, namely, the consensus about the rules, e.g., about transaction dissemination and validation, the universality of the blockchain state and financial value that the digital token carries. Then, the studies on the Nakamoto protocol's incentive compatibility can also be categorized according to these three aspects. Since the introduction of ASIC devices and pool mining for PoW-based blockchain networks, concerns have been raised about the nodes' incentive to fully abide by the protocol~\cite{kroll2013economics, Eyal2014, 6824541, courtois2014longest}. Due to the explosion of network-level hashrates (see Figure~\ref{fig_hashrate}), most of the practical blockchain networks, i.e., cryptocurrency networks, are nowadays dominated by the proxies of mining pools~\cite{rosenfeld2011analysis} (see Figure~\ref{fig_pool_share}). An individual node in a mining pool is known as a mining worker, since it no longer performs the tasks of transaction validation or propagation and does not even keep any blockchain data. On the contrary, only the proxy of the pool, i.e., the pool server/task operator maintains the replica of the blockchain. The pool server divides the exhaustive preimage search for PoW solution into a number of sub-tasks and outsources them to the mining workers\footnote{According to the Stratum mining protocol~\cite{recabarren2017hardening}, the pool server only needs to send a miner the Merkle root of the transactions in the block (see Figure~\ref{fig_blockchain}) and a difficulty level to complete the puzzle solving sub-task.}. In this sense, only the pool server can be considered as a node in the blockchain network. Studies have shown that joining a mining pool has become the more plausible strategy than working as an individual consensus node, since such a strategy reduces the income variance and secures stable profits~\cite{Eyal2014, 7163021}. However, this leads to the formation of mining-pool Cartel~\cite{Eyal2014} and is against the design goal of Nakamoto consensus in~\cite{nakamoto2008bitcoin}, that ``the network is robust in its unstructured simplicity''.

A further study in~\cite{Babaioff:2012:BRB:2229012.2229022} reveals that under the current framework of Nakamoto protocols, no incentive is provided for nodes to propagate the transactions that they are aware of. The study considers the situation when transaction fees dominate the block rewards~\cite{Laszka2015}. The analysis in~\cite{Babaioff:2012:BRB:2229012.2229022} models the paths of transaction dissemination as a forest of $d$-ary directed trees, where each transaction issuer considers its peer nodes as the tree roots and the nodes on the far end of the network as the leafs. During transaction dissemination, a consensus node can add any number of pseudo-identities (a.k.a., fake identities) before selectively relaying the transaction to any of its neighbors. It is shown that a consensus node tends to not broadcast any transaction that offers a fee. By doing so, it reduces the number of nodes that are aware of the transaction and hence the competition of mining that transaction. An improved protocol is proposed in~\cite{Babaioff:2012:BRB:2229012.2229022} by introducing a broadcasting incentive mechanism. More specifically, the proposed mechanism requires that each relaying node in the path of transaction propagation shares a uniform portion of reward with the root (i.e., mining) node, when the height of the relaying node is small than a predetermined threshold in the directed tree.  The analysis of the new protocol is based on the formulation of a normal-form game~\cite{maschler_solan_zamir_2013}, and thus the equilibrium strategy of each node can be obtained through iterative removal of dominated strategies. The designed incentive mechanism is shown to guarantee that only the non-Sybil and information propagating strategies survive in the iterated removal of weakly dominated strategies, as long as the miners are connected to sufficient many peers.

Similar studies to enforce honest block/transaction propagation can also be found in~\cite{abraham2016solidus, ersoy2017information}. The study in~\cite{ersoy2017information} casts the problem of incentivizing block propagation into the framework of routing in $k$-connected networks, where each rational node can freely choose between relaying and mining (or both). A protocol of transaction fee-sharing is designed therein to guarantee that the rational strategy of honest nodes in the network is to propagate the received transactions.  It is required that a mining node shares the reward of a new transaction with the relaying nodes in one path between itself and the client which issues that transaction. According to~\cite{Babaioff:2012:BRB:2229012.2229022}, creating pseudo-identities does not increase the connectivity of a node. From such an observation, it is proved in~\cite{ersoy2017information} that assigning the propagation reward of each relaying node as a decreasing function of the hop count guarantees transaction propagation, as long as the computing power (or other resources for mining) controlled by each node does not dominate the network. Comparatively, the study in~\cite{abraham2016solidus} ensures that the payment made to the transaction-relaying nodes cannot be denied by the miners of the new blocks. With the proposed propagation protocol in~\cite{abraham2016solidus}, each intermediate hop adds its own signature to the transaction before sending it to the next hop. While working on their own PoW-puzzle solution, the relaying nodes freely charge their descendants at least a minimum fee for propagation. The miner whose block finally gets confirmed by the blockchain will pay for the propagation fees to one selected path of nodes. As in~\cite{Babaioff:2012:BRB:2229012.2229022} the process of transaction propagation and relaying price competition is modeled as a non-cooperative game in~\cite{abraham2016solidus}. It is proved that with the proposed propagation protocol based on the chain of signatures, a rational miner's equilibrium strategy is to always choose the shortest path, and a rational intermediate node' equilibrium strategy is to always charge its descendants the minimum fees for relaying transactions.

When block creation reward dominates the mining reward, incentive incompatibility may appear in different forms. Intuitively, it is plausible for a rational miner to pack up a proper number of transactions with decent fees in the new block for profit maximization. However, empty blocks with only coinbase transaction or blocks with a tiny number of transactions can be frequently observed in the practical blockchain networks\footnote{See Blocks \#492972 in Bitcoin and \#3908809 in Ethereum for examples.}. An informal game theoretic analysis in~\cite{stone2015examination} indicates that the consensus nodes tend to ignore the received blocks of large size in a flat network and relay the smaller competing blocks instead. The reason is that large blocks result in longer delay due to transaction validation, hence increasing the probability of orphaning any blocks that are mined based on them. Although mining empty block does not violate the current Nakamoto protocol, it results in the same situation as a Distributed Denial of Service (DDoS) attack~\cite{Baqer2016} by blocking the confirmation of normal transactions.

Furthermore, the statistical studies in~\cite{pappalardo2017blockchain, Möser2015} have shown that the consensus nodes behave rationally and are prone to prioritize the transactions with higher transaction fees during block packing. However, when the coinbase reward dominates the block mining reward, the miners are yet not incentivized to enforce strictly positive fees~\cite{Möser2015}. In the case study of Bitcoin network, extra delays for the small-value transactions are identified ranging from 20 minutes~\cite{Möser2015} to as long as 30 days~\cite{pappalardo2017blockchain}. Also, it is observed in~\cite{Möser2015} that most of the lightweight nodes still set an arbitrary transaction fee in the real-world scenarios. It is unclear whether the miners or the transaction issuers adopt best-response strategies systematically. The study in~\cite{houy2014economics} simplifies the consensus process as a supply game subject to the trade of a specific type of physical goods. In the considered scenario, the miners essentially become the follower players in a two-level hierarchical/Stackelberg game\footnote{A Stackelberg game is characterized by the sequential play of leaders and followers, where the leaders may expect better equilibrium payoffs~\cite{maschler_solan_zamir_2013}.} led by the blockchain network, which is assumed to be able to set the transaction prices.
Then, they are expected to have an incentive for including all transactions if there exists no block-size limit. On the other hand, it is pointed out in~\cite{rizun2015transaction} that, since the block orphaning probability exponentially grows with the block size, a healthy transaction fee market does not exist for unlimited block size due to the physical constraint of link capacity in the network.

Finally, it is worth noting that most of the existing studies are based on the presumption that the tokens carried by a blockchain have monetary value and their exchange rate volatility is small. An optimistic prediction is provided in~\cite{athey2016bitcoin} based on an assumption excluding any state variables on the user sider except the belief in ``proper functioning of a cryptocurrency''. In the absence of investors and when the blockchain is used only for the purpose of remittance, it is shown in~\cite{athey2016bitcoin} that the tokens of a blockchain network admit a unique equilibrium exchange rate in each period of the belief evolution. Conditioned on the survival of a cryptocurrency, the equilibrium state depends on the excess in users' valuation of the blockchain over the other payment options as well as the supply of the tokens in the market. Together with the Stackelberg game-based interpretation in~\cite{houy2014economics}, it is reasonable to consider that the equilibrium price of a blockchain token is determined by the demand-supply relation in the market. It is worth noting that the data security is only guaranteed by sufficient PoW computation power in the blockchain network. Currently, except for a few studies such as~\cite{Feng2018Cyber}, it is generally unclear how the impact of security issues is reflected in the users' valuation of the blockchain. As a result, whether the security requirement of the Nakamoto protocol is compatible with the market clearing price remains an open question.

\subsection{Resource Investment and Transaction Selection for Mining under Nakamoto Protocols}
\label{subsub_resource_allocation}
According to (\ref{eq_winning_prob}), an honest consensus node has to invest in the mining resources, e.g., hashrates, disk space, etc, to win the puzzle solution competition under Nakamoto consensus protocols. Intuitively, the more resources a miner casts into the network, the higher chance the miner has to win the puzzle competition and obtain the mining reward. However, the success is not guaranteed because this also depends on the mining resources of other miners. Since mining resources are usually expensive, how to properly invest in the mining resources to maximize the profit is a big concern of the miners.

The study in~\cite{dimitri2017bitcoin} abstracts the mining investment in the Bitcoin network as the energy consumption cost. It is assumed that $N$ active miners in the network are competing in the ``all-pay contest'' for block-mining rewards. The cost of presenting a unit mining resource by each miner may be different, e.g., with different electricity prices in different areas. The miners determine how much to invest in mining resources (hashrates) such that the expected profit is maximized. This forms a non-cooperative game among the miners. Analysis of the game's unique Nash equilibrium in~\cite{dimitri2017bitcoin} shows that the decision of a miner to participate in the mining process or not solely depends on its individual mining cost, as long as the block reward is positive. Meanwhile, the structure of the formulated mining game prevents the emergence of a monopolistic mining activity. Namely, it is guaranteed that at least two miners will remain active in the game with positive expected profits.

By (\ref{eq_avg_block_delays}) and (\ref{eq_orphan_prob}), even if a miner succeeds in the puzzle solution competition, it is still possible for the proposed block to get orphaned due to the propagation delay. For ease of exposition, we can assume that all transactions in a block set the same amount of transactions fee $F$. Let $R$ denote the fixed reward for block generation and $m$ denote the number of transactions in the block. Then, the revenue to mine this block is $R+mF$. Apparently, a rational miner expects to include as many as possible transactions in a block to maximize the received reward. However, due to the risk of block orphaning, a miner also has to carefully balance the tradeoff between the mining reward and the risk of block orphaning. In~\cite{rizun2015transaction}, the author proposes a mining profit model by assuming the propagation delay of a block to follow a Poisson distribution. Thus, the orphaning probability can be approximated by (\ref{eq_orphan_prob}). Let $\eta$ denote the monetary cost per hash query and $\psi$ denote the probability for the miner being the leader (see also (\ref{eq_prob_success})). Then, for an average block arrival duration $T$ and block propagation time $\tau$, a miner's profit can be modeled as follows:
\begin{equation}
U=(R+F) \psi e^{-\frac{\tau}{T}} - \eta h T.
\label{eq:miner's profit}
\end{equation}
The profit model in (\ref{eq:miner's profit}) is capable of reflecting the impact of miners' strategies in both resource investment and transaction selection. Therefore, this model is especially appropriate for game-theoretic formulation of mining resource management problems. Recently, (\ref{eq:miner's profit}) and its variation have been adopted to construct the payoff function of miners by a series of studies, which propose to use different game-based models, e.g., evolutionary game~\cite{8326513}, hierarchical game~\cite{xiong2017optimal} and auctions~\cite{jiao2017social}, to capture the rational behaviors of individual miners in different network setups.

In~\cite{houy2014bitcoin}, an alternative model of winning probability is proposed to explicitly capture the influence of the adversary miners' strategy of block-size selection. We denote $s_i$ as block size of miner $i$ in a blockchain network and $w_i$ as its computational power. Then, the block winning probability of miner $i$ can be expressed by~\cite{houy2014bitcoin}:
\begin{equation}
\label{eq:p_win}
\Pr\nolimits_{i}^{\textrm{win}} = \frac{w_i}{T} \Bigg[ \prod_{j \neq i} \Big( e^{-\frac{ w_j ( t + \tau (s_i) - \tau (s_j) )}{T}} 	\Big)  \Bigg]  		,
\end{equation}
where $t$ is the time when all miners start mining a new block and $\tau (s_i)$ is the time needed for a block with size $s_i$ to reach consensus. In~(\ref{eq:p_win}), the first and second terms represent the probability for miner $i$ to first solve the puzzle based on its block, for this block to be the first one reaching the consensus across the network, respectively. (\ref{eq:p_win}) implies that the strategy of mining a large block may have positive externalities to other miners in the network. By analyzing the Nash equilibrium of the non-cooperative mining game with two miners, the author of~\cite{houy2014bitcoin} shows an interesting result, namely, the miner with higher computational power will prefer blocks of larger sizes. Meanwhile, the author also discusses the scenarios in which the Nash equilibrium is a breaking point, i.e., miners adopt the strategy of including no transaction in their proposed blocks.

The studies in~\cite{rizun2015transaction} and~\cite{houy2014bitcoin} essentially assume that the mining process is synchronized and all miners honestly follow the rules of block/transaction propagation in Nakamoto protocols. However, such assumptions may not be met in practical scenarios. Thus, related strategies may not be the miners' best response and further investigation is needed on this topic.

\subsection{Rational Mining and Exploitation of Nakamoto Protocols}
The discussions on the incentive compatibility of Nakamoto protocols and the strategies of resource investment lead to the following question: is it possible for a rational miner to exploit the vulnerability of Nakamoto Protocols and find a strategy leading to the reward more than that in proportion to the devoted resources? In this section, we will further devote our survey on the existing analysis of this problem.

\subsubsection{Selfish Mining Strategy}
The study in~\cite{Eyal2014} shows that selfish miners may get higher payoffs by violating the information propagation protocols and postponing their mined blocks. Specifically, a selfish miner may hold its newly discovered block and continue mining on this block secretly. Thereby, the selfish miner exploits the inherent block forking phenomenon of Nakamoto protocols. In this case, honest miners in the network continue their mining based on the publicly known view of the blockchain, while the selfish miners mine on their private branches. If a selfish miner discovers more blocks in the same time interval, it will develop a private longer branch of the blockchain. When the length of the public chain known by honest miners approaches that of the selfish miner's private chain, the selfish miner will reveal its private chain to the network. According to the longest-chain rule, the honest nodes will discard the public chain immediately when they learn the longer view of the chain from the selfish miner. Such a strategy of intentionally forking results in the situation of wasted computation by the honest miners, while the revenue of the selfish miner can be significantly higher than strictly following the block revealing protocol. More seriously, if selfish miners collude and form a selfish mining pool with a sufficiently large amount of computational power, other rational miners will be forced to join the selfish mining pool, which can devastate the blockchain network~\cite{Eyal2014}.
\begin{figure}[t]
	\begin{center}
		\includegraphics[width=.36\textwidth]{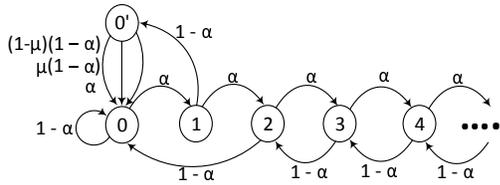}
		\caption{Blockchain state transition in the presence of a selfish pool (adapted from~\cite{Eyal2014}).}
		\label{fig:StateMachine1}
	\end{center}
\end{figure}

In~\cite{Eyal2014}, the authors introduce an approach based on the Markov chain model to analyze the behavior as well as performance of a selfish mining pool. Figure~\ref{fig:StateMachine1} illustrates the progress of the blockchain as a state machine. The states of the system, i.e., the numbers in the circles represent the lead of the selfish pool in terms of the difference in block number between the private branch and the public branch. In Figure~\ref{fig:StateMachine1}, state $0$ is the original state when the selfish pool has the same view as the public chain. State $0'$ indicates that two branches of the same length are published in the network by the selfish pool and the honest miners, respectively. The transitions in Figure~\ref{fig:StateMachine1} correspond to the mining event, i.e., a new block is mined either by the selfish pool or the honest miners. $\alpha$ in Figure~\ref{fig:StateMachine1} represents the computational power of the selfish mining pool. Note that the transition from state $0$ to state $0'$ depends on not only the computational power of the selfish pool, but also the fraction, i.e., $\mu$ of honest miners that mine on the selfish pool's branch. In~\cite{Eyal2014}, the analysis on the steady state probability of the Markov chain leads to the following two important observations:
\begin{itemize}
	\item For a given $\mu$, a selfish pool of size $\alpha$ obtains a revenue larger than its relative size in the range of $\frac{1-\mu}{3-2 \mu} < \alpha < \frac{1}{2}$.
	\item A threshold on the selfish-pool size exists such that each pool member's revenue increases with the pool size.
\end{itemize}

\begin{figure}[t]
	\begin{center}
		\includegraphics[width=.40\textwidth]{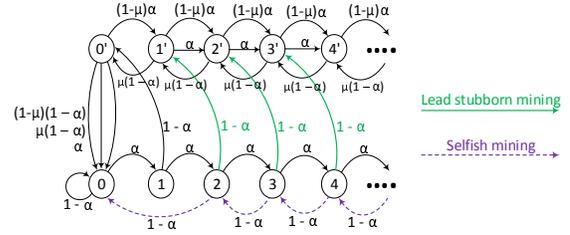}
		\caption{Lead-stubborn mining. The black and purple transitions together define the selfish mining state machine. The black and green transitions define the stage machine of lead-stubborn mining (adapted from~\cite{7467362}). }
		\label{fig:Stubborn}
	\end{center}
\end{figure}

Extended from~\cite{Eyal2014}, the study in~\cite{7467362} introduces a new mining strategy known as the {stubborn mining strategy}, which is supposed to outperform the typical selfish mining strategy. The key idea behind the stubborn mining strategy is that the selfish miner is stubborn and may only publish part of the private blocks even when it loses the lead to the honest nodes. As shown in Figure~\ref{fig:Stubborn}, the major difference between the two selfish strategies lies in how the selfish miner publishes the private blocks. For example, at state 2, the typical selfish miner will immediately publish all the private blocks once the lead to the honest miners decreases by one block (see Figure~\ref{fig:StateMachine1}). Then, the system transits to state 0. In contrast, every time when the honest miners mine a new block, the stubborn miner will stubbornly reveal one block of the private chain, even by doing so it will lose the lead. Simulations in~\cite{Eyal2014} show that stubborn mining achieves up to 13.94\% higher gains than selfish mining strategy.

\begin{figure}[t]
	\begin{center}
		\includegraphics[width=.48\textwidth]{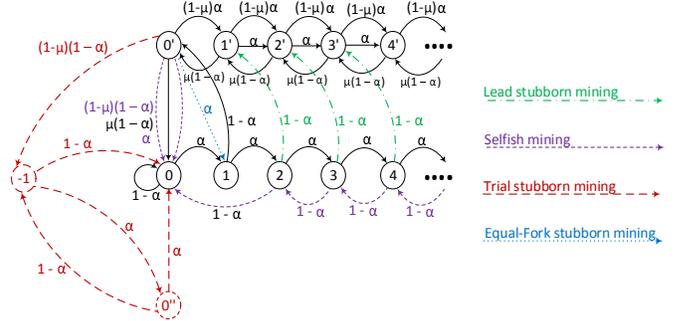}
		\caption{Lead, Equal-Fork, and Trail Stubborn mining. Black and purple transitions denote selfish mining. Black and green transitions denote lead-stubborn mining. Black and blue transitions denote Equal-Fork stubborn mining. Black and brown transitions denote Trail-stubborn mining (adapted from~\cite{7467362}).}
		\label{fig:ForkStubborn}
	\end{center}
\end{figure}

Furthermore, the study in~\cite{7467362} also introduces another two extensions of the stubborn mining strategy, namely, the Equal-Fork Stubborn (EFS) and the Trail Stubborn (TS) mining strategies (see Figure~\ref{fig:ForkStubborn}). In Figure~\ref{fig:ForkStubborn}, state -1 indicates that the public chain is one block longer than the private chain. As indicated by the transitions from other states to state -1, the TS miner is more stubborn and keeps mining on the secret branch even when it is one block behind the public chain. From state -1, when the TS miner finds one new block ahead of the honest miners, the system will transit to state $0^{''}$. Namely, the private chain catches up with the public chain and the block numbers on both chains are equal. In contrast, if the honest miners find a new block ahead of the ST miner, the system transits to state $0$. Namely, the ST miner starts to mine new blocks based on the public chain. Here, the difference between state $0^{''}$ and state $0'$ lies in that only the ST miner knows the existence of the private chain in state $0^{''}$, while in state $0'$ the honest miners can freely choose to mine on one of the two chains. The comparisons between the three stubborn mining strategies are given in Figure~\ref{fig:ForkStubborn}. Simulations in~\cite{7467362} show that stubborn mining strategies can improve the profit by up to 25\% than the original selfish mining strategy proposed in~\cite{Eyal2014}.

%
%

\begin{figure}[t]
	\begin{center}
		\includegraphics[width=.36\textwidth]{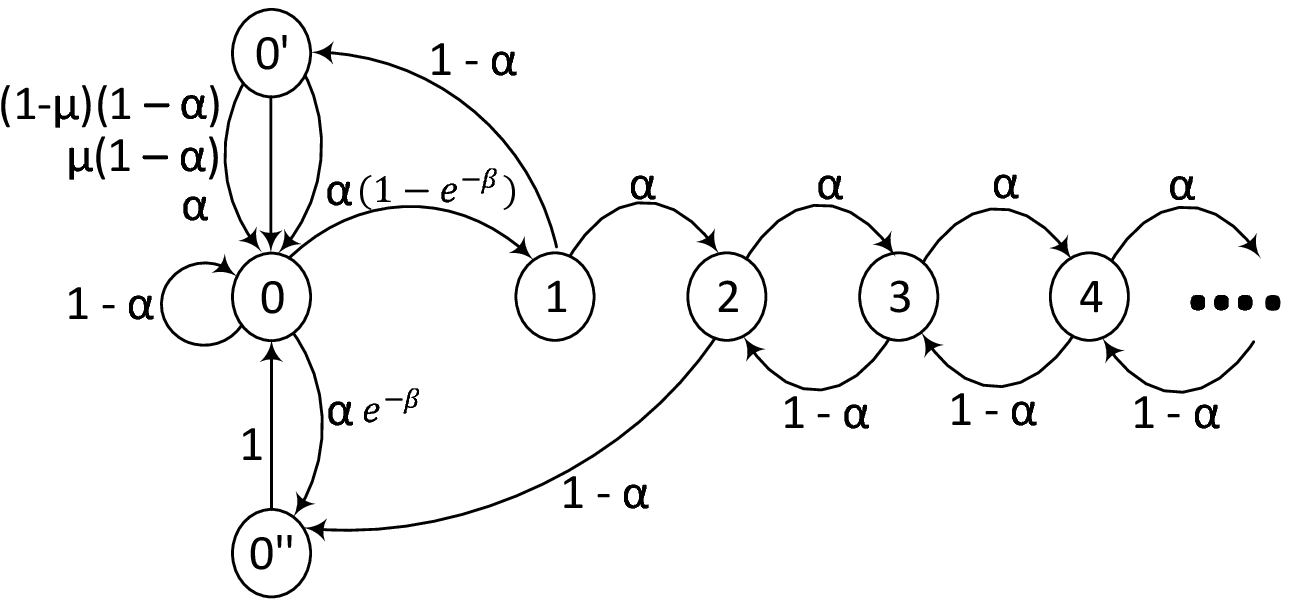}
		\caption{Improved Markov model for selfish mining with transaction fees (adapted from~\cite{carlsten2016impact}).}
		\label{fig:StateMachine2}
	\end{center}
\end{figure}

The author in~\cite{carlsten2016impact} studies the impact of transaction fees on selfish mining strategies in the Bitcoin network. Note that due to the inherent design of the token issuing scheme in Bitcoin, the constant mining reward of each block halves every time when a fixed interval of blocks, i.e., every 210,000 blocks, is generated. Then, it is natural to increase the transaction fee to compensate for the mining cost of the consensus nodes. The arbitrary levels of transaction fees lead to a situation where some hidden blocks may have very high values. As a result, selfish miners want to publish it immediately due to the risk of orphaning. Hence, in the revised Markov chain model for selfish mining in Figure~\ref{fig:StateMachine2}, the author introduces a new state $0^{''}$. State $0^{''}$ is almost identical to state $0$, except that, if the selfish miner mines on the next block in state $0^{''}$, it will immediately publish that block instead of holding it. Compared with the original selfish mining model in Figure~\ref{fig:StateMachine1}, state $0$ transits to state $1$ with probability $\alpha (1-e^{-\beta})$ and to state $0^{''}$ with probability $\alpha e^{-\beta}$, where $\beta$ is the size of the mining block. The new factor $\beta$ is introduced to model the impact of transaction fees on the miner's decisions. With the revised transition probability, if the selfish miner finds a block of high value in state $0$, it may publish the block (i.e., transiting to state $0''$) instead of holding it (i.e., transiting to state $1$). The analysis in~\cite{carlsten2016impact} shows that this improved selfish mining strategy leads to positive profit for all miners regardless of their hashrates.

From the aforementioned Markov models, we note that the selfish miner may adopt various policies by choosing to release an arbitrary number of block in each state. In~\cite{Sapirshtein2017, sompolinsky2016bitcoin, Gervais:2016:SPP:2976749.2978341}, a Markov Decision Process (MDP) model is proposed to generalize such a process of policy derivation. As an example, the study in~\cite{Sapirshtein2017} considers the honest miners as non-adaptive players following the Nakamoto protocol. Then, the problem of searching optimal selfish-mining strategy can be modeled as a single-player MDP. Four actions are considered to control the state transitions in the MDP:
\begin{itemize}
	\item \emph{Adopt}: the selfish miner accepts the honest network's chain and all private blocks are discarded;
	\item \emph{Override}: when taking the lead, the selfish miner publishes its private blocks such that the honest network discards its current view;
	\item \emph{Match}: the selfish miner publishes a conflicting branch of the same height. A fraction of the honest network will fork on this branch;
	\item \emph{Wait}: the selfish miner does not publish new blocks and keeps working on its private branch.
\end{itemize}
The state the MDP is defined by the difference in block lengths between the selfish miner and the honest network as well as the situation of computation forking among the honest miners. By controlling the maximum difference in block lengths, it is possible to obtain a finite-state MDP. Using standard MDP solution techniques, an $\epsilon$-optimal policy for selfish mining can be obtained based on such a truncated-state MDP.

In~\cite{GOBEL201623}, the authors consider a similar mining competition between a selfish mining pool and the honest nodes. The study in~\cite{GOBEL201623} extends the model of selfish mining by considering the propagation delay between the selfish mining pool and the honest community. The delay is assumed to be exponentially distributed with rate $\mu$. The block-mining Markov model in~\cite{GOBEL201623} adopts a 2-dimensional state of $(k,l)$, which denotes the length of blocks built by the pool and the community upon the common prefix blocks, respectively. Let  $\lambda_1$ and $\lambda_2$ denote the block-arrival rate for the pool and the community. The authors then derive the following transition rates of the block mining system:
\begin{equation}
\label{eq:transaction_rate}
\begin{aligned}
q \big( (k,l),(k+1,l) \big) 	=& 	\lambda_1, 	&	k \geq 0 , l \geq 0 	,\\
q \big( (k,l),(k,l+1) \big) 	=& 	\lambda_2, 	&	k \geq 0 , l \geq 0 	,\\
q \big( (k,l),(0,0) \big) 		=& 	\mu, 		&	k < l 					,\\
q \big( (k,k-1),(0,0) \big) 	=& 	\mu, 		&	k \geq 2 				,\\
q \big( (k,l),(k',l') \big) 	=& 	0,			&	\text{otherwise} 		.
\end{aligned}
\end{equation}
Based on this transition map, the authors in~\cite{GOBEL201623} propose to detect selfish mining behaviors by monitoring the proportion of orphaned blocks. Specifically, if there is a significant increase in the fraction of orphaned blocks, it is highly possible that selfish mining exists in the network.

\begin{figure}[t]
	\begin{center}
		\includegraphics[width=.45\textwidth]{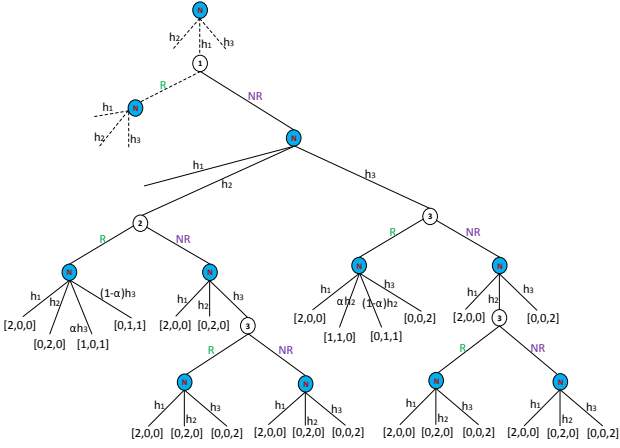}
		\caption{An illustration of the Bayesian mining game (adapted from~\cite{RePEc:chc:wpaper:0060}). Miner 1 believes that its is the real leader of the puzzle solving competition and decides to take action NR. Here $\alpha$ is the probability for miners to mine on the first block when they receive two blocks in a short time.}
		\label{fig:SequentialGame}
	\end{center}
\end{figure}

In~\cite{RePEc:chc:wpaper:0060}, the authors adopt a more general assumption of multiple selfish miners in a Bayesian game-based formulation\footnote{A Bayesian game~\cite[Chapter 4]{han2012game} describes the situation when players are of incomplete information. The players' payoffs are determined not only by their strategies but also by their types, which they may not be fully aware during the play.}. In the considered game, miners decide on whether to report a new block (R), i.e., to mine honestly, or not (NR), i.e., to mine selfishly. When a miner makes a decision, it does not know whether it is the real leader of the mining competition, or whether some other miners have secretly started mining on their private blocks. To ease the analysis of this mining game with incomplete information, the authors assume that a miner always reports when it finds two successive blocks. With this extra assumption, a decision tree can be constructed (see Figure~\ref{fig:SequentialGame}), and the backward induction approach is adopted to find the miners' equilibrium strategies. Figure~\ref{fig:SequentialGame} presents the decision tree in a case of three miners. In the presented subgame, miner 1 believes that it is the real leader of the mining competition. Here, let $h_i$ denote the normalized computational power of miner $i$, and $\mu_i(h_i)$ denote miner $i$'s belief of being the leader of the puzzle solution competition. From the decision tree and following the Bayesian rule, we can obtain the information about the states, transition probabilities, and expected payoffs after miner 1 takes the action of NR. The authors provide the condition on the fraction of computational power for action NR to become the optimal mining strategy.

\subsubsection{Block Withholding in Pool-Based Mining}
Block withholding (BWH) is a mining strategy used by selfish miners to increase their revenues through diminishing the winning probability of honest miners in mining pools~\cite{7973732, 7243747}. In~\cite{7243747}, the authors study the impact of BWH on the Bitcoin network. It is assumed that a selfish miner is able to split the computational power into different mining pools. It may spend most of its computational power to honestly mine on one pool, and use the rest computational power to perform BWH on the other pools. The mining pools are supposed to adopt the pay-per-share protocol~\cite[Section 2.2]{rosenfeld2011analysis}.
In the victim mining pools, the selfish miner submits all shares\footnote{A share is a preimage solution for a block that meets the relaxed (i.e., approximated) difficulty requirement set by the pool. A miner receives its reward in proportion to the number of shares that it submits to the pool.
} to the pool operators except the valid puzzle solutions. Although this mining strategy reduces the attacker's revenue in the attacked pools, it will increase the attacker's revenue in the pool that it chooses to mine honestly. A computational power splitting game with multiple players is formulated in~\cite{7243747}. In the game, one selfish miner adopts BWH and all the other miners mine honestly. The selfish miner chooses which pools to attack and how much computational power to allocate in the targeted pools. It is shown that the attacker always gains positive reward by mining dishonestly regardless of its mining power. This finding implies a risk for big mining pools to dominate the network through BWH attacks on smaller mining pools.

\begin{figure}[t]
	\begin{center}
		\includegraphics[width=.36\textwidth]{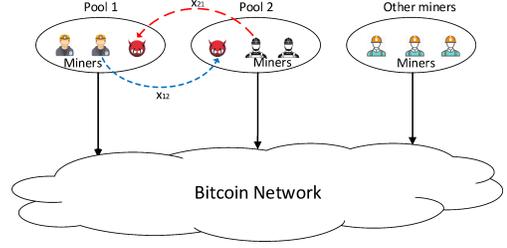}
		\caption{Block withholding attacks between two miners.}
		\label{fig:BWH_Attack}
	\end{center}
\end{figure}

The study in~\cite{7163020} considers a more complicated case where mining pools attack each other with BWH. The author of~\cite{7163020} considers a scenario of two mining pools which attempt to send their miners to each other to diminish their opponents. As illustrated in Figure~\ref{fig:BWH_Attack}, pool $P_1$ uses $x_{12}$ out of the $m_1$ computational power to attack pool $P_2$. Meanwhile, pool $P_2$ uses $x_{21}$ out of the $m_2$ computational power to attack pool $P_1$. Then, the revenue of each pool can be derived as follows:
\begin{equation}
\begin{aligned}
\label{eq:win_prob_BWH}
R_1 & = \frac{m_1 - x_{12}}{m- x_{12} - x_{21}} 	,\\
R_2 & = \frac{m_2 - x_{21}}{m- x_{12} - x_{21}} 	,
\end{aligned}
\end{equation}
where $m$ is the total mining power in the blockchain network. By~\cite{7163020}, the revenues of the pools can be expressed as the functions of $x_{12}$ and $x_{21}$:
\begin{equation}
\begin{aligned}
\label{eq:revenue_BWH}
r_1(x_{12},x_{21}) & = \frac{m_2 R_1 + x_{12} (R_1 + R_2)}{m_1 m_2 + m_1 x_{12} + m_2 x_{21}} ,\\
r_2(x_{21},x_{12}) & = \frac{m_1 R_2 + x_{21} (R_1 + R_2)}{m_1 m_2 + m_1 x_{12} + m_2 x_{21}} 	.
\end{aligned}
\end{equation}
Thus, by observing the attack rate of its opponent, a mining pool can adjust its attack rate in the next round to maximize its long-term revenue through repeated plays. The analysis of this repeated game reveals that the game admits a unique equilibrium, and the pool size will be the main factor that determines the attacking rates of each pool. A similar conclusion about the impact of the pool size on BWH attacks between two pools can also be found in~\cite{Laszka2015}.

Extended from the studies in ~\cite{7243747,7163020}, it is found out in~\cite{7728010} that when a mining pool performs a BWH attack to a victim mining pool, the other mining pools will benefit from this attack even if they do not adopt BWH. Thus, the other pools are interested in sponsoring the attacker to launch the BWH attack to the victim pool. Consequently, the expected gain of the attacker will be greater than the case in~\cite{7243747}. This implies that miners have more incentives to perform BWH attacks with the Nakamoto consensus protocols.

To alleviate the impact of BWH attacks, modifications to the Nakamoto protocol and the pool-mining protocols are suggested in the literature. The author in~\cite{rosenfeld2011analysis} proposes that the pool operator should insert mining tasks for which the solutions are known in advance, and tag the miners that do not submit the results. Since it is difficult to find puzzles with expected solutions, the author suggests that some new data fields should be added to the conventional block data structure (see Figure~\ref{fig_blockchain}). These fields enable the pool operator to allocate mining tasks to its miners, but the miners are unable to know the exact puzzle solutions. Alternatively, in~\cite{Bag2016}, the authors propose to give an extra reward to the miners that find the valid blocks, hence reducing the revenue of selfish miners and discouraging BWH attacks.

\begin{table*}[!thb]
  \scriptsize
	\caption{Summary of Selfish Mining Strategies and their Incurred Risks in Blockchain Networks} 
	\centering 
	\begin{tabular}{|l|c|c|c|c|}
		\hline
		\textbf{Attacks} & \textbf{Selfish mining} & \textbf{Block withholding} & \textbf{Lie-in-wait} & \textbf{Pool hopping} \\
		\hline
		References & \cite{Eyal2014, carlsten2016impact, GOBEL201623, 7467362, RePEc:chc:wpaper:0060, Sapirshtein2017} & \cite{7243747, 7163020, Laszka2015, 7728010,Bag2016} & \cite{rosenfeld2011analysis} &	\cite{rosenfeld2011analysis} \\
		\hline
		Concept & After finding a new block,  & After finding a new block & After finding a new block in &	The attacker moves to  \\
		& the attacker hides the block  &  in the victim pool, the  &  a mining pool, the attacker &  another pool or start mining  \\
		& and continues mining on & attacker discards that block  &   holds the block and uses &  by himself when the  \\
		& the mined block secretly.  & and continues mining on its &  all the computational power &	 mining time at its current \\
		&   &  block in another pool. &  to mine on that pool. &  pool reaches a threshold. \\
		\hline
		Risks  & A new attacker's found block & The attacker loses its  & The attacker can lose its &	There is no risk and   \\
		of &  can be discarded if one of & reward at the victim pool &  reward for its mined block & loss for the attacker if   \\
		attackers &  other miners finds a new block & if it finds a new block &  and all computational power & its mining pools use  \\
		& before it finds a next new block. &  in this pool. &  at the pool it found the block. & pay-per-share protocol. \\
		\hline
		Risks  & Lose their rewards for  & Lose their rewards & Can lose their rewards if the & Their profits will be  \\
		of & their mined blocks. & for blocks found  & block found by the attacker & reduced if they are in  \\
		honest &  & by attackers. & in their mining pool is & mining pools using  \\
		miners &  &   &  discarded from the network.  & pay-per-share protocol. \\
		\hline
		Suggested  & Modification to the mining protocol, & Modification to the task assignment & Modification to the task assignment & Change the payment  \\
		solutions &  e.g., blockchain propagation &   protocol in pools such that  & protocol in pools such that & method for mining pools. \\
		& method and blockchain update rule.  &  miners do not know real & miners do not know real &   \\
		&  & results of their mining tasks. & results of their mining tasks.  & \\
		\hline
	\end{tabular}
	\label{tab:setup_parameters}
\end{table*}
\subsubsection{Lie-in-Wait Mining in Pools}

Lie-in-wait (LIW) is a strategic attack where a selfish miner postpones submitting the block that it finds to a mining pool, and uses all of its computational power resources to mine on that pool~\cite{rosenfeld2011analysis}. In this case, an attacker is assumed to first split its computational power to mine in different pools. Then, if it finds a block in a pool, instead of submitting the block to get the reward from the pool, the attacker holds the block, and concentrates all of its computational power in other pools to mine on the pool where it finds the block. However, the attacker may take a risk by not releasing the block immediately and concentrating all the computational resources on the target pool. The reason is that if one of other pools finds a new block before this block is published, the selfish miner will lose its reward as well as suffer from the cost of mining in the target pool. It is shown in~\cite{rosenfeld2011analysis} that the success of attacks follows an exponential distribution, and the maximum expected gain of the LIW attacker is solely determined by the pool numbers and block interval in the network.


\subsubsection{Pool Hopping Strategy}

With the strategy of pool hopping, the miners exploit the vulnerability of the payment mechanism of mining pools to increase their own profits. With the pay-per-share protocol, the number of submitted shares in one block competition round follows a geometric distribution with success parameter $\delta$ and mean $D$~\cite{rosenfeld2011analysis}. For $I$ shares submitted to a pool, the pool still needs $D$ more shares on average to mine the block. When ignoring the transaction fees, the more shares submitted to a pool in a round, the less each share is worth. Since a miner immediately receives the payment for the submitted share, this implies that a share submitted early may have a higher reward. Therefore, a selfish miner can benefit by mining only at the early stage of a round, and then hop to other pools to increase his revenue. The study in~\cite{rosenfeld2011analysis} shows that there exists a critical point measured in the number of submitted shares. The best strategy of a selfish miner is to mine on a pool until this point is reached, then hop to another pool or mine by himself.

One straightforward way to address the block hopping problem in pay-per-share mining pools is to increase the value of shares at the end of each round. The pool
operator may score the shares according to the elapsed time since the beginning of each round. A share can be scored by an exponential score function $s(t)=e^{t/\delta}$, where $t$ is the time stamp of the submitted share and $\delta$ is a parameter controlling the scoring rate of shares. With the help of share scoring, we can handle pool hopping attacks in mining pools by decreasing the score of shares at the beginning and increasing the score of shares later. Such score-based method is also known as Slush's method and has been implemented in the mining pools such as Slushpool~\cite{Slushpool}. In~\cite{rosenfeld2011analysis}, other incentive mechanisms such as pay-per-last-N-shares and payment-contract-based methods are also sketched. However, analytical studies on these mechanisms are missing and their effectiveness in preventing pool hopping attacks still remain an open issue.

\section{Virtual Block Mining and Hybrid Consensus Mechanisms beyond Proof of Concepts}
\label{sec_consensus_II}
With the consensus protocols and the related issues reviewed in Sections~\ref{sec_consensus} and~\ref{sec:AMS}, a natural question arises regarding whether it is possible to simulate the random leader-election process among permissionless nodes in an approach other than under the framework of Nakamoto-like protocols. To answer this question, we focus on the designing methodology of the virtual-mining protocols in this section. Then, we further introduce a category of protocol design aiming at performance improvement by combining the properties of both the permissionless protocols and the classical BFT protocols.

\subsection{Proof of Stake and Virtual Mining}\label{sub_sec_POS}
The concept of PoS was first proposed by Peercoin~\cite{king2012ppcoin} as a modified PoW scheme to reduce the energy depletion due to exhaustive hash queries. Peercoin proposes a metric of ``coin age'' to measure the miner's stake as the product between the held tokens and the holding time for them. Miner $i$ solves a PoW puzzle as in (\ref{eq_puzzle_pow}) with an individual difficulty $D(h_i)$. The Peercoin kernel protocol allows a miner to consume its ``coin ages'' to reduce the difficulty i.e., $h_i$, for puzzle solution. The public verification of the ``coin ages'' is done through empirically estimating the holding time of the miner's Unspent Transaction Output\footnote{A UTXO is a transaction output whose value has not been spent by the receiver. It can be used as the input of a new transaction. Bitcoin-like networks sum up all the existing UTXOs of an account to recover its balance state.} (UTXO) based on the latest block on the public chain.

By completely removing the structure of PoW-based leader election, the protocols of pure PoS are proposed in~\cite{Bentov2016, Bentov:2014:PAE:2695533.2695545, kiayias2016provably, Kiayias2017}. To simulate a verifiable random function following the stake distribution (see also (\ref{eq_winning_prob})), an algorithm, follow-the-coin (a.k.a., follow-the-satoshi), has been proposed by~\cite{Bentov:2014:PAE:2695533.2695545} and widely adopted by these works\footnote{A reference implementation in Python (see also~\cite{Bentov:2014:PAE:2695533.2695545}) can be found at \url{http://www.cs.technion.ac.il/~idddo/test-fts.py}.}. Here, the terms ``coin'' or ``satoshi'' are used to indicate the minimum unit of the digital tokens carried by the blockchain. Briefly, all the tokens in circulation are indexed, for example, between 0 and the total number of available coins in the blockchain network. A simplified PoS protocol can use the header of block $t-1$ to seed the follow-the-coin algorithm and determine the random mining leader for block $t$. Specifically, the hash function $\mathcal{H}(\cdot)$ is queried with the header of block $t-1$, and the output is used as the random token index to initialize the searching algorithm. The algorithm traces back to the minting block (i.e., the first coinbase transaction~\cite{Bentov2016}) for that token or the UTXO account that currently stores it~\cite{Bentov:2014:PAE:2695533.2695545}. Then, the creator or the holder of the token is designated as the leader for generating block $t$. To enable public verification of the block, the valid leader is required to insert in the new block its signature, which replaces the data field ``nonce'' for PoW-based blockchains.

It is worth emphasizing that the pure PoS protocols do not rely on a Poisson process-based puzzle solution competition to simulate the random generator of the block leader. Therefore, the ZK puzzle-solving process can be simply replaced by the process of asymmetric key-based signing and verification, and the proof of resource is no longer needed. For this reason, PoS is also known as a process of ``virtual mining''~\cite{7163021} since the block miners do not consume any resources. In the literature, a number of protocol proposals are claimed to be able to (partially) achieve the same purpose. However, these protocols either need special hardware support, e.g., Intel SGX-enabled TEEs for proof of luck/elapsed-time/ownership~\cite{Milutinovic:2016:PLE:3007788.3007790, Chen2017}, or are still under the framework of PoW, e.g., Proof of Burn (PoB)~\cite{P4Titan2014}, Proof of Stake-Velocity (PoSV)~\cite{ren2014proof} and ``PoS'' using coin age~\cite{king2012ppcoin}. Strictly speaking, they cannot be considered as the real virtual mining schemes in permissionless blockchain networks.

Compared with the PoX-based protocols, PoS keeps the longest-chain rule but adopts an alternative approach for simulating the verifiable random function of block-leader generation. For this reason, the same framework for analyzing the properties of Byzantine agreements in PoW-based blockchain networks~\cite{Garay2015} can be readily used for the quantitative analysis of PoS protocols. For example, the investigations in~\cite{bentov2016snow, Kiayias2017} mathematically evaluate the properties of common prefix, chain quality and chain growth based on the same definition in Table~\ref{table_pox_properties}. The authors propose in~\cite{Kiayias2017} the ``Ouroboros'' protocol, and consider that the stakes are distributed at the genesis block by an ideal distribution functionality. By assuming an uncorrupted ideal sampling functionality, Ouroboros guarantees that a unique leader is elected in each block generation round following the stake distribution among the stakeholders (see also (\ref{eq_winning_prob})). With Ouroboros, forking no longer occurs when all the nodes are honest. However, when adversary exists, forking may be caused by the adversarial leader through  broadcasting multiple blocks in a single round. The study in~\cite{Kiayias2017} shows that the probability for honest nodes to fork the blockchain with a divergence of $k$ blocks in $m$ rounds is no more than $\exp(-\Omega(k)+\ln(m))$ under the condition of honest majority. It is further shown that the properties of chain growth and chain quality are also guaranteed with negligible probability of being violated.

The studies in~\cite{Bentov:2014:PAE:2695533.2695545, bentov2016snow} introduce the mechanism of epoch-based committee selection, which dynamically selects a committee of consensus nodes for block generation/validation during an epoch (i.e., a number of rounds). Compared with the single-leader PoS protocol, i.e., Ouroboros~\cite{Kiayias2017} and its asynchronous variation~\cite{10.1007/978-3-319-78375-8_3}, the committee-based PoS gears the protocol design toward the leader-verifier framework of traditional BFT protocols (see also Figure~\ref{fig_blockchain_PBFT}). In~\cite{Bentov:2014:PAE:2695533.2695545}, the scheme of Proof of Activity (PoA) is proposed with the emphasis that only the active stake-holding nodes get rewarded. The PoA is featured by the design that the leader is still elected through a standard PoW-based puzzle competition, and is only responsible for publishing an empty block. Using the header of this block to seed the follow-the-coin algorithm, a committee of $N$ ordered stakeholders is elected and guaranteed to be publicly verifiable. The first $N-1$ stakeholders work as the endorsers of the new empty block by signing it with their private keys. The $N$-th stakeholder is responsible for including the transactions into that block. The transaction fees are shared among the committee members and the block miner. In this sense, PoA can be categorized as a hybrid protocol that integrates both PoW and PoS schemes.

In~\cite{bentov2016snow}, the authors propose a protocol called ``Snow White'', which uses a similar scheme to select a committee of nodes as in~\cite{Bentov:2014:PAE:2695533.2695545}. However, only the selected committee members are eligible for running for the election of the block generation leader. Under the Snow White protocol, the leader of an epoch is elected through a competition based on repeated preimage search with the hash function. At this stage, the difference of Snow White from the standard PoW puzzle in (\ref{eq_puzzle_pow}) is that the hash function is seeded with the time stamp instead of an arbitrary nonce. Like PoA, Snow White also pertains the characteristics of a hybrid protocol. The analysis in~\cite{bentov2016snow} shows that the proposed protocol supports frequent committee reconfigurations and is able to tolerate nodes that are corrupted or offline in the committee.

The recent proposal by Ethereum, Casper~\cite{buterin2017casper} provides an alternative design of PoS that is more similar to traditional BFT protocols. The current proposal of Casper does not aim to be an independent blockchain consensus protocol, since it provides no approach of leader election for block proposal. Instead, the stakeholders join the set of validators and work as the peer nodes in a BFT protocol. The validators can broadcast a vote message specifying which block in the blockchain is to be finalized. The validator's vote is not associated with its identity, but with the stake that it holds. According to~\cite{buterin2017casper}, Casper provides plausible liveness (instead of probabilistic liveness with PoW) and accountable safety, which tolerate up to $1/3$ of the overall voting power (weighted by stake) that is controlled by the Byzantine nodes.

\subsection{Issues of Incentive Compatibility in PoS}\label{subsec_issue_pos}
Regarding the incentive compatibility of PoS, an informal analysis in~\cite{Kiayias2017} shows that being honest is a $\delta$-Nash equilibrium\footnote{At a $\delta$-NE, the payoff of each player is within a distance of $\delta>0$ from the equilibrium payoff.} strategy when the stakes of the malicious nodes are less than a certain threshold and the endorsers are insensitive to transaction validation cost. However, a number of vulnerabilities are also identified in PoS. In~\cite{Li2017}, the nothing-at-stake attack is considered. In order to maximize the profits, a block leader could generate conflicting blocks on all possible forks with ``nothing at stake'', since generating a PoS block consumes no more resource than generating a signature. A dedicated digital signature scheme is proposed to enable any node to reveal the identity of the block leader if conflicting blocks at the same height are found. Alternatively,  a rule of ``three strikes'' is proposed in~\cite{Bentov2016} to blacklist the stakeholder who is eligible for block creation but fails to properly do so for three consecutive times. In addition, an elected mining leader is also required to sign an auxiliary output to prove that it provides some extra amount tokens as the ``deposit''. In case that this node is malicious and broadcasts more than one block, any miner among the consecutive block creation leaders can include this output as an evidence in their block to confiscate the attacker's deposit. Such a scheme is specifically designed to disincentivize block forking by the round leader.

Grinding attack is another type of attacks targeting PoS~\cite{Kiayias2017}. With PoS, the committee or the leader is usually determined before a round of mining starts. Then, the attacker has incentive to influence the leader/committee election process in an epoch to improve its chances of being selected in the future. When the verifiable random generator takes as input the header of the most recent block for leader/committee election, the attacker may test several possible block headers with different content to improve the chance of being selected in the future (e.g.,~\cite{Kiayias2017,Bentov:2014:PAE:2695533.2695545}). It is expected to use an unbiased, unpredictable random generator to neutralize such a risk~\cite{Kiayias2017}. In practice, the protocol usually selects an existing block that is a certain number of blocks deep to seed the random function instead of using the current one~\cite{Bentov:2014:PAE:2695533.2695545,bentov2016snow}.

With all the aforementioned studies, a significant limit of the existing analyses about PoS-based protocols lies in the simplified assumption that ignores the stake trade outside the blockchain network (e.g., at an exchange market)~\cite{poelstra2014distributed}. A study in~\cite{houy2014will} provides a counterexample for the persistence of PoS in such a situation. The study in~\cite{houy2014will} assumes no liquidity constraint in a blockchain network, where nodes own the same stake at the beginning stage. The author of~\cite{houy2014will} considers a situation where a determined, powerful attacker attempts to destroy the value of the blockchain by repeatedly buying the stake from each of the other nodes at a fixed price. After taking into account the belief of the nodes that the attacker will buy more tokens, the interaction between the attackers and the stakeholders is modeled as a Bayesian repeated game. The study concludes that the success of the attack depends on two factors, namely, the attacker's valuation of the event ``destroying the blockchain'' and the profit (e.g., monetary interest) that the nodes can obtain from holding the stake. When the former factor is large and the latter is small, the nodes in the network will end up in a competition to sell their stakes to the attackers. As a result, the blockchain can be destroyed at no cost.

\subsection{Hybrid Consensus Protocols}\label{subsec_hybrid}
Despite the unique characteristics of permissionless consensus protocols, public blockchain networks are known to be limited in performance (e.g., transaction throughput) due to the scalability-performance tradeoff~\cite{Vukolic2016}. To boost permissionless consensus without undermining the inherent features such as scalability, a plausible approach is to combine a permissionless consensus mechanism (e.g., Nakamoto protocol) with a fast permissioned consensus protocol (e.g., BFT). Following our previous discussion (cf. PoA~\cite{Bentov:2014:PAE:2695533.2695545} and Casper~\cite{buterin2017casper}), we study in this subsection how a standard permissionless consensus protocol can be improved by incorporating (part of) another consensus protocol in the blockchain networks.

In~\cite{194906}, the protocol ``Bitcoin-NG'' is proposed to extend the PoW-based Nakamoto protocols. The prominent feature of Bitcoin-NG is to decouple the consensus process in a blockchain network (e.g., Bitcoin network) into two planes: leader election and transaction serialization. To bootstrap the transaction throughput, the protocol introduces two types of blocks, namely, the key blocks that require a PoW puzzle solution for leader election and the microblocks that require no puzzle solution and are used for transaction serialization. The time interval between two key blocks is known as an epoch. In an epoch, the same leader is allowed to publish microblocks with the limited rate and block size. Although operation decoupling in Bitcoin-NG does not ensure strong consistency, it paves the way for incorporating additional mechanisms on the basis of standard Nakamoto protocols.

Following the methodology of~\cite{194906}, hybrid consensus mechanisms atop Nakamoto protocols are proposed in~\cite{Decker:2016:BMS:2833312.2833321, pass_et_al:LIPIcs:2017:8004} with the goal of providing strong consistency and immediate finality. In~\cite{Decker:2016:BMS:2833312.2833321}, the ``PeerCensus'' protocol is proposed by decoupling block creation and transaction committing/confirmation. PeerCensus consists of two core components, namely, a PoW scheme named as BlockChain (BC) and a BFT-based scheme named as Chain Agreement (CA). With the proposed BC protocol, nodes acquire the voting right of the CA protocol when they propose new blocks through PoW and are approved by the committee of CA. The CA protocol is adapted from BFT protocols such as PBFT~\cite{Castro:2002:PBF:571637.571640} and the Secure Group Membership Protocol (SGMP)~\cite{481515}. Through the four stages of propose, pre-prepare, prepare, and commit of BFT protocols (cf. Figure~\ref{fig_blockchain_PBFT}), CA designates the miner of the newest block in the chain as the leader for the next block proposal. The leader proposes one from the multiple candidate blocks obtained in BC. The peer nodes in the committee extend the pre-prepare stage with an operation of block validation. The design of PeerCensus ensures that committing transactions (i.e., CA) is independent of block generation (i.e., BC). Therefore, no forking occurs in the condition of honest majority and strong consistency is guaranteed.

\begin{figure}[t]
\centering     
\includegraphics[width=.35\textwidth]{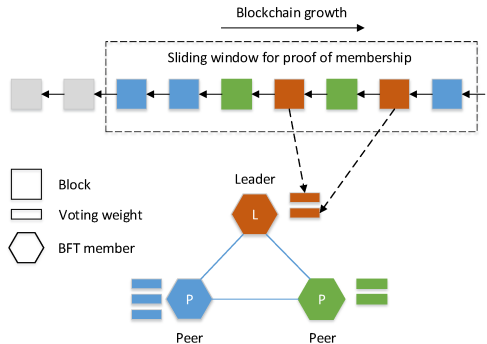}
\caption{Illustration of BFT-committee formation with weighted voting power. Valid weights are only credited to the miners of the blocks in the sliding window (adapted from~\cite{kogias2016enhancing}).}
\label{fig_sliding_window_blockchain}
\end{figure}

In~\cite{pass_et_al:LIPIcs:2017:8004}, a hybrid consensus protocol is proposed by combining the data framework of two-type blocks in Bitcoin-NG and the hybrid PoW-BFT design in PeerCensus. As in PeerCensus, the Nakamoto protocol is used to construct a ``snailchain'', which is allowed to commit transactions from a specific mempool of outstanding transactions known as the ``snailpool''. Following the quantitative analysis of the common prefix blocks in a chain in~\cite{Garay2015}, only a fixed number of miners whose recently minted blocks are a certain number of blocks deep in the chain can be used to form the committee for the BFT protocol. In contrast to PeerCensus, the BFT committee of miners in the proposed protocol has no influence on how the next block on the snailchain is determined. Instead, it is responsible for committing transactions from an independent mempool known as the ``txpool''. For this reason, the transactions approved by the BFT protocol are committed off the snailchain without relying on any mining mechanism. In this sense, these transactions can be considered similar to those in the microblocks of Bitcoin-NG. The hybrid consensus protocol in~\cite{pass_et_al:LIPIcs:2017:8004} explicitly addresses the problem of BFT-committee scalability in PeerCensus and provides a secured (with theoretical proof) consensus property of immediate finality. Namely, the transaction confirmation time from the txpool only depends on the network's actual propagation delay. The method of using Nakamoto protocols to select nodes into a BFT committee is also known as the proof of membership mechanism~\cite{kogias2016enhancing}. A sliding-window mechanism is proposed in~\cite{kogias2016enhancing} to generalize the mechanisms of dynamic BFT-committee selection in~\cite{pass_et_al:LIPIcs:2017:8004, Decker:2016:BMS:2833312.2833321}. As illustrated in Figure~\ref{fig_sliding_window_blockchain}, the BFT committee is maintained by a fixed-size sliding window over the PoW-based blockchain. The sliding window moves forward along the blockchain as new blocks are appended/confirmed. Consensus nodes minting multiple blocks in the window are allowed to create the same number of pseudo-identities in the BFT consensus process to gain the proportional voting power.

For hybrid consensus using BFT protocols to guarantee strong consistency, a natural thinking is to replace the Nakamoto protocols with virtual mining (e.g., PoS) for selecting the leader or committee in BFT-consensus processes. A typical example for such an approach can be found in the ``Tendermint'' protocol~\cite{kwon2014tendermint}, where a node joins the BFT committee of block validators by posting a bond-deposit transaction. The validator no longer needs to prove its membership by competing for the PoW-puzzle solution. Alternatively, its voting power is equal to the amount of stake measured in bonded tokens. Meanwhile, instead of randomly electing the leader of block proposal in the committee (cf.~\cite{194906}), Tendermint adopts a round-robin scheme to designate the leader in the committee. The similar design can be found in a number of recent proposals such as Proof of Authority (PoAu)~\cite{PoAu} and Delegated Proof of Stake (DPoS)~\cite{larimer2014delegated}. To generalize the mechanisms of BFT-committee selection based on virtual mining, the authors in~\cite{Gilad:2017:ASB:3132747.3132757} further propose a consensus protocol called ``Algorand''. Like the other hybrid protocols, Algorand relies on BFT algorithms for committing transactions. It assumes a verifiable random function to generate a publicly verifiable BFT-committee of random nodes, just as in~\cite{Bentov:2014:PAE:2695533.2695545}. The probability for a node to be selected in the committee is in proportion to the ratio between its own stake and the overall tokens in the network. For leader election, Algorand allows multiple nodes to propose new blocks. Subsequently, an order of the block proposals is obtained through hashing the random function output with the nodes' identities specified by their stake. Only the proposal with the highest priority will be propagated across the network.

In Table~\ref{table_hybrid_comparison}, we provide a summary of the virtual-mining mechanisms and the hybrid consensus protocols discussed in this section.

\begin{table*}[t]
  \centering
  \scriptsize
  \caption{Summary of Virtual Mining and Hybrid Consensus Protocols for Permissionless Blockchains}
  \renewcommand{\arraystretch}{1.3}
 \begin{tabular}{|p{2.4 cm} | p{1.0cm}| p{1.2cm} | p{2.5 cm}| p{1.5 cm}| p{2.6cm} | p{3.0cm}|}
 \hline
 \makecell{Protocol Name} & \makecell{Virtual \\ Mining} & \makecell{Hybrid \\ Consensus} & \makecell{Simulating Leader \\ Election with} & \makecell{Rule of\\ Longest Chain} &\makecell{Decoupling Block\\ Proposal from \\ Transaction Commitment\\ } & \makecell{Featured Consensus \\ Properties} \\
 \hline
 Proof of stake~\cite{Bentov2016, kiayias2016provably, Kiayias2017} & Yes & No & Verifiable random function, e.g., follow-the-coin & Yes & N/A & No resource consumption \\
 \hline
 Proof of luck, elapsed-time and ownership~\cite{Milutinovic:2016:PLE:3007788.3007790, Chen2017}  & Yes & No &  Trusted random function implemented by Intel-SGX-protected enclave & Yes & N/A &  No resource consumption. Special hardware support is needed\\
 \hline
 Proof of burn~\cite{P4Titan2014}  & Partially & No &  PoW puzzle competition & Yes & N/A & Reduced resource consumption \\
 \hline
 Proof of stake-velocity~\cite{ren2014proof}  & Partially & No & PoW puzzle competition & Yes & N/A & Reduced resource consumption \\
 \hline
 Snow White~\cite{bentov2016snow}  & Partially & PoS-PoW & Modified preimage search with the hash function & Yes & N/A & Robust consensus through reconfigurable PoS committee \\
 \hline
 Proof of activity~\cite{Bentov:2014:PAE:2695533.2695545}  & Partially & PoW-PoS &  PoW puzzle competition for empty block proposal & Yes & Transactions are committed by a random group of stakeholders & Higher cost for  attackers to compromise the network consensus than PoW/PoS \\
 \hline
 Casper~\cite{buterin2017casper}  & No & PoW-PoS &  PoW puzzle competition &  Yes & N/A & Validators use BFT protocols to anchor checkpoint blocks in the block tree\\
 \hline
 Bitcoin-NG~\cite{194906}  & No & Partially & PoW puzzle competition & Yes & Proposals of microblocks do not need PoW solutions & Leader election is only performed at key blocks \\
 \hline
 PeerCensus~\cite{Decker:2016:BMS:2833312.2833321} & No & PoW-BFT &  PoW puzzle competition & N/A & Yes, Blocks are committed by BFT committees  & Strong consistency without blockchain forking \\
 \hline
 Hybrid consensus protocol~\cite{pass_et_al:LIPIcs:2017:8004} & No & PoW-BFT &  PoW-puzzle competition in the snailchain &  Yes & Partially, only the transactions in txpools are committed following BFT protocols & Immediate finality \\
 \hline
 Tendermint~\cite{kwon2014tendermint}, Proof of authority~\cite{PoAu} and delegated proof of stake~\cite{larimer2014delegated}  & Yes & PoS-BFT & Verifiable random function or deterministic mechanism & N/A &Yes, following typical BFT protocols  & Deterministic consensus properties \\
 \hline
 Algorand~~\cite{Gilad:2017:ASB:3132747.3132757} & Yes & PoS-BFT & Verifiable random function & N/A & Yes, following typical BFT protocols & Safety and liveness are guaranteed under strong synchrony \\
 \hline
\end{tabular}
\label{table_hybrid_comparison}
\end{table*}

\section{Relaxed and Parallel Consensus Protocols for Performance Scalability}
\label{sec_consensus_III}
\begin{figure}[t]
\centering     
\includegraphics[width=.35\textwidth]{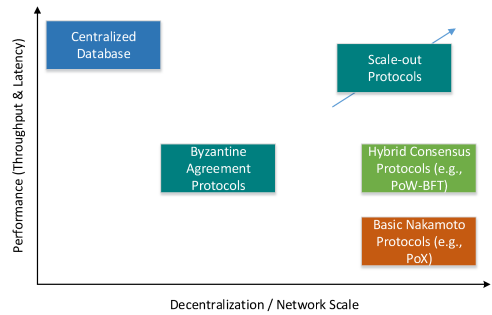}
\caption{ Illustration of performance and scalability of different consensus protocol families (see also the discussion in~\cite{Vukolic2016}).}
\label{fig_performance_vs_scalability}
\end{figure}

So far, we have surveyed the design methodologies of various consensus protocols, especially for permissionless blockchains. As our discussion indicates, the BFT-based consensus mechanisms achieve high transaction throughput with immediate finality at the cost of high message complexity. Thus, they are restricted to small numbers of replicas and offer limited network scalability in terms of the number of consensus nodes. In contrast, the permissionless protocols surveyed in Sections \ref{sec_consensus} and \ref{sec_consensus_II} provide good network scalability with low message complexity. However, most of the Nakamoto-like protocols (except the hybrid protocols guaranteeing immediate finality~\cite{Decker:2016:BMS:2833312.2833321, pass_et_al:LIPIcs:2017:8004}) provide only probabilistic consensus finality. As a result, consistency of replicas across the entire network (cf. the consistency condition for the PoW-based protocol in (\ref{eq_consistency_condition})) is maintained at the cost of low transaction throughput and high latency. Figure~\ref{fig_performance_vs_scalability} provides a descriptive illustration of the scalability levels of different protocol families with respect to both performance and network size. For the protocols surveyed in our previous sections, network scalability and transaction throughput are generally considered as two performance indices that can only be attained at the cost of each other. In this section, we aim to review the solutions that scale out the throughput of a permissionless blockchain as the size of the network increases.

\subsection{Off-chain and Side-chain Techniques}\label{sub_sec_off_chain}
For cryptocurrencies, one popular and straightforward approach to throughput enhancement is to adjust the parameters, e.g., the block size and confirmation time in Nakamoto-like protocols. A typical example of this approach can be found in the Segregated Witness proposal (SegWit)~\cite{segwit2015} for Bitcoin soft fork, which lifted the block-size limit from 1MB to 4MB. However, the study in~\cite{Croman2016} points out that such a reparameterization approach is constrained by the network's bandwidth (e.g., for block size) as well as the blockchain's security requirement (e.g., confirmation time). Thus, such an approach does not really scale out the throughput as the network size increases. With the emphasis on compatibility to the existing consensus protocol or network realization, alternative approaches, e.g., the Lightning network~\cite{poon2016bitcoin}, that aim to lower the frequency of global block validation/synchronization, are proposed by the development communities, specifically for value transfer networks.

The Lightning network~\cite{poon2016bitcoin} and its variations such as Blind Off-chain Lightweight Transactions (Bolt)~\cite{Green:2017:BAP:3133956.3134093} and the TEE-based Teechain~\cite{lind2017teechain} introduce the concept of (bidirectional) micro-payment channels between two nodes via untrusted intermediary relays. Specifically, the payment channels are realized as logical channels overlaying on the existing blockchains (e.g., on Bitcoin~\cite{poon2016bitcoin} or on ZCash~\cite{Green:2017:BAP:3133956.3134093}) and therefore do not modify the underlying consensus protocols. The value transfer between the two end nodes on each channel is kept ``off-chain'' as a local sequence of mutually-agreed balance-state updates, also known as commitment transactions~\cite{poon2016bitcoin}. In other words, the sequence of transactions on an established channel are not broadcast to the entire network and kept locally between the two end nodes as well as the intermediaries when needed. Then, transactions of value transfer over a channel are not confirmed as normal transactions and cannot be spend until the ``closure'' of the channel. When closing the channel, only the most recent commitment transaction is broadcast and needs to be mined by the blockchain network. By doing so, the requirement of validating/synchronizing every transaction across the network is relaxed and the number of transactions to be mined is greatly reduced, hence making the underlying blockchain network more throughput-scalable.

Due to the lack of trust, simply relaxing the consensus requirement and keeping transactions in local payment channels will incur the risk of double spending. To address this problem, the technique of $2$-of-$2$ multisignature\footnote{An $m$-of-$n$ ``multisig'' transaction requires the verification of a tuple of at least $m$ signatures for the same text from $n$ corresponding public keys~\cite{okupski2014bitcoin}.} is enforced in the Lightning networks and a number of specifically designed smart contracts (i.e., scripts in Bitcoin) are introduced. To establish a channel, a funding transaction has to be created jointly by the end parties and broadcast to the network in order to lock their submitted tokens in escrow. An order of broadcast is defined by creating for each party a different version of every subsequent commitment transaction, i.e., in the form of a half-signed transaction containing only the signature of the counterparty, with the same balance outputs. An accompanying revocable transaction\footnote{A revocable transaction has two payout paths. If both parties of it agree, its output can be spent immediately. Otherwise if after a certain waiting time one or both parties do not broadcast, the fund can be redeemed. It is revoked only when both parties agree to update with a superseding transaction.} is also created to enable updating the balance changes. It also provide a means of revoking transactions in case a violation occurs or a waiting time limit is reached. In normal scenarios, only the latest commitment transaction is broadcast to close the channel. Otherwise, by broadcasting the right version of revocable transactions, one end node is able to provide the publicly verifiable proof of recognizing a malicious behavior by the counterparty, and claim all of its deposit in the funding transaction as a punishment.

Other than the off-chain schemes that aim to reduce the amount of transactions over the network, an alternative design is to extend an existing blockchain-based value transfer network with multiple ``side-chains''~\cite{back2014enabling}. A side-chain is an independent blockchain network that validates a subset of transactions and keeps track of the corresponding assets.  Such a design introduces parallelism into the existing network and each side-chain is only responsible for validating a fraction of the total amount of transactions in the network. Therefore, it is able to increase the transaction throughput by adding more side-chains. As in the off-chain techniques, side-chains do not modify existing consensus protocols. Instead, the fundamental goal is to enable bidirectional atomic value transfer between side-chains. More specifically, any value transaction between side-chains is either completely confirmed by both side-chains or not at all. Meanwhile, the value carried by the transaction can be imported from and returned to a side-chain with no risk of double spending. To achieve such a goal (also known as ``two-way peg'' in~\cite{back2014enabling}), special proofs of value locking and redeeming are needed whenever inter-chain transfer happens. Especially, since the consensus nodes of the receiving side-chain usually do not track the state changes of the sending side-chain, providing a compact, non-interactive proof of events occurring on side-chains becomes the utmost concern of the network designers.

\begin{figure}[t]
\centering     
\includegraphics[width=.40\textwidth]{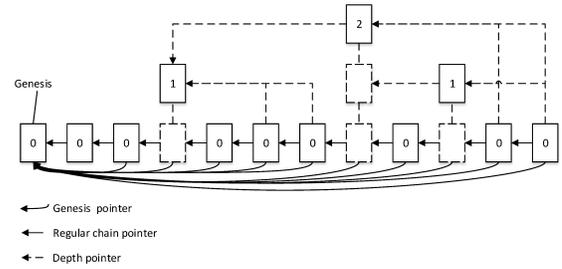}
\caption{A graphical example of the hierarchical blockchain with levels 0, 1 and 2. A block with header $bh$ is of level $\mu$ if $bh < D(h)/2^{\mu}$ (see also (\ref{eq_puzzle_pow})). Besides the regular hash pointer to the previous block, a block of level  $\mu$ also maintains a list of hash pointers (interlinks) to the most recent preceding blocks in every level $\mu'$ such that $\mu'>\mu$. The genesis block is defined to be of infinite level and hence every other block has to include a pointer to it.}
\label{fig_super_chain}
\end{figure}

In~\cite{back2014enabling}, the Simplified Payment Verification (SPV) proof is adopted from~\cite{nakamoto2008bitcoin} based on the proof-of-inclusion path in Merkle trees to provide compact proofs of value locking for atomic transfer (cf. Figure~\ref{fig_Merkle_tree}). Further enhancement of the proof is also proposed in~\cite{back2014enabling} by introducing a trusted cross-chain federation of mutually distrusting functionaries (i.e., approving nodes). Out of the federation, the majority vote in the form of an $m$-of-$n$ multisignature is used to replace the SPV proof for locking/redeeming a cross-chain pay-to-contract transaction. Furthermore, an SPV proof is accompanied by an array of block headers, whose parent is the block containing the SPV-locked transaction on the sending side-chain. This can be informally considered as a ``proof of PoW'' shown to the receiving side-chain that the transaction in concern is sufficiently deep in the sending side-chain and thus safely locked (see also our discussion about (\ref{eq_consistency_condition})). In~\cite{kiayias2017non}, a formal primitive called Non-Interactive-Proofs-of-Proof-of-Work (NIPoPoW) is proposed to fill the gaps of compactness and non-interactiveness in the proposal of~\cite{back2014enabling} for PoW-based side-chain networks. To avoid tracking/validating every block on the sending side-chain, the study in~\cite{kiayias2017non} proposes to replace the linear list-based blockchains with a skiplist-like data structure called interlink (see Figure~\ref{fig_super_chain} and also~\cite{10.1007/978-3-662-53357-4_5}). As with SPV, a valid NIPoPoW of transaction confirmation also contains an array of blocks (i.e., suffix proof) preceded by the block in concern as a stability proof of that block in the chain. Instead of validating the entire source side-chain, NIPoPoW only has to include $2m$ blocks in expectation from each level of the hierarchical blockchain in the proof. Here, $m$ is a system-determined security parameter to ensure that for every level $\mu$, the proof only needs to include a number of blocks from the tail of level $\mu$ to span the last $m$-size suffix of blocks in the higher level $\mu+1$. Compared with a secured SPV proof for inter-chain transaction, with NIPoPoW the number of source-chain blocks tracked by the receiving side-chain is only a polylogarithmic function of the source side-chain's length.

\subsection{Sharding for Scale-out Throughput}\label{sub_sec_sharding}
Inspired by the infrastructures of distributed database and cloud, the concept of ``sharding''~\cite{Croman2016} is also applied to the blockchain networks. As in side-chain networks, the approach of sharding partitions the global blockchain state into parallel subsets (i.e., shards), and each shard is maintained by a sub-group (i.e., committee) of nodes instead of the entire network. To improve the transaction throughput as well as retain the open-membership nature of permissionless blockchains, multiple BFT committees can be constructed following a similar procedure of the hybrid protocols (cf. Section~\ref{subsec_hybrid}). As a result, the sharding protocols generally face the same challenges as in side-chain networks and hybrid consensus protocols, i.e., in providing secured shard formation to guarantee permissionless decentralization and in providing cross-shard synchronization to guarantee atomic transactions.

The study in~\cite{ren2018scale} adopts the UTXO structure from Bitcoin and proposes the ``spontaneous sharding'' mechanism specifically for value transfer networks. Spontaneous sharding introduces a level of individual (spontaneous) chains for each node to maintain its own transactions of interest in a first-in-first-out fashion. It keeps a globally shared main chain, which only records the signed abstracts (i.e., header) of the blocks on each individual chain using a BFT-based consensus protocol. In this sense, spontaneous sharding is considered to be a transitional design from micro-payment channels to sharding, since it admits only the transaction-sharding process but not the validator-sharding process. The validity of the proposed mechanism is built upon the assumption that all nodes in the network are rational. Namely, a node is interested in inspecting a transaction only if it needs that transaction to validate a subsequent transaction output that it receives. Only the rational owner of an unspent transaction is responsible for providing the proof to the validators (i.e., receivers). However, due to the existence of sharded individual chain, the protocol in~\cite{ren2018scale} faces an unresolved problem of lacking compact proof (cf.~\cite{kiayias2017non}), since for every proof, the validators have to trace back to the genesis block of each related individual chain.

\begin{figure}[t]
\centering     
\includegraphics[width=.40\textwidth]{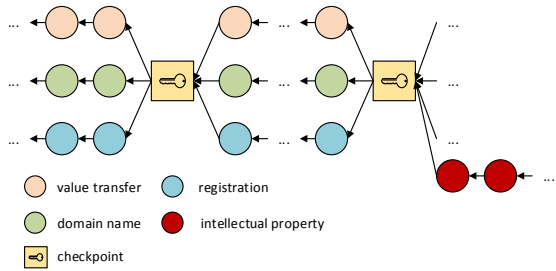}
\caption{Service oriented sharding with multi-chain structure (adapted from~\cite{10.1007/978-3-319-70972-7_22}). PoW solution is required for generating a checkpoint. Users are able to propose new services by posting transactions to register the corresponding channels in a checkpoint block (see the sub-blockchain for the ``intellectual property'' service).}
\label{fig_shard_service}
\end{figure}

In~\cite{10.1007/978-3-319-70972-7_22}, a different approach of transaction sharding is proposed under the name of ``Aspen''. Instead of maintaining an individual chain for each node, Aspens organizes transactions into sub-blockchains (see Figure~\ref{fig_shard_service}) based on the type of services related to each transaction. It introduces periodic checkpoint blocks for synchronizing sub-blockchains (cf. the anchor points in Casper~\cite{buterin2017casper}). Aspen is instantiated on Bitcoin-NG~\cite{194906} and only requires the checkpoint blocks to be generated upon PoW-puzzle solution to determine the proposal leaders of micro-blocks in each service channel (i.e., sub-blockchain). To avoid designing complex proofs of cross-chain transactions (cf.~\cite{back2014enabling,ren2018scale}), Aspen does not allow two-way transfer between channels and requires that each fund is only spendable in a specific channel.

In~\cite{Luu:2016:SSP:2976749.2978389}, a different sharding protocol named ``Elastico'' is proposed with the emphasis on the process of validator sharding through dynamically forming multiple BFT-committees. Elastico organizes the transaction approving process by epochs, and in each epoch a number of committees are formed in parallel based on the PoW-puzzle solution in a similar way to the proof of membership in~\cite{kogias2016enhancing}. The study in~\cite{Luu:2016:SSP:2976749.2978389} proposes a mechanism of generating distributive epoch randomness by using one network-level BFT committee, which determines a subset of hash values randomly provided by its members. The committee can run any non-leader interactive consistency protocol, e.g.,~\cite{pease1980reaching} to reach an agreement on such a single set to generate the public random number. In an epoch, the candidates of the committees have to solve the PoW puzzle based on the public random number. Elastico also uses the least-significant bits of the PoW solution (i.e., the hash value) to group the candidate nodes into different committees. Thus, this procedure guarantees that the committees are randomly formed and unpredictable. Meanwhile, to avoid designing complex proofs of cross-shard transactions (cf.~\cite{ren2018scale}), Elastico relies on the network-level committee to merge the locally agreed values in each committee into a single chain. The network-level committee first checks whether the values received from each local committee are signed by their majority members. If so, it merges the received values into an ordered union and runs a similar BFT protocol to approve the final result with signatures by the committee majority. By limiting the burden of quadratic message complexity within shard committees of small size, Elastico is able to achieve roughly $O(n)$ message complexity and provide almost linear throughput scalability in terms of the hash power in the network. Also, compared with the aforementioned throughput-scalable protocols, e.g.,~\cite{ren2018scale, back2014enabling, okupski2014bitcoin}, Elastico does not limit itself to value transfer networks and can be applied to generic data services with non-spendable transactions.

By enabling parallelization of both data storage and network consensus, protocols aiming at ``full sharding'' are proposed in~\cite{KokorisKogias:255586, cryptoeprint:2018:460}. In~\cite{KokorisKogias:255586}, a protocol named ``OmniLedger'' is designed to provide ``statistically representative'' shards for permissionless transaction processing. As in~\cite{Luu:2016:SSP:2976749.2978389}, OmniLedger is built upon two levels of epoch-based Byzantine agreement processes, with the network level being responsible for epoch randomness generation and the shard level for intra-committee consensus. In the network level, a global identity blockchain is adopted and can only be extended by the network-level leaders. Any node that wants to join a committee has to register to this global blockchain through a Sybil-proof identity establishment mechanism. Especially, such a mechanism is not limited to PoW and can be replaced by other means, e.g., PoS. At the beginning of an epoch, all the nodes with established identities are required to run an interactive consistency protocol by sharing with each other a ``ticket'' based on a gossip protocol. The ticket is generated as the hash value of the node's address and the header of the identity blockchain. The node that generates the smallest ticket will be elected as the network-level leader. The leader is expected to run a verifiable random function (e.g., RandHound~\cite{7958592}) and generate a global random string with a valid proof. Upon reception of this random string, other registered nodes are able to compute a permutation based on this string as well as their own identity, and then finish the assignment of shard committees by subdividing their results into equally-sized buckets. In addition, OmniLedger proposes to swap gradually in-and-out committee members per epoch. This design not only allows bootstrapping new nodes joining the network, but also avoids excessive message overhead and latency due to complete shard reconstruction (cf. Elastico). In the shard level, a committee can employ any leader-based BFT protocol (e.g., ByzCoin~\cite{kogias2016enhancing}) to provide intra-shard consensus.

In~\cite{cryptoeprint:2018:460}, another epoch-based, two-level-BFT protocol for full sharding is proposed under the name ``RapidChain''. In the network level, RapidChain requires a reference BFT-committee to run a distributed randomness generation protocol similar to~\cite{Luu:2016:SSP:2976749.2978389} and generate a public random string to initialize the formation of shard-level committees. As in~\cite{KokorisKogias:255586}, the shard-level committee reconfiguration in RapidChain only reorganizes a subset of committee members at each epoch to ensure operability during committee transition. At the bootstrapping stage in a network of $n$ nodes, the established identity of a node is mapped to a random position in the range $[0,1)$ by using the hash function. Then, with some constant $k$ (i.e., committee size), the range is partitioned into $n/k$ regions, and the shard-level committees are consequently formed based on this region partition. At the reconfiguration stage, RapidChain defines the set of the first half shard-level committees with more active members as the ``active committee set''. The network-level committee is responsible for assigning new nodes into the active shard-level committees uniformly at random. After that, it shuffles a constant number of members from every existing committee and randomly reassign them to other committees. On the shard level, RapidChain requires the members of each BFT-committee to run also the distributed randomness generation protocol and generate a local random string. Then, the committee members compete for the leader election through solving the standard PoW puzzle based on the local random string. The members elect a node with the smallest PoW solution by gossiping their votes with signatures to each other. Then, the BFT protocol will be led by that node to reach the intra-shard consensus for transaction commitment.

\begin{figure}[t]
\centering     
\subfigure[]{\label{fig_tx_omniledger}\includegraphics[width=.46\textwidth]{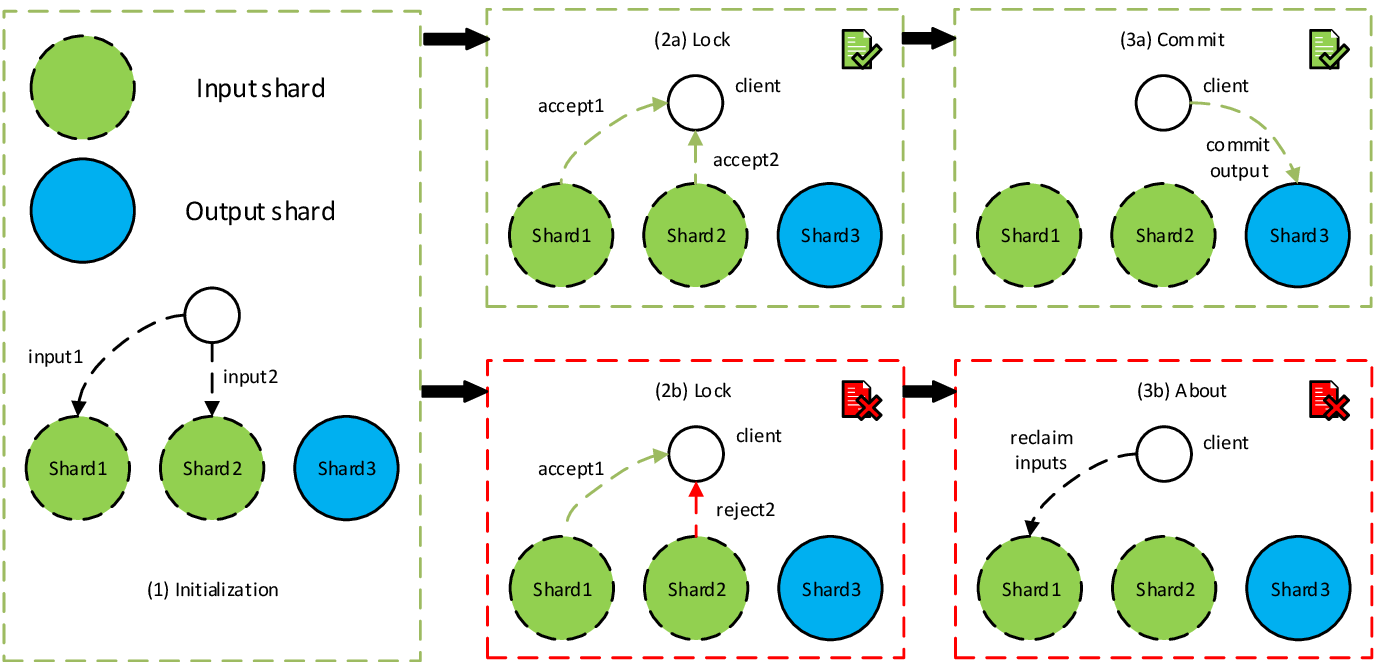}}
\subfigure[]{\label{fig_tx_rapidchain}\includegraphics[width=.46\textwidth]{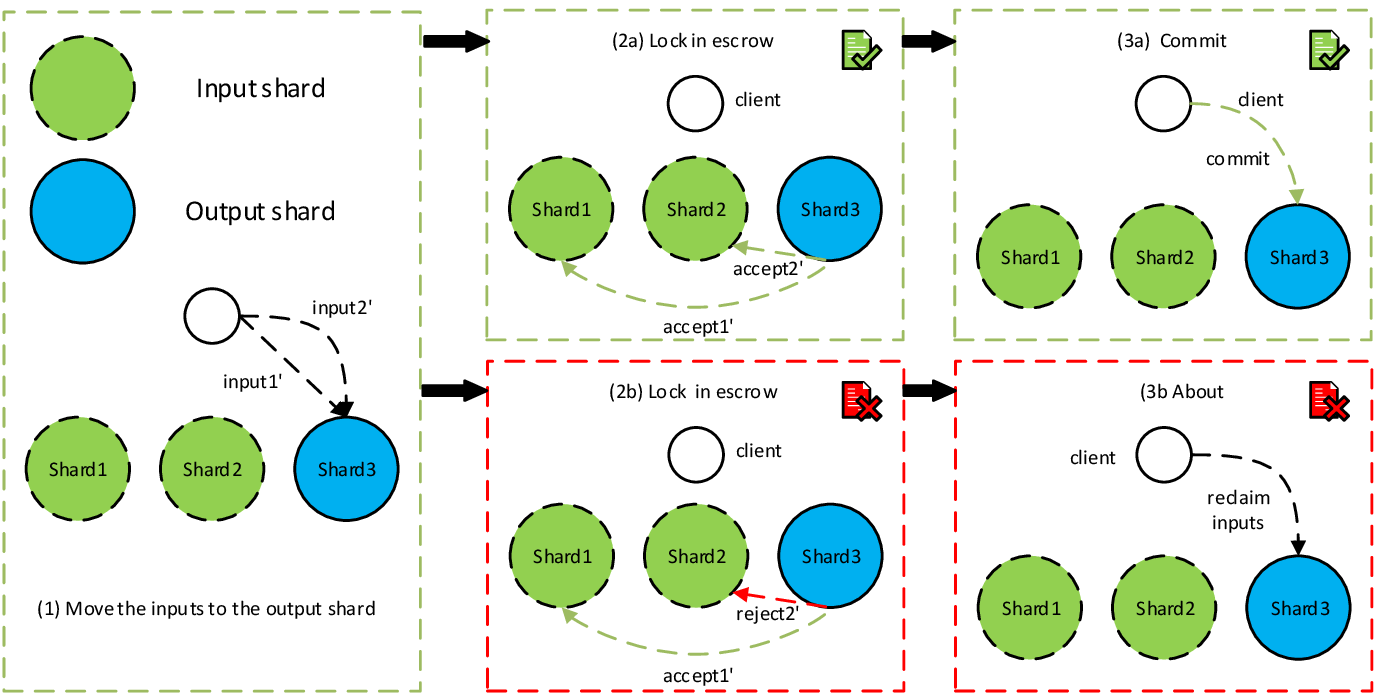}}
\caption{Atomic cross-shard transaction protocols in (a) OmniLedger~\cite{KokorisKogias:255586} and (b) RapidChain~\cite{cryptoeprint:2018:460}. In the two protocols, different parties are responsible for collecting input-shard approvals for committing transactions.}
\label{fig_cross_shard_tx}
\end{figure}

As in~\cite{ren2018scale, 10.1007/978-3-319-70972-7_22}, full sharding also partitions the storage of the blockchain state into multiple shards (e.g., local ledgers). Then, the full sharding protocols~\cite{KokorisKogias:255586, cryptoeprint:2018:460} are characterized by their ways of handling cross-shard transactions to guarantee atomic transaction commitment. In~\cite{KokorisKogias:255586}, OmniLedger uses UTXO to represent the client's balance state. Therefore, a cross-shard transaction is always associated with at least an input shard as well as an output shard (see Figure~\ref{fig_tx_omniledger}). OmniLedger adopts a lock-unlock-abort mechanism by requiring the input shard of a cross-shard transaction to ``lock'' the input first. Namely, the leader of the input shard has to provide a proof-of-acceptance in the form of Merkel proof before the corresponding transaction can be committed. If the transaction is found to be invalid, the input shard creates a proof-of-rejection in a similar form by using a designated bit to indicate an acceptance or rejection. Even with a proof-of-acceptance, the receiving client still cannot freely spend the UTXO. The receiver is required to send an unlock-to-commit transaction with that proof to the output-shard committee. Until the output shard validates this special transaction and includes it into the new block, the receiver is able to spend the UTXO of the original transaction.

In~\cite{cryptoeprint:2018:460}, RapidChain proposes a different approach of committing cross-shard transaction, which does not require a receiver to collect proofs from the input shards. Instead, for any input value of a transaction from a different shard, the output-shard leader is required to create a single-in-single-out transaction where the output is equal to the input of the original transaction. By doing so, the output committee tries to create a local record of the input and holds the input value in escrow. To confirm the escrow, the output-shard leader is responsible for sending this new transaction back to the input-shard committee for approval. After the input committee adds this transaction into its ledger, the output-shard leader will create a final transaction, with the UTXO of the escrow transaction being the input and the same outputs of the original transaction. After the output-shard committee adds the final transaction to its ledger, the transfer process is finished and the corresponding UTXO becomes spendable by the receivers. An illustrative comparison between the protocols of cross-shard transactions in OmniLedger and RapidChain is given by Figure~\ref{fig_cross_shard_tx}.

\subsection{Nonlinear Block Organization}
\begin{figure}[t]
\centering     
\includegraphics[width=.28\textwidth]{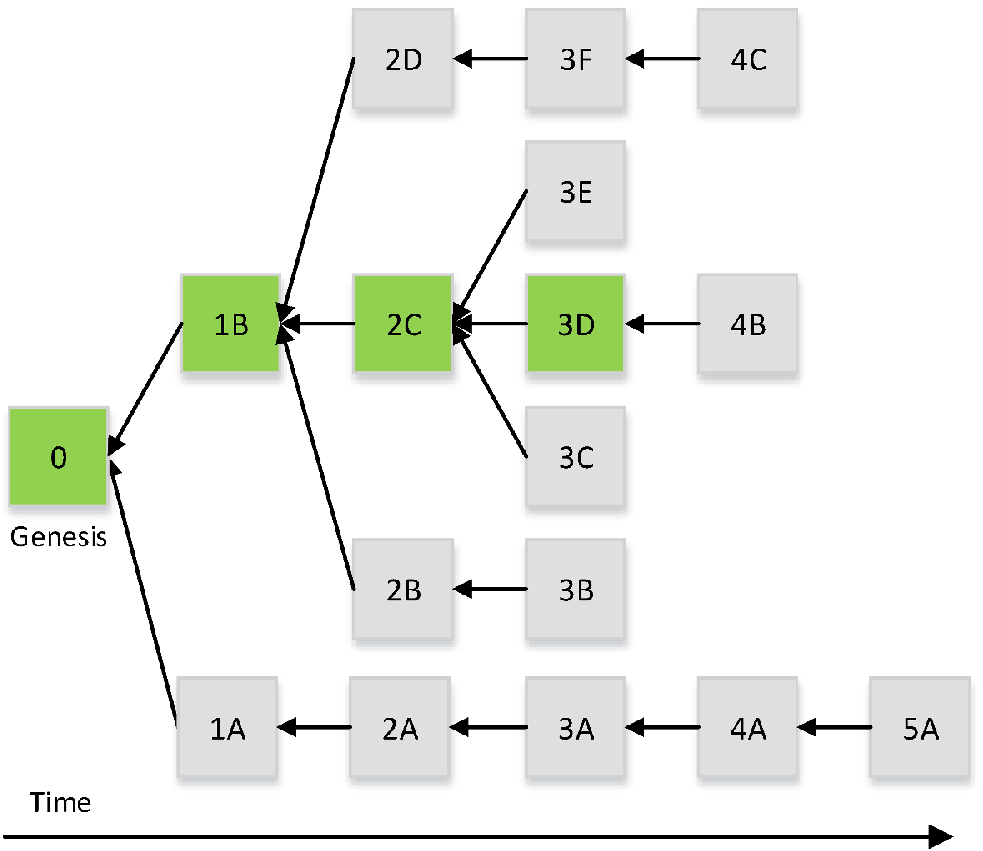}
\caption{A tree of blocks. Instead of choosing the longest chain (Blocks 1A to 5A), Block 1B with a subtree weight 11 is selected into the main chain. Consequently, Blocks 2C (with a subtree weight 5) and 3D (with a subtree weight 2) are selected into the main chain of the current view.}
\label{fig_spectre}
\end{figure}

Another approach aimed at improving the network throughput focuses on the design of transaction data organization. As we briefly introduce in Section~\ref{subsec_data_org}, instead of organizing block in a linear list, the approaches of nonlinear block organization are able to (partially) address the scalability problem by changing the mechanism of transaction validation in the consensus layer. The earliest scheme of nonlinear block organization can be found in~\cite{Sompolinsky2015} as the protocol of Greedy Heaviest-Observed Sub-Tree (GHOST). In a GHOST-based network, nodes store all the locally observed valid blocks and consequently maintain a tree of their respective forks. As an alternative to the longest-chain rule, GHOST extends the canonical chain of PoW-generated blocks by the block with the heaviest subtree, i.e., the subtree with the largest number of tree-nodes (see Figure~\ref{fig_spectre}). In~\cite{kiayias2016trees}, a unified security description of GHOST and the Nakamoto protocol is established by slightly modifying the $K$-consistency property in~\cite{pass2017analysis} (see also Section~\ref{sub_sec_POW}) into a new property of $K$-dominance, which measures the discrepancy in the weights between sibling subtrees. As pointed out in~\cite{Sompolinsky2015}, the rate of main-chain growth of GHOST is lower than that of the longest-chain rule when the block generation rate and the network delay are the same. However, since GHOST relaxes the block-generation constraint for the same level of security requirement against 51\% attacks, it is able to shorten safely the waiting time for block confirmation and thus has a limited ability of improving the network throughput.

A further step toward nonlinear block organization is proposed in~\cite{10.1007/978-3-662-47854-7_33}, where blocks are ordered in a DAG and each block is allowed to have multiple predecessors (cf. single parent block in GHOST~\cite{Sompolinsky2015}). Namely, the header of each block may contain more than one pointer to the precedent blocks to indicate the pairwise order. The DAG-based protocol in~\cite{10.1007/978-3-662-47854-7_33} also selects a main chain (cf. GHOST) of linear order from the DAG. To form such a linear order on the blocks at the current view, a node runs for each block a postorder traversal algorithm on the DAG and checks if the transactions in the current block are consistent with the visited one. Compared with the longest-chain rule or GHOST, the DAG-based rule of chain expansion allows the non-conflicting, off-chain blocks to be selectively included into the ledger view. For example, from the perspective of a main-chain block, its off-chain descendant blocks can still be included into the ledger as long as they are not far away from the main chain as both predecessors and descendants. Then, by including the discarded (i.e., off-chain) blocks, the proposed protocol possesses a limited ability of increasing the network throughput.

\begin{figure}[t]
\centering     
\includegraphics[width=.44\textwidth]{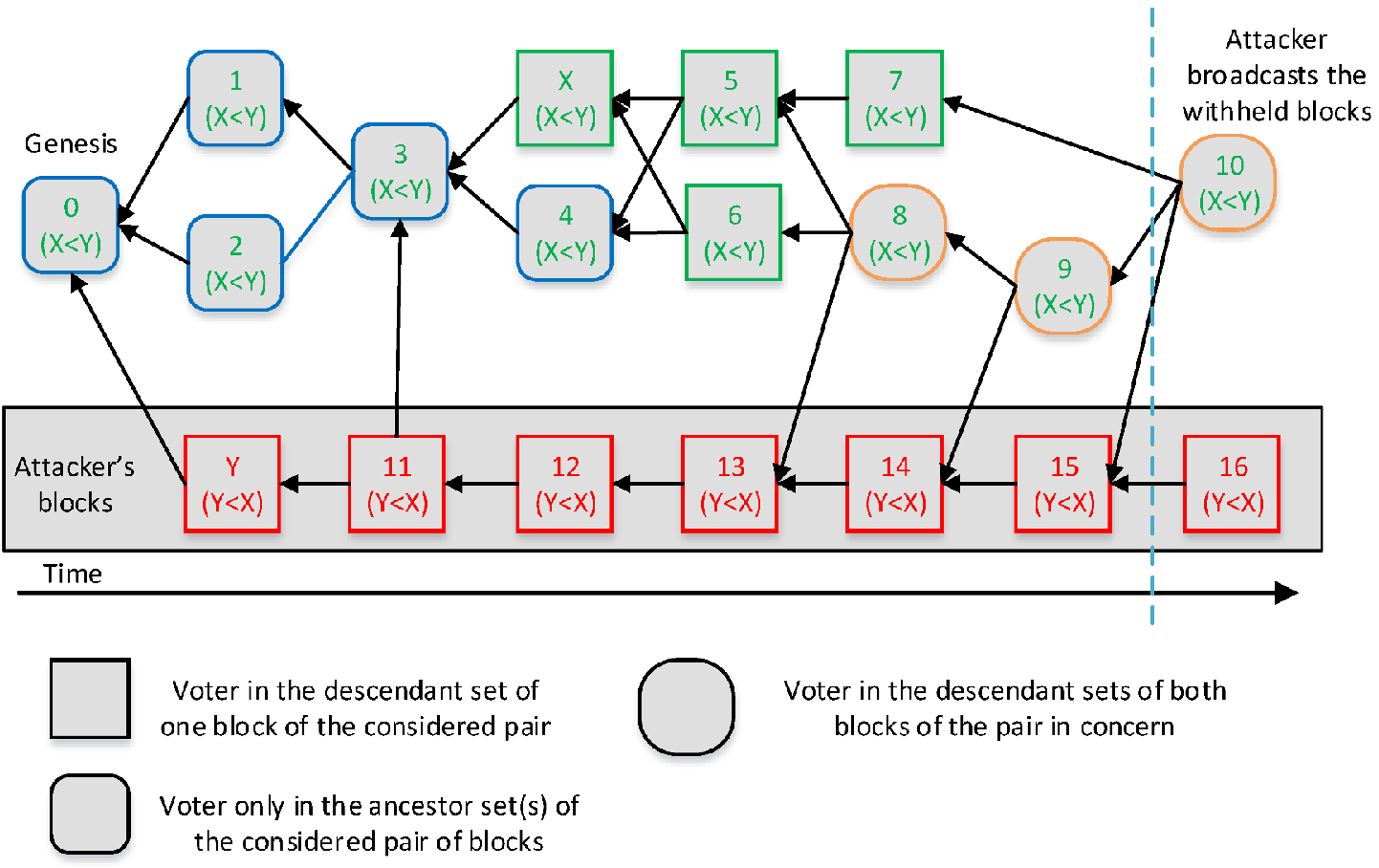}
\caption{An example of the virtual voting procedure on the order of blocks $X$ and $Y$ in a DAG with block withholding attacks. Blocks (voters) in the descendant set of $X$ will vote $X<Y$ (i.e., $X$ preceding $Y$) since they only see $X$. Blocks $0$-$4$ will vote $X<Y$ since they see more $X<Y$ votes in their sets of descendant block. Blocks $8$-$10$ which have both $X$ and $Y$ as the ancestors run an recursive query to their predecessor sets and use the majority voting results as their own votes.}
\label{fig_ghost}
\end{figure}

To further improve the network throughput, the protocol proposed in~\cite{10.1007/978-3-662-47854-7_33} is later extended to the protocol ``SPECTRE'' in~\cite{sompolinsky2016spectre}. SPECTRE relaxes the requirement on node synchronization, and allows blocks to concurrently grow on the ledger without specifying a main branch. To define the rule of ledger extension, SPECTRE introduces a virtual pairwise voting mechanism to determine the order of any pairwise blocks in the DAG. In brief, each block in the DAG contributes to the vote on the relative order of not only its preceding blocks but also its descendant blocks according to the topology of the DAG. Compared with the main chain-based rules, SPECTRE is shown to be robust to block-withholding attacks (cf.~\cite{7243747}). The reason is that with vote-based pairwise ordering, secret chains published by the attackers cannot win the votes by existing blocks from the honest nodes due to the lack of connections in the DAG (see Figure~\ref{fig_ghost}). Without undermining the network security, i.e., increasing the transaction reversal probability, SPECTRE admits faster commitment time as the block creation process is accelerated. By (\ref{eq_poisson_rate}), the more nodes in the network, the higher the expected block generation rate is given a fixed PoW difficulty. As indicated by~\cite{sompolinsky2016spectre}, for a target transaction-reversal probability, a known propagation delay and a fixed PoW-difficulty level, SPECTRE is able to increase the transaction throughput as the network size increases.

Based on the aforementioned protocols, a number of DAG-based schemes have been proposed with a variety of emphasis on different performance indeces. For example, Byteball~\cite{churyumov2016byteball} adopts the concept of main chain/tree (see also~\cite{Sompolinsky2015, kiayias2016trees, 10.1007/978-3-662-47854-7_33}) but uses authenticated witnessing nodes to determine the partial order of blocks at each user's view. Another DAG-based protocol, i.e., Conflux~\cite{li2018scaling} modifies GHOST by adding in each new block the reference pointers to all existing blocks without descendants at the current DAG view. Compared with~\cite{Sompolinsky2015, 10.1007/978-3-662-47854-7_33}, Conflux is claimed to provide 100\% utilization of the off-chain blocks and thus is able to improve the network scalability. Furthermore, a similar protocol to SPECTRE is proposed in~\cite{popov2016tangle,popov2017equilibria} as IoTA Tangle. The major difference of IoTA Tangle lies in that it discards the data structure of block as a package of transactions. Instead, it requires nodes to publish directly transactions onto the transaction DAG. A node is enforced by the protocol to approve/reference more than two transactions by linking their hash values in the header of its new transaction to expand the DAG. By doing so, the node expects to accumulate sufficient weight (cf. votes on the partial orders in SPECTRE) for this transaction from the future transactions\footnote{As in SPECTRE, an IoTA transaction (indirectly) approves/references an earlier transaction if it can reach that transaction via directed links.} by other nodes to finally confirm it. So far, complete theoretical proof of the liveness property of IoTA Tangle is still an open issue~\cite{popov2016tangle,popov2017equilibria}. However, the study in~\cite{popov2017equilibria} implies that, if self-interested nodes have the same capability of information acquisition and transaction generation as the other nodes, they will possibly reach an ``almost symmetric'' Nash equilibrium. Namely, they will be forced to cooperate with the network by choosing the default parent-selection strategy followed by the honest nodes.

\section{Emerging Applications and Research Issues of Blockchains with Public Consensus}\label{sub_sec_outlook}
In the previous sections, we have provided an in-depth survey on three main categories of permissionless consensus protocols for blockchain networks, namely, the Nakamoto-like protocol based on PoX puzzles, the virtual mining and hybrid protocols and the emerging open-access protocols emphasizing the scale-out performance.
On top of the consensus provided by these protocols, the blockchain is able to fully exert its functionalities such as smart contracts for a wide range of applications. In general, we can divide the studies on the emerging blockchain-based applications into two categories: the service provision atop the blockchain consensus layer and the consensus provision to existing blockchain frameworks. The former category of studies usually exploit special characteristics of blockchain networks, e.g., self-organization and data security, to guarantee target features in their respective services. In contrast, the latter emphasizes the P2P or decentralized characteristics of blockchain networks. Hence, most of them focus on rational nodes' strategies or the overlaid incentive mechanism design of resources allocation in the consensus process. In this section, we provide an extensive review on the properties of blockchain networks and the applications exerting mutual influence on each other. Meanwhile, a series of open research issues are also identified.

\subsection{General-Purpose Data Storage}\label{sub_sec_storage}
The Cambridge's 2017 annual blockchain benchmarking study identified that the majority of blockchains use cases are still dominated by the capital market sectors~\cite{hileman20172017}. Nevertheless, significant effort has recently been put into the study of using blockchains for storage of generic data, which aims at preserving the properties of data immutability and trackability in a decentralized environment. A naive approach is to ``piggy-back'' arbitrary data (e.g., non-transferable metadata) onto transactions in established public blockchains\cite{10.1007/978-3-319-70278-0_14}. For example, in the Bitcoin network, nodes can use the special script instruction OP\_RETURN to indicate that the transaction output is unspendable and expected to be removed from the UTXO. Then, the transaction is allowed to carry a limited length of arbitrary data onto the chain. Typical examples of directly storing metadata onto blockchains can be found in asset ownership registration, e.g., Namecoin\footnote{\url{https://namecoin.org}.}~\cite{kalodner2015empirical} as a blockchain-based namespace system. Note that the direct on-chain storage is limited by the message length and naturally requires full replication of each data object over the network. Then, this solution needs to be improved to lift the data-length constraint and reduce the synchronization cost. In~\cite{196208}, where a naming system is constructed on top of Namecoin, the data storage is decoupled from the block serialization (i.e., name registration) process. In order to achieve this, the authors of~\cite{196208} adopt a ``virtualchain'' to process registration/modification operations of names (e.g., domain names or IP addresses). Only the minimal metadata, i.e., the hashes of the name-payload pairs and state transitions are stored on the blockchain. The third party storage is connected by virtualchain to store the payload of arbitrary length with digital signatures from the data owner.

The same idea of decoupling the storage layer from the main blockchain can also be found in works such as~\cite{Shafagh:2017:TBA:3140649.3140656, 8404099, mcconaghy2016bigchaindb}. The studies in~\cite{Shafagh:2017:TBA:3140649.3140656, 8404099} focus on data storage and sharing for large-scale IoTs. Therein, two similar blockchain frameworks are proposed by introducing the off-chain storage. In brief, the data generated by IoT devices is stored in DHTs, and only the pointer to the DHT storage address needs to be published onto the blockchain. The DHT-based storage is provided by an off-chain layer of decentralized DHT nodes. Upon seeing that transactions of storing/accessing requests are confirmed by the blockchain, the DHT nodes are responsible for accordingly storing or sending the data from/to the legitimate IoT nodes. In~\cite{mcconaghy2016bigchaindb}, further discussion is provided regarding the issue of how to control the data replication factor in the network. Instead of using an off-chain storage layer, the design in~\cite{mcconaghy2016bigchaindb} compromises the property of decentralization in exchange for a stronger control of replication synchronization. In the proposed framework of blockchain-like database, i.e., BigchainDB, P2P communication protocols are replaced by the built-in broadcasting protocol, and a committee (i.e., federation) of voting nodes are designated for block validation and ordering. Such a permissioned design shares a certain level of similarity with the framework of HyperLedger~\cite{cachin2016architecture}. By doing so, it is possible for the federation nodes to control where to store a submitted transaction and flexibly determine the replication factor (i.e., the number of shards/replicas) per table in the underlying distributed database. Such design avoids the issue of full data replication over the network and makes it possible for constructing a large-scale, high-throughput database directly on a blockchain network.

\subsection{Access Control and Self-Organization}\label{sub_sec_access_control}
\begin{figure}[t]
\centering     
\includegraphics[width=.32\textwidth]{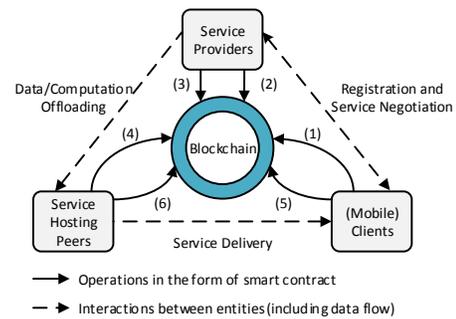}
\caption{A generic framework of using blockchains as system integrators for self-organization. The operation flow is realized as a sequence of smart contracts: (1) service registration/requests by the clients, (2) access/certificate granting by the providers, (3) requesting service hosting (e.g., auction for computation/storage offloading) by the providers, (4) peers answering (e.g., bidding for) the hosting requests, (5) delivery negotiation between hosting peers and clients and (6) service completion with proofs of delivery.}
\label{fig_blockchain_intermedeiate}
\end{figure}
The most popular design approach sees blockchains as enabling technologies for implementing accountable and secure services in a decentralized fashion. In other words, blockchains are utilized as a decentralized intermediary for channeling/accounting services upon demands as well as for guaranteeing data security and confidentiality. In Figure~\ref{fig_blockchain_intermedeiate}, we describe a generic framework of decentralized service provision built upon blockchains. The most prominent feature of this framework lies in that the interactions between different entities in the system are all tunneled autonomously in the form of smart contracts. Such a framework has been adopted by a wide range of service provision systems including P2P file sharing based on InterPlanetary File System\footnote{\url{https://github.com/ipfs/ipfs}.} (IPFS)~\cite{Filecoin}, decentralized content delivery~\cite{wenbo, goyal2018secure}, access control in telecommunication networks~\cite{Identity2017blockchain, 8315203} and various missions for access and permission management, e.g., in IoTs~\cite{nikouei2018real} and clouds~\cite{8260822}. For different task requirements, this application framework can be expanded by including additional entities, e.g., third-party auditors~\cite{7973733}, as well as new operations, e.g., Hierarchical Identity Based Encryption (HIBE, see also~\cite{lewko2011unbounded})~\cite{fotiou2016decentralized}. To provide a better idea on how this emerging framework can be shaped in recent studies, we categorize the blockchain-based proposals for self-organization according to the areas or context that they are applied in.

\subsubsection{Access Control in Wireless Networks}\label{sub_sub_sec_WN}
In~\cite{Identity2017blockchain}, the authors propose to use blockchains for providing Identity and Credibility Service (ICS) in cloud-centric Cognitive Radio (CR) networks. The CR users utilize their pseudonymous identities on the blockchain to negotiate with the network operator, i.e., the spectrum owner, for granting opportunistic access and settling payment. According to~\cite{Identity2017blockchain}, the ICS can be provided by either the blockchain itself or a third-party entity registered on-chain, and the network access negotiation is automated by smart contracts. Meanwhile, it is pointed out in~\cite{Identity2017blockchain} that
the blockchain's consensus mechanism can be employed for coordinating spectrum sensing among the distributed CR users. However, it is not clear how the CR-user consensus can be achieved on top of the ledger consensus as with the classical methods~\cite{5464224} in CR networks.

In another study~\cite{raju2017design}, the same authors propose to use a permissioned blockchain to handle the network access exchange, i.e., the spectrum handoffs. The CR users and their base station controller submit the information of spectrum and network utilization as metadata onto the blockchain. Then, the CR network responds by updating the smart contracts and publishing the new access prices and number of network access units allocated to each CR onto the blockchain for execution. A similar design with more technical details can be found in~\cite{8269834}. Therein, a blockchain based on the Nakamoto protocol with its embedded tokens and smart contract layer is adopted as a spectrum auction platform. More specifically, multiple primary users as providers sell their unused bands at a certain price with smart contracts and allocate them to responding CR users when the contracts are executed upon certain conditions. It is claimed in~\cite{8269834} that the blockchain-based spectrum allocation outperforms the conventional medium-access protocols such as Aloha. However, technical details are missing about how the issue of high transaction latency is addressed to satisfy the CR network's constraint due to the timescale of small-scale fading in wireless channels.

Blockchains are also introduced into vehicular ad-hoc networks (VANETs) to address the issues of network volatility due to high mobility. For Vehicles-to-Infrastructure (V2I) communication, the study in~\cite{lei2017blockchain} uses the Nakamoto-based blockchain as a secure key-delivery channel to handle the access of a moving vehicle to groups of Road Side Units (RSUs) in different regions. By encapsulating the key information in a blockchain transaction, the security manager of one region is responsible for issuing the transactions to that of the new region as well as mining the new blocks onto the blockchain. Comparatively, the study in~\cite{8358773} focuses more on the ad-hoc nature of VANETs and employs the blockchain to collect the trustworthiness rating on messages sent to each other by the peer vehicles. The RSUs do not only work as the consensus nodes in the blockchain network but also work as the decentralized storage hosting peers of the trust rating data (cf. Figure~\ref{fig_blockchain_intermedeiate}). It is worth noting that in~\cite{8358773} the transactions carrying vehicle reports are essentially unspendable. The RSUs employ weighted average to the rating scores to estimate the quality of the received reports. Then, they use the estimation results as the difficulty parameter for PoW-based mining in a similar manner of the Peercoin-like protocols (see also Section~\ref{sub_sec_POS}).

In the existing studies on blockchains-based network access control, the study in~\cite{8470085} is among the few to explicitly address the issue of high signaling latency over the blockchain due to the adoption of Nakamoto protocols. In~\cite{8470085}, the process of authentication transfer for User Equipments (UEs) in a 5G ultra dense network is handled by a blockchain in a similar way as in~\cite{raju2017design}. Instead of delegating the transaction/contract execution process to a dedicated overlay blockchain, it is proposed in~\cite{8470085} that the Access Points (APs) use the PBFT protocol within a dynamic consensus committee to handle the requests of authentication by UEs in the form of transactions or smart contracts. In order to implement the PBFT protocol, a local server center is introduced as the primary peer (i.e., leader) of the committee. Nevertheless, we note that any non-leader consistency protocol can be adopted in this framework to preserve the property of complete decentralization (see also Section~\ref{sub_sec_sharding}). According to~\cite{8470085}, the PBFT-based blockchain is able to keep the transaction delay around 100ms. Compared with the standard Nakamoto protocols, it is more practical to deploy network control mechanisms over PBFT-based blockchains for delay-critical tasks such as access hand-over. However, how to find a balance between the required levels of latency and decentralization (e.g., with hybrid consensus protocols) still remains an open question.

\subsubsection{Self-Organization and Security Enhancement under Various Network Architectures}
Apart from network access control, blockchains have also been applied to various scenarios as a decentralized platform for self-organization. As briefly shown in Section~\ref{sub_sub_sec_WN}, blockchains can also be used for security enhancement with its embedded cryptographic functionalities. Typical examples for the former applications can be found in proactive caching and Content Delivery Networks (CDNs)~\cite{herbaut2017model, wenbo, goyal2018secure}. In~\cite{goyal2018secure}, a decentralized CDN platform is established with the help of blockchains among the three parties of content providers, content serving peers and clients (cf. Figure~\ref{fig_blockchain_intermedeiate}). With smart contracts, the content providers offload the tasks of content delivery to multiple content serving peers. It is suggested in~\cite{goyal2018secure} that the content providers use smart contract prices to control the file placement on multiple serving peers according to the demand frequency and achievable QoS at the peers. Furthermore, the work in~\cite{wenbo} mathematically formulates the pricing-response interaction between the providers and the serving peers as a potential game~\cite[Chapter 3.4]{han2012game}. Then, it designs a series of smart contracts for automatically matching the peers to the providers under the same CDN framework. A modified PoS protocol is subsequently proposed to incentivize the serving peers to work as the consensus nodes of the blockchain without consuming significant computational power.

In~\cite{herbaut2017model}, the authors design a blockchain-based brokering platform for video delivery in a user-centric CDN ecosystem. The proposed platform is built upon three independent blockchains for content brokering, delivery monitoring and delivery provisioning, respectively. The content broking blockchain handles the content requests and matches the clients' requests to the providers' offers in a series of smart contracts among the three parties. The delivery monitoring blockchain records proofs of delivery and finalizes the payment and refund between the providers and the clients. The delivery provisioning blockchain provides smart contacts for content dissemination between the providers and the serving peers. In such a framework, the decentralized entities in the CDN treat the blockchain as a ready-to-use service offered by a third party. Therefore, any form of blockchains (e.g., the permissioned HyperLedger) can be employed as long as the requirement of transaction throughput and latency is met.

In various applications of edge/fog/cloud computing, more and more attempts are also found to use blockchains for providing services such as trusted auditing and secured data delivery in addition to autonomous brokering. In~\cite{7973733}, the blockchain is used as a tamper-proof provenance database on the cloud server side to record the history of the creation and operations performed on a cloud data object. By adopting a public blockchain, any node in the blockchain network is able to perform data auditing. By using pseudonymous identities on blockchains, the proposed auditing mechanism reduces the probability that auditors can correlate the real identity of a specific user with the operations. In other works such as~\cite{8053750, nikouei2018real}, the blockchain is introduced into the three-layer paradigm of edge-fog-cloud computing. In~\cite{nikouei2018real}, the blockchain is used as a connector to provide encrypted channel by using the public key functionality for data delivery from the edge devices to the fog and cloud. More specifically, the study in~\cite{nikouei2018real} considers a smart video surveillance network, where the preprocessing tasks such as object tracking are handled at the edge devices, and the more sophisticated tasks of data aggregation and decision making are performed in the fog/cloud based on the data filtered at the edge. To prevent malicious modification on video frames in the untrusted fog layer, the cloud layer deploys smart contracts on the blockchain to provide an indexing service and generate unique index for every video frame with transactions published onto the blockchain. The work in~\cite{8053750} adopts the same data-processing flow from the edge to the cloud as in~\cite{nikouei2018real}. In contrast to~\cite{nikouei2018real}, the blockchain is used to provide automatic matching between the data-service requests and the providers in the cloud's service provider pool. In this sense, the blockchain is again used to provide the broking service as in~\cite{goyal2018secure, wenbo}.

\subsubsection{Trusted Broking Services in Cyber-Physical Systems}
In the context of crowdsourcing (e.g., crowdsourcing of mobile sensors, a.k.a., crowdsensing), permissionless blockchains are also found to be especially appropriate for providing non-manipulable brokering services between clients (i.e., task requesters) and service providers (i.e., crowdsourcing workers). In~\cite{licrowdbc}, a purely decentralized crowdsourcing system for general purpose is proposed following the paradigm described by Figure~\ref{fig_blockchain_intermedeiate}. In the proposed framework, the procedures of identity registration, task/receiving, reputation rating and reward assignment are all automated in the form of smart contracts. Following the approaches described in Section~\ref{sub_sec_storage}, the blockchain network delegates the data storage to an independent storage layer and only keeps the metadata on-chain. Similar blockchain-based frameworks are also adopted for crowdsensing in recent studies such as~\cite{Feng1812:Competitive, 8306424}, where additional functionalities are adopted in the blockchain networks to address different performance requirement such as throughput scalability~\cite{Feng1812:Competitive} and anonymity enhancement~\cite{8306424}.

In the context of IoTs, blockchain-based infrastructure is also envisioned as a promising alternative of the centralized one for data management, trading automation and privacy protection. In~\cite{aitzhan2016security}, the authors introduce the micro-payment channels (see also Section~\ref{sub_sec_off_chain}) based on a Bitcoin-like blockchain to conduct energy trading in a decentralized smart grid without relying on trusted third parties. In~\cite{kang2017enabling}, a P2P surplus-energy trading mesh of the plug-in hybrid electric vehicles is built on a Nakamoto protocol-based blockchain. In the proposed framework, a number of authorized nodes are responsible for processing and recording the transactions and an iterative double auction mechanism is deployed based on the transactions published on the blockchain. This framework of blockchains as a P2P trading mediator is also adopted in~\cite{8234700, 8457186}, where the PBFT protocol is used to replace the Nakamoto protocol and form a consortium blockchain. Furthermore, the mathematical tool of contract theory (see~\cite{7856876} for more details) is adopted to determine the optimal prices and requested utility in the relevant smart contracts.

\subsection{Consensus Provision and Computation Offloading under Nakamoto Protocols}
In contrast to the studies that we review in Sections~\ref{sub_sec_storage} and~\ref{sub_sec_access_control}, another line of works focus on (decentralized) resource allocation for consensus provision in the Nakamoto-based blockchain networks. In other words, these studies view the consensus in blockchain networks of a given protocol as the goal to be achieved instead of a ready-to-use service. Recall that the Nakamoto protocols require consumption of certain resources in the PoW-like puzzle solution competition for new block proposing (see also Section~\ref{sec_consensus}). With this property in mind, a plethora of works, e.g.,~\cite{xiong2017optimal, jiao2017social, luong2017optimal, xiong2017mobile, 8440765, suankaewmanee2017performance}, are devoted to the studies of resource allocation in the block mining process in exchange for monetary rewards (i.e., mining reward in tokens) offered by the blockchain. In~\cite{xiong2017optimal, 8440765, xiong2017mobile}, a scenario of deploying blockchains on the mobile edge devices is considered. Due to the intensive resource consumption for PoW solution, it is difficult to directly migrate blockchain networks to the mobile environment~\cite{xiong2017mobile}. Therefore, the computation offloading schemes are proposed in these studies by either formulating the problems in a nonlinear/binary programming framework~\cite{8440765} or as a hierarchical (i.e., Stackelberg) game~\cite{xiong2017optimal, xiong2017mobile}.

We use~\cite{xiong2017optimal} as an example to explain how the PoW-work offloading process can be formulated as a conventional optimization or game theoretic problem. To offload the tasks of PoW-solution searching from mobile devices to the edge/fog/cloud, a series of factors including transaction transmission delay and blockchain-forking probability need to be considered when constructing the utility model of the mobile node at the edge. Considering that the computation providers at the edge/fog are able to control the price of the offered computational resource, the offloading process is modeled in~\cite{xiong2017optimal} as a two-stage Stackelberg game. In brief, the cloud/fog providers act as the leader to set the resource price, and the edge devices acts as the follower to determine the share of resource to purchase for offloading the mining tasks. According to the various assumptions about the offloading scenarios (e.g., multi-leader vs. single leaders), different approaches such as nonlinear optimization formulation or best response-based equilibrium searching are applied to each layer's sub-problem in the manner of backward induction~\cite[Chapter 3.4.2]{han2012game}. Extending from the basic scenarios in~\cite{xiong2017optimal, 8440765}, various tools of mechanism design, e.g., auctions~\cite{jiao2017social, luong2017optimal}, can be further applied into the similar offloading problems for resource allocation in the blockchain consensus process.

\subsection{Some Open Issues and Potential Directions}
In the existing literature on blockchains, a number of open issues have been discussed regarding the non-consensus layers in blockchains, e.g., the issues of security and privacy~\cite{conti2017survey} and quantitative analysis of smart contract performance~\cite{tsankov2018securify}. In the following, we discuss issues and emerging research directions that have not been covered in the surveyed works.

\subsubsection{Cost of Decentralization}
The properties of permissionless blockchains such as trustlessness and self-organization have been widely recognized as the advantage over the conventional ledger/brokering systems. However, decentralization with blockchain networks is not ``at no cost''. As we have partly discussed in Section~\ref{sec_consensus_III}, even the scalable consensus protocols do not completely solve the problem of balancing between the requirement of security and resource efficiency. For instance, hwo to
adaptively control the replication factor in shards still remains an open issue.

Furthermore, consider that historical data such as spent transactions become huge as the blockchain grows. With the current design of append-only chains, it seems inevitable for ordinary nodes to eventually run out of storage and for the blockchain network to be controlled by a few powerful nodes. Then, it is plausible to seek an approach for ``pruning'' the blockchain data without undermining its immutability. Although hard forks such as SegWit~\cite{segwit2015} can be considered a manual pruning process, it is better expected that the out-of-date blocks ``have the right to be forgotten''~\cite{7871020}. Unfortunately, except a handful of experimental proposal~\cite{8038518, brucemini}, the issues of data pruning, e.g., how to delete obsolete transactions and migrate UTXOs buried in the chain, also remains an open issue.

\subsubsection{Support for Secure Big-Data Computation}
In the existing research, privacy concerns for blockchains are mostly placed on the levels of identity registration and encrypted data delivery (see Section~\ref{sub_sec_access_control}). With more and more demands for big-data processing in various fields~\cite{7473815, 8373692}, the question arises regarding whether it is also possible to provide on- or off-blockchain support for secure Multi-Party Computation (MPC). For example, hospitals may want to learn patterns for diagnosis by using the private electronic medical records from the patients without seeing the raw data. In such a scenario, the existing privacy policies offered by blockchains (e.g., access authentication) turn out to be insufficient. This issue is partially touched in~\cite{kim2018device} for mobile federated learning, where each node connected to the blockchain trains on the same structure of deep neural network with the local data. Then, they only exchange the locally trained model for global model aggregation~\cite{konecny2016federated}. Note that in~\cite{kim2018device} the blockchain is merely used to conduct a convoy of the locally trained parameters as in~\cite{nikouei2018real}. Following such design arises a natural question, namely, how can we directly offer general-purpose, privacy-preserving MPC on-chain (e.g., in blocking mining work) or off-chain with decentralized providers (cf. Figure~\ref{fig_blockchain_intermedeiate})?

The question above generally remains unaddressed, and only a few works~\cite{zyskind2015enigma, 8360354} can be found in the literature with limited strength for specific-purpose MPC provision. These works are based on the framework of cryptographic MPC techniques and allow mutually trustless parties to compute
a joint function directly on their encrypted inputs to obtain the right outcome. In~\cite{zyskind2015enigma}, the multi-parties store their public-key-encrypted data on an off-chain storage plain as in~\cite{Shafagh:2017:TBA:3140649.3140656}, while in~\cite{8360354} the encrypted data is stored directly on a permissioned blockchain (e.g., HyperLedger). However, due to the quadratic message complexity of the existing MPC protocols~\cite{zyskind2015enigma}, only a small number of computation parties can be supported on-chain~\cite{8360354}. Moreover, only a limited number of mathematical operations (e.g., polynomial functions) are supported by the protocols, and the MPC-based blockchain framework is still far from matured.

\section{Conclusions}
\label{sec_conclusion}
In this paper, we have provided a comprehensive survey on the recent development of blockchain technologies, with a specific emphasis on the designing methodologies and related studies of permissionless, distributed consensus protocols. We have provided in the survey a succinct overview of the implementation stacks for blockchain networks, from where we started our in-depth investigation into the design of consensus protocols and their impact on the emerging applications of blockchain networks. We have examined the influence of the blockchain consensus protocols from the perspective of three different interested parties, namely, the deployers of blockchain networks, the consensus participants (i.e., the consensus nodes) in the blockchain networks and the users of blockchain networks.

We have provided a thorough review of the blockchain consensus protocols including BFT-based protocols, Nakamoto protocols, virtual mining and hybrid protocols, for which we highlighted the link of permissionless consensus protocols to the traditional Byzantine agreement protocols and their distinctive characteristics. We have also highlighted the necessity of incentive compatibility in the protocol design, especially for the permissionless blockchain networks. We have provided an extensive survey on the studies regarding the incentive mechanism embedded in the blockchain protocols. From a game-theoretic perspective, we have also investigated their influence on the strategy adoption of the consensus participants in the blockchain networks.

Based on our comprehensive survey of the protocol design and the consequent influence of the blockchain networks, we have provided an outlook on the emerging applications of blockchain networks in different areas. Our focus has been put upon how traditional problems, especially in the areas of telecommunication networks, can be reshaped with the introduction of blockchain networks. This survey is expected to serve as an efficient guideline for further understanding about blockchain consensus mechanisms and for exploring potential research directions that may lead to exciting outcomes in related areas.

\bibliography{Reference}
\bibliographystyle{IEEEtran}

\end{document}